\documentclass{article}
\usepackage[version=4]{mhchem} 
\usepackage{xr}
\usepackage{graphicx}
\usepackage{siunitx}
\graphicspath{ {./Figures/} }
\usepackage[section]{placeins} 

\usepackage{float} 
\usepackage[sorting=none,style=chem-angew,autocite = superscript]{biblatex} 
\usepackage[singlelinecheck=false]{caption} 
\usepackage{float}
\floatstyle{plaintop}
\restylefloat{table}
\usepackage{multicol}
\usepackage{geometry}
\graphicspath{ {Figures/} }
\usepackage[hidelinks]{hyperref}

\newcommand{\eqlink}[1]{\hyperref[#1]{Eq.~(\ref*{#1})}}
\usepackage{amsmath}
\usepackage{amssymb}
\usepackage{booktabs}
\usepackage{multirow}
\usepackage{listings}
\makeatletter
\newcommand*{\addFileDependency}[1]{
  \typeout{(#1)}
  \@addtofilelist{#1}
  \IfFileExists{#1}{}{\typeout{No file #1.}}
}
\makeatother
\newcommand*{\myexternaldocument}[1]{
    \externaldocument{#1}
    \addFileDependency{#1.tex}
    \addFileDependency{#1.aux}
}
\IfFileExists{SI.aux}{\myexternaldocument{SI}}{%
  \IfFileExists{SI-labels.tex}{
\newlabel{fig:si_train_a100}{{S13}{S20}{Parallel training on 1--16 A100 GPUs, batch 16 structures per rank, on the cleaned MPtrj database. (a) Speedup $X(P)/X(1)$ against rank count, with the ideal line. (b) Parallel efficiency. (c) Global throughput in structures per second. (d) Where one training step goes at $P=16$: data loading and neighbour-list construction, host-to-device transfer, forward and backward, the gradient all-reduce, and gradient clipping plus the optimizer update. The all-reduce share is what separates the curves in (b)}{}{}}
\newlabel{sec:si_agent}{{S2.9}{S5}{Agent interface}{}{}}
\newlabel{sec:si_config}{{S3}{S6}{Default trainer configuration}{}{}}
\newlabel{sec:si_impl}{{S2}{S3}{Framework implementation details}{}{}}
\newlabel{sec:si_ptables}{{S16}{S23}{Appendix: Periodic-table statistics for each database}{}{}}
\newlabel{tab:hparams}{{S2}{S8}{Settings of the training of each architecture. $\ell _\mathrm {max}$ is the highest spherical harmonic degree the features carry and $m_\mathrm {max}$ the highest order kept at each degree; NequIP, MACE and Allegro build full $O(3)$ irreps, so $m_\mathrm {max}=\ell _\mathrm {max}$, whereas the Equiformers truncate below $\ell _\mathrm {max}$ to cut the cost of the SO(2) convolution. $^\ast $Equivalent values: PaiNN exposes neither setting, but its Cartesian scalars and vectors are equivariant to $\ell \leq 1$ with all orders present}{table.2}{}}
\newlabel{tab:si_cost}{{S6}{S15}{Inference cost and accuracy of different architectures on one Tesla V100-PCIE-32GB, float32, 5.5\,\AA {} cutoff for energy+forces on bulk fcc Pt supercells}{table.6}{}}
\newlabel{tab:si_finetune}{{S4}{S13}{Data efficiency of fine-tuning on the AdsTraj dataset}{}{}}
\newlabel{tab:si_lmp_scaling}{{S11}{S21}{The performance of the IANN LAMMPS interface of \texttt {pair\_style iann/multi\_gpu} on NVIDIA A100-80GB GPUs in Figure~7 in the manuscript is tabulated here. The step time per step on 1 to 8 GPUs, the corresponding efficiencies derived from them, and the $Y_8$ throughput at eight GPUs are listed at each system size. For 8 GPUs, four GPUs sit in one node (NVLink), and eight span two (Slingshot). The last column is the exact serial run of LAMMPS interface with \texttt {pair\_style iann} on a single GPU}{table.11}{}}
\newlabel{tab:si_train_a100}{{S9}{S19}{Training efficiency of different architectures on MPtrj on 1 to 16 NVIDIA A100-SXM4-80GB GPUs. $B$ is each run's own per-rank batch size, so the global batch is $P \times B$ and the work per GPU is fixed; $X$ is the global throughput in structures per second and the efficiency is $X(P)/(P\,X(1))$. The all-reduce column is the part of the gradient reduction that is not hidden behind the backward pass}{}{}}
\newlabel{tab:si_train_sweep}{{S10}{S19}{Training global throughput $X(P)$ in structures per second, and parallel efficiency $X(P)/(P\,X(1))$, at every allocation from 1 to 16~A100 GPUs. $B$ is the per-rank batch size, so the global batch at each column is $P\times B$. Every world size times an identical set of structures}{}{}}
}{%
    \GenericWarning{}{WARNING: no SI.aux and no SI-labels.tex -- SI refs will be ??}}}
\DeclareSIUnit\bar{bar}
\DeclareSIUnit\angstrom{\text {Å}}
\usepackage{soul}
\usepackage{color}
\usepackage{authblk}

\definecolor{codekw}{rgb}{0.13,0.29,0.53}
\definecolor{codestr}{rgb}{0.60,0.20,0.10}
\definecolor{codecom}{rgb}{0.35,0.45,0.35}
\definecolor{codebg}{rgb}{0.975,0.975,0.975}
\definecolor{codeframe}{rgb}{0.75,0.75,0.75}
\lstdefinestyle{iannpy}{
  language=Python,
  backgroundcolor=\color{codebg},
  basicstyle=\ttfamily\scriptsize,
  keywordstyle=\color{codekw}\bfseries,
  stringstyle=\color{codestr},
  commentstyle=\color{codecom}\itshape,
  showstringspaces=false,
  breaklines=true,
  frame=single,
  rulecolor=\color{codeframe},
  framesep=4pt,
  columns=fullflexible,
  literate={-}{-}1,
}
\lstdefinestyle{iannsh}{
  language=bash,
  backgroundcolor=\color{codebg},
  basicstyle=\ttfamily\scriptsize,
  keywordstyle=\color{codekw},
  commentstyle=\color{codecom}\itshape,
  showstringspaces=false,
  breaklines=true,
  frame=single,
  rulecolor=\color{codeframe},
  framesep=4pt,
  columns=fullflexible,
}
\title{IANN: A Unified Framework for Equivariant InterAtomic Neural Networks with User-Friendly Implementation}

\author[1,2,4]{{Changzhi Ai}\thanks{Corresponding author: changzhi@stanford.edu}}
\author[3]{Xueping Qin}
\author[4]{Dimosthenis Sokaras}
\author[2]{Frank Abild-Pedersen}
\author[2]{Johannes Voss}

\affil[1]{Department of Chemical Engineering, Stanford University, Stanford, California 94305, United States}
\affil[2]{SUNCAT Center for Interface Science and Catalysis, SLAC National Accelerator Laboratory, Menlo Park, California 94025, United States}
\affil[3]{Center for Nanoscale Materials, Argonne National Laboratory, Lemont, Illinois, United States}
\affil[4]{Stanford Synchrotron Radiation Lightsource, SLAC National Accelerator Laboratory, Menlo Park, California 94025, United States}

\date{}

\begin{document}

\maketitle

\begin{abstract}

Machine-learning interatomic potentials (MLIPs) can now achieve near-density functional theory (DFT) accuracy in predicting energies and forces at a fraction of the computational cost, but the practical benefit is limited by software fragmentation: each state-of-the-art architecture is distributed as its own codebase, with its own data format, configuration system, training loop, checkpoint layout and molecular-dynamics interface. Comparing two architectures on the same dataset, moving a trained model from a training script into a production simulation, or extending either one with a new architecture or component therefore requires substantial re-engineering. Here, we introduce the IANN (InterAtomic Neural Network) framework, an open-source framework built on PyTorch that reimplements a range of equivariant graph neural network architectures — including PaiNN, NequIP, MACE, Allegro, EquiformerV2, EquiformerV3, and UMA — within a unified framework for data handling, model training, and deployment. Every model consumes and returns the same data structure, so the choice of certain architecture is a one-word change in a configuration dictionary rather than a change of code base. Forces, stresses, and virials are obtained analytically by differentiating the predicted energy with respect to the edge vectors of the atomic graph, which is exact under periodic boundary conditions and shared by all architectures. Distributed data-parallel training is configured automatically from the SLURM, PBS, or MPI environment, so the same training script runs unchanged in various computing environments, such as a multi-GPU workstation and a multi-GPU/node supercomputer. Trained model checkpoints are self-describing and can be directly used as ASE calculators, combined into ensembles that report energies, forces, and their associated uncertainties, or compiled into TorchScript and executed within LAMMPS through bundled C\texttt{++} pair styles supporting single- and multi-GPU inference and on-the-fly ensemble variance estimation. We demonstrate the framework by training different architectures on a common dataset with millions of structures, enabling meaningful comparisons through consistent training, graph construction, and evaluation procedures. We further provide PaiNN foundation models trained on 41.9 million curated DFT structures from the open-source MPtrj, Alexandria, OMat24, MatPES, OC20, OC22, OC25, MPr$^2$SCAN, and MatPESr$^2$SCAN datasets, spanning three exchange-correlation functionals (PBE, RPBE, and r$^2$SCAN). Additionally, IANN provides an agent interface to facilitate integration with autonomous agents. As open-source, MIT-licensed  software, IANN provides an accessible and unified platform for developing, benchmarking, and deploying advanced machine-learning interatomic potentials, accelerating their broader application across computational catalysis, energy storage and battery design, molecular therapeutics, and broader materials discovery.

\end{abstract}

\section{Introduction}

Atomistic simulations based on density functional theory (DFT) have become a cornerstone of modern computational materials science and molecular science. By establishing fundamental relationships between atomic structures and physical properties from first principles, DFT enables the prediction and characterization of materials and molecules before their experimental synthesis, transforming materials discovery and design from empirical trial-and-error approaches into rational, knowledge-driven processes. In catalysis, DFT enables the calculations of adsorption energies, reaction free energies, and activation barriers for elementary reaction steps, including the potential dependence in electrocatalytic systems, where the stability, selectivity and activity of the catalyst can be evaluated through thermodynamic and kinetic analyses, including the phase and Pourbaix diagrams, scaling relationships, and volcano plots.\cite{Norskov2011,Medford2015,Ai2022,Ai2022_2} In battery materials, formation energies, ion-migration barriers, and interfacial decomposition energies encode key factors governing capacity, rate capability, and cycle life.\cite{Leung2012,CamachoForero2020,Qin2019} In drug discovery, DFT provides insights into conformational energetics, protonation states, and molecular interactions that govern the binding affinity of a candidate molecule for its biological target.\cite{Raha2007} Despite its broad applicability, DFT is fundamentally bottlenecked by its high computational cost. This computational burden makes a wide range of applications, including geometry optimization of complex systems, high-throughput screening, large-scale simulations, long-timescale molecular dynamics (MD), and extensive finite-temperature sampling, prohibitively expensive for routine use.

Machine-learning interatomic potentials (MLIPs) were developed to address this bottleneck by learning the potential-energy surface from a finite set of reference calculations and then evaluating it at a cost that is orders of magnitude lower.\cite{Behler2007,Bartok2010,Deringer2019} Early high-dimensional neural network potentials described each atomic environment through hand-crafted invariant descriptors,\cite{Behler2007} and were then followed by learned invariant representations built on message passing.\cite{Gilmer2017,Schutt2018} The current generation is explicitly equivariant: internal features transform under rotation in the same way as the physical tensors they represent, which markedly improves data efficiency and accuracy. Specifically, PaiNN extends scalar node features with Cartesian vector features,\cite{Schutt2021} while NequIP and Allegro employ spherical harmonics to construct equivariant representations of higher angular orders.\cite{Batzner2022,Musaelian2023} MACE combines equivariant message passing with a systematic many-body expansion,\cite{Batatia2022} whereas the Equiformer family introduces equivariant attention mechanisms, with the latest EquiformerV3, which supports higher-degree spherical harmonics while selectively truncating the interaction orders retained at each degree.\cite{Liao2023,Liao2024,Liao2026} UMA replaces the SO(3) tensor product itself: each edge is rotated to align with a common axis, allowing the convolution to be decomposed into independent SO(2) operations on spherical harmonic features. Its context-conditioning mechanism further enables a single set of model parameters to be shared across multiple domains.\cite{UMA2025} Most recently, large-scale models trained on millions of atomic structures, such as UMA\cite{UMA2025} and MACE-MP\cite{Batatia2024}, have emerged as foundation models designed for zero-shot prediction or fine-tuning across diverse atomistic systems.

Despite this rapid progress, an important practical challenge remains largely overlooked. Each architecture is typically released as an independent research codebase, with its own conventions for dataset format, configuration systems, training workflows, logging, checkpoint structures, and integration with production MD engines (sometimes even missed). These inconsistencies create substantial practical barriers for researchers seeking to develop machine learning potentials for specific chemical systems. First, an honest comparison between two architectures on \emph{one's own} data, with identical data splits, loss weights, optimizer, and stopping criteria, requires reimplementing at least one of them or accepting confounded results. Second, the cost of switching architecture after the fact is high, which biases groups towards whichever model they adopted first. Third, model deployment remains architecture-specific, as integrating a trained potential into LAMMPS\cite{Thompson2022} typically requires dedicated wrappers and pair-style implementations for each architecture. Fourth, high-performance-computing (HPC) concerns leak into user workflows, as distributed training must be adapted to the launcher and scheduler of each machine. Fifth, extensibility is limited: adding a custom architecture or modifying existing model components often requires substantial changes to the framework-specific code, making it difficult to experiment with new ideas without becoming tied to a particular implementation. In addition, most existing frameworks lack an agent-friendly interface, which would enable autonomous agents to orchestrate model training, inference, and deployment.

IANN (InterAtomic Neural Network) framework addresses these challenges by implementing different neural network architectures as configurable options within a unified PyTorch\cite{Paszke2019} framework. Its equivariant models use the unified data object and trainer, allowing users to compare inductive biases on their own data while holding the splits, loss weights, optimiser, and stopping criteria fixed. The common interfaces also make it easier to add or modify architectures without rebuilding the surrounding workflow. Beyond training, checkpoints can be directly utilized with an ASE\cite{Larsen2017} calculator, assembled into ensembles for uncertainty quantification, or compiled to TorchScript for use in LAMMPS\cite{Thompson2022} through bundled pair styles. Training, inference, and deployment are accessible non-interactively, enabling agent-driven workflows.

Key and specific contributions of this work include:

\begin{itemize}
  \item \textbf{A unified data representation.} All architectures consume and return \texttt{AtomsData}, a typed, TorchScript-compatible data object built directly, so that datasets, batching and neighbour-list construction are shared across models rather than reimplemented per model.
  \item \textbf{A unified trainer.} One unified dictionary configures the model, loss, optimizer, learning-rate schedule, data splitting and normalisation, early stopping, checkpointing and distributed execution, with roughly ten lines of user code.
  \item \textbf{Uniform analytic derivatives.} Forces, stresses, and virials are obtained for each architecture by differentiating the predicted energy with respect to the graph edge vectors through a single unified implementation.
  \item \textbf{Launcher-agnostic parallelism.} Distributed data-parallel training detects SLURM, PBS, OpenMPI, MPICH/Intel MPI, and MVAPICH2 environments and derives rank, world size, local rank, and rendezvous address automatically; the same script runs unchanged from a workstation to a multi-node GPU allocation.
  \item \textbf{A production path to LAMMPS.} A generic TorchScript wrapper plus C\texttt{++} pair styles (\texttt{pair\_style iann} and \texttt{pair\_style iann/multi\_gpu}) run any IANN model inside LAMMPS, and a companion compute style exposes ensemble variance as a thermodynamic quantity for on-the-fly reliability monitoring.
  \item \textbf{Curated databases at three levels of theory.} We curated, standardized, and reorganized public DFT datasets into three functional databases (PBE, RPBE, and $\mathrm{r^2SCAN}$), comprising 41.9 million structures that were cleaned of geometric outliers and unreliable labels and packaged in a unified format.\cite{IANNdata}
  \item \textbf{Provided foundation models.} Multiple foundation models trained on the curated open-source databases are provided,\cite{IANNmodels} serving either directly for inference or as starting checkpoints for fine-tuning on a small in-house dataset.
  \item \textbf{An agent interface.} The framework provides an interface with machine-readable output, a Model Context Protocol server, and a set of agent skills, enabling large-language-model (LLM) agents to directly train, inspect, and deploy models.
\end{itemize}


\section{Theoretical background}
\label{sec:background}

\subsection{Locality and the atomic graph}

All models implemented in IANN exploit the locality of atomic interactions through the standard decomposition of the total potential energy into atomic contributions,
\begin{equation}
  E = \sum_{i=1}^{N} E_i ,
  \label{eqn:energy_decomposition}
\end{equation}
where $E_i$ depends only on the local environment of atom $i$ within a cutoff radius $r_\mathrm{c}$. This locality assumption is physically motivated by the short-range nature of atomic interactions and enables efficient scaling to large systems. This locality assumption naturally represents each atomic configuration as a graph $\mathcal{G} = (\mathcal{V}, \mathcal{E})$, where the vertex set $\mathcal{V} = \{1, \ldots, N\}$ corresponds to atoms and the edge set $\mathcal{E}$ connects atom pairs within the cutoff distance,
\begin{equation}
  \mathcal{E} = \left\{ (i,j) \;\middle|\; \lVert \vec{d}_{ij} \rVert < r_\mathrm{c} \right\}, \qquad
  \vec{d}_{ij} = \vec{r}_j - \vec{r}_i + \mathbf{S}_{ij}\mathbf{h},
  \label{eqn:graph}
\end{equation}
where $\mathbf{h}$ is the $3 \times 3$ cell matrix whose columns define the periodic lattice vectors, and $\mathbf{S}_{ij} \in \mathbb{Z}^3$ is the integer image shift vector that selects the periodic replica of atom $j$ nearest to $i$. Working with edge vectors $\vec{d}_{ij}$ rather than with absolute positions ensures exact translational invariance by construction and encodes periodicity as a graph property rather than requiring special treatment in the network architecture.

\subsection{Equivariant message passing}

A function $f$ acting on graph features is equivariant with respect to rotations if
\begin{equation}
  f\!\left(\mathbf{R}\vec{d}_{ij}\right) = \mathbf{D}(\mathbf{R})\, f\!\left(\vec{d}_{ij}\right)
  \label{eqn:equivariance}
\end{equation}
for every rotation $\mathbf{R}$, where $\mathbf{D}(\mathbf{R})$ is the representation of $\mathbf{R}$ acting on the output space. Scalars transform with $\mathbf{D} = 1$ (invariance) and Cartesian vectors with $\mathbf{D} = \mathbf{R}$ (covariance). Models built on spherical harmonics generalize this to irreducible representations of arbitrary degree $\ell \le \ell_\mathrm{max}$, where $\mathbf{D}$ is the corresponding Wigner $\mathcal{D}$-matrix. Equivariance enables the network to represent directional information---such as bond orientations and local symmetry breaking---without learning rotational symmetries from data, thereby improving sample efficiency and generalization.

All architectures in IANN follow the general message-passing framework
\begin{equation}
  \vec{m}_i^{(t)} = \bigoplus_{j \in \mathcal{N}(i)} M_t\!\left( h_i^{(t)}, h_j^{(t)}, R(\lVert \vec{d}_{ij} \rVert),\, Y(\hat{d}_{ij}) \right),
  \qquad
  h_i^{(t+1)} = U_t\!\left( h_i^{(t)}, \vec{m}_i^{(t)} \right),
  \label{eqn:message_passing}
\end{equation}
where $h_i^{(t)}$ are the node features of atom $i$ at layer $t$ (in general containing mixed irreducible representations), $\mathcal{N}(i)$ denotes the neighbors of atom $i$ within the cutoff radius, $R$ is a radial basis function multiplied by a smooth cutoff envelope, $Y$ denotes spherical harmonics evaluated on the normalized edge direction $\hat{d}_{ij} = \vec{d}_{ij}/\lVert \vec{d}_{ij} \rVert$, and $\bigoplus$ represents permutation-invariant aggregation---implemented in IANN as a scatter-add operation over edges. The architectures differ in how the message function $M_t$ mixes irreducible representations: PaiNN uses gated products of scalar and vector channels; NequIP, Allegro, and MACE employ tensor products of irreducible representations; the Equiformer family implements equivariant attention mechanisms; UMA utilizes a spherical-channel backbone. Architectures also differ in whether they include explicit many-body expansions, but all share a common interface defined by Eq.~\eqref{eqn:message_passing}. After $T$ message-passing layers, a readout network maps the invariant ($\ell=0$) component of $h_i^{(T)}$ to the atomic energy $E_i$ in \eqlink{eqn:energy_decomposition}.

Radial information is encoded using a Bessel function basis with a polynomial cutoff envelope~\cite{Gasteiger2020}
\begin{equation}
  R_n(d) = \sqrt{\frac{2}{r_\mathrm{c}}}\,\frac{\sin\!\left(n \pi d / r_\mathrm{c}\right)}{d}\, u(d),
  \qquad n = 1,\ldots,N_\mathrm{basis},
  \label{eqn:bessel}
\end{equation}
where $d = \lVert \vec{d}_{ij} \rVert$ is the interatomic distance, and the envelope function $u(d)$ and its first $p-1$ derivatives vanish at $d = r_\mathrm{c}$. This ensures that energies and forces remain continuous as atoms enter or leave the cutoff sphere---a prerequisite for stable molecular dynamics simulations.

\subsection{Analytic derivatives}

Forces are not predicted by a separate network head but obtained as exact analytic derivatives of the predicted energy with respect to atomic positions, which guarantees that the resulting force field is conservative. IANN computes these derivatives with respect to the edge vectors rather than the atomic positions directly,
\begin{equation}
  \vec{F}_i = -\frac{\partial E}{\partial \vec{r}_i}
  = \sum_{j \in \mathcal{N}(i)} \frac{\partial E}{\partial \vec{d}_{ij}}
  - \sum_{j \,:\, i \in \mathcal{N}(j)} \frac{\partial E}{\partial \vec{d}_{ji}},
  \label{eqn:forces}
\end{equation}
which follows from $\partial \vec{d}_{ij}/\partial \vec{r}_i = -\mathbf{I}$ and $\partial \vec{d}_{ij}/\partial \vec{r}_j = +\mathbf{I}$. In the implementation, a single call to automatic differentiation produces $\partial E / \partial \vec{d}_{ij}$ for all edges simultaneously, and the two sums in \eqlink{eqn:forces} are evaluated as scatter-add operations onto the center and neighbor indices of each edge (\autoref{fig:mechanism}c). This route is preferred over differentiating with respect to $\vec{r}_i$ directly because the periodic image shifts $\mathbf{S}_{ij}\mathbf{h}$ of \eqlink{eqn:graph} are absorbed into the graph construction phase. Consequently, the same code path correctly handles isolated molecules, slab geometries, and fully periodic bulk systems without modification. The virial tensor and stress tensor are computed from the same edge derivatives:
\begin{equation}
  \mathbf{W} = \sum_{(i,j) \in \mathcal{E}} \vec{d}_{ij} \otimes \frac{\partial E}{\partial \vec{d}_{ij}},
  \qquad
  \boldsymbol{\sigma} = \frac{1}{V}\,\mathbf{W},
  \label{eqn:virial}
\end{equation}
where $\mathbf{W}$ is the $3\times 3$ virial tensor, $V$ is the simulation cell volume, and $\boldsymbol{\sigma}$ is the stress tensor. Because the virial is constructed from the same per-edge gradients $\partial E/\partial\vec{d}_{ij}$ as the forces, no additional backward pass through the network is required. In molecular dynamics simulations, the kinetic contribution to the pressure is added by the integrator rather than computed by the model.

 \subsection{Training objective}

Models are trained by minimizing a weighted combination of energy, force, virial, and stress residuals:
\begin{equation}
  \mathcal{L} = w_E\,\mathcal{L}_E + w_F\,\mathcal{L}_F + w_W\,\mathcal{L}_W + w_\sigma\,\mathcal{L}_\sigma ,
  \qquad
  w_E = 1 - w_F - w_W - w_\sigma ,
  \label{eqn:loss}
\end{equation}
where the energy weight $w_E$ is determined automatically from the user-specified derivative weights to ensure $\sum w = 1$. The energy loss term uses mean-squared error normalized per atom:
\begin{equation}
  \mathcal{L}_E = \frac{1}{B}\sum_{b=1}^{B}\left( \frac{\hat{E}_b - E_b}{N_b} \right)^{\!2},
  \label{eqn:loss_energy}
\end{equation}
where $B$ is the batch size, $\hat{E}_b$ and $E_b$ are the predicted and reference energies for structure $b$, and $N_b$ is the number of atoms in structure $b$. This per-atom normalization prevents large structures from dominating batches with heterogeneous system sizes. The force loss term uses the mean Euclidean norm of the per-atom force error:
\begin{equation}
  \mathcal{L}_F = \frac{1}{N}\sum_{i=1}^{N} \left\lVert \hat{\vec{F}}_i - \vec{F}_i \right\rVert_2 ,
  \label{eqn:loss_forces}
\end{equation}
where $N$ is the total number of atoms in the batch, and $\hat{\vec{F}}_i$ and $\vec{F}_i$ are the predicted and reference forces on atom $i$. This $L_1$ norm in force space is rotationally invariant and less sensitive to outliers than the squared ($L_2$) norm. Optionally, training targets are standardized using the mean and standard deviation of the per-atom training energies, and predictions are de-standardized within the model before output. This ensures that saved model checkpoints always predict energies in physical units regardless of the standardization used during training.

\section{Related work}
\label{sec:related}

Several mature frameworks address different aspects of the MLIP workflow. SchNetPack\cite{Schutt2019,Schutt2023} pioneered the concept of a modular atomistic deep-learning toolbox with reusable data pipelines, training loops, and ASE integration, establishing a benchmark for usability. Its model zoo focuses on the SchNet, PaiNN, and FieldSchNet architectures. DeePMD-kit\cite{Wang2018,Zeng2023} is widely deployed in the production environment, offering highly optimized descriptors, extensive MD engine integration, and mature HPC support. However, it is built around the Deep Potential model family rather than the equivariant graph neural networks that currently define the state-of-the-art in accuracy. GPUMD\cite{Fan2022} achieves exceptional molecular dynamics throughput using the neuroevolution-potential (NEP) model, though it is similarly limited to a single architecture family. TorchMD-NET\cite{Tholke2022} implements equivariant transformer architectures with specialized support for biomolecular simulations. The reference implementations of NequIP~\cite{Batzner2022}, Allegro~\cite{Musaelian2023}, MACE~\cite{Batatia2022}, and the Equiformer family~\cite{Liao2023,Liao2024} are well-engineered within their respective scopes. However, each constitutes a separate package with its own data format, configuration schema, and deployment workflow, preventing interoperability: models cannot generally be exchanged or compared directly across these frameworks. Workflow-level tools address complementary challenges: CURATOR~\cite{Yang2024} and DP-GEN~\cite{Zhang2019} automate active learning and dataset generation, while AdsorbML~\cite{Lan2023} provides an efficient screening protocol that combines pre-trained potentials with selective DFT refinement.

IANN occupies the intermediate layer between individual model implementations and high-level workflow tools. Rather than proposing novel architectures or active-learning strategies, it standardizes the software interface---encompassing data objects, training loops, model checkpoints, ASE calculators, and MD engine exports---enabling architectures with fundamentally different inductive biases to be trained, compared, and deployed through identical code paths. Three key features distinguish IANN from existing frameworks. First, architecture selection becomes a controllable experimental variable that can be systematically varied while holding all other factors constant---precisely what is required for rigorous benchmarking on domain-specific datasets. Second, deployment is fully model-agnostic: because all architectures expose a unified tensor-level interface, a single TorchScript wrapper and a single pair of LAMMPS pair styles support all models, including ensemble predictions. Third, HPC parallelism is managed by the framework rather than requiring user intervention. This design choice is practically significant because training workflows must often execute seamlessly across diverse computing environments, from local workstations to HPC clusters with varying job schedulers and MPI implementations.

\section{Software structure}
\label{sec:software}

\begin{figure}[!htbp]
    \centering
    \includegraphics[width=\textwidth]{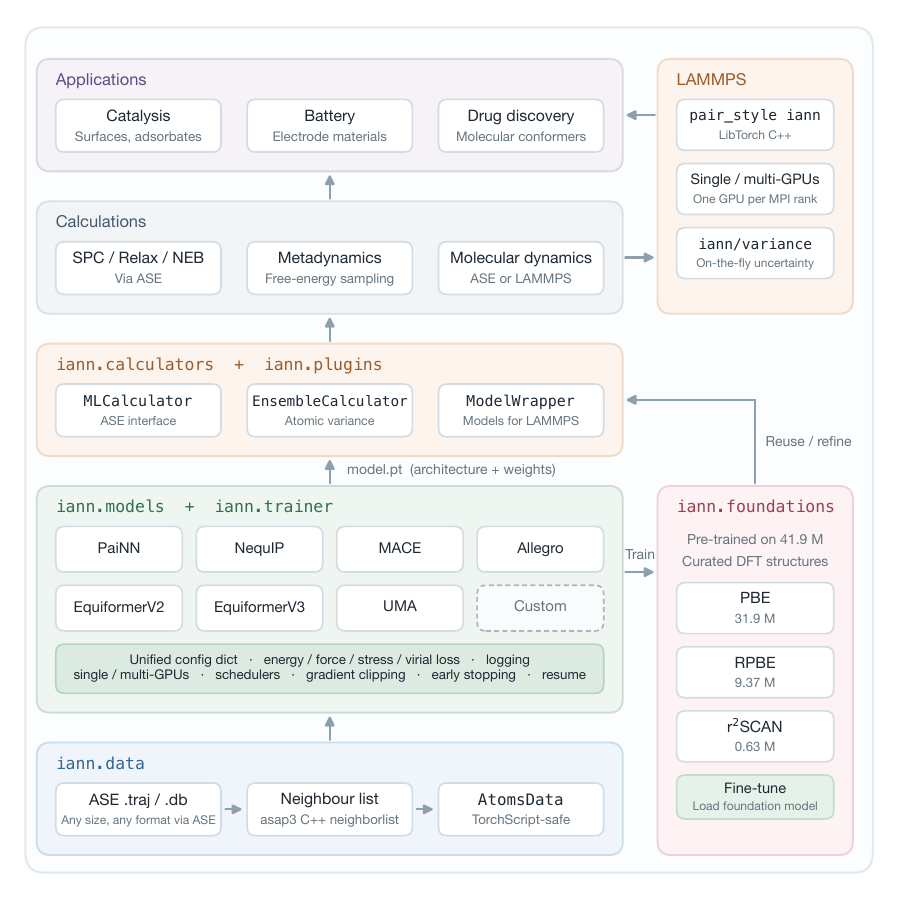}
    \caption{Software structure of the IANN framework: five main layers with two auxiliary layers.}
    \label{fig:framework}
\end{figure}

\autoref{fig:framework} shows the structure of IANN, which consists of five main layers with two auxiliary ones. It starts from first layer, data layer (\texttt{iann.data}) as shown in the bottom in \autoref{fig:framework}, which serves unified data to the second layer, models (\texttt{iann.models}) and trainer (\texttt{iann.trainer}) layer. There are plenty of MLIPs architectures that can be chosen and trained on the unified trainer on a single GPU or in parallel across many. Training writes a self-describing checkpoint that the third layer consumes either as an ASE calculator via \texttt{iann.calculators} (with optional ensemble support) or as a TorchScript model compiled for LAMMPS via \texttt{iann.plugins}. Two auxiliary layers attach to these: \texttt{iann.foundations} supplies pre-trained potentials at three levels of DFT theory, to be reused directly or refined by the trainer, and the bundled pair styles execute the compiled model inside LAMMPS. The fourth layer is the calculations those two interfaces drive, such as surrogate DFT single points, relaxations and nudged-elastic-band searches, enhanced sampling, and molecular dynamics. The final layer (fifth layer) encompasses the application domains served by IANN: catalysis, battery materials, and drug discovery. The subsections that follow describe these architectural layers in sequence.

\FloatBarrier

\subsection{Data layer}

In order to deal with various structure formats, and to turn them into the graph that a model can consume, IANN provides a unified data layer (\texttt{iann.data}). Trajectory or database are read through ASE that can support various formats. The neighbour lists that define the graph nodes and edges are built by a C\texttt{++} compiled asap3 whose cost grows linearly with the number of atoms. Both are packed into a unified data object, \texttt{AtomsData}, which is natively accepted by every architecture and readily compiled via TorchScript. This allows the same object to serve both training in Python and inference inside LAMMPS in C\texttt{++}. More detailed construction can be found in Section~\ref{sec:si_impl} of the Supporting Information.

\subsection{Models and trainer layer}

Based on the unified data object, the second layer implements several state-of-the-art MLIPs architectures alongside a template for custom models (\texttt{iann.models}), all managed through a unified trainer (\texttt{iann.trainer}). The trainer handles the training loop, the objective on energies, forces, stresses and virials, logging, checkpointing, learning-rate scheduling, early stopping, and restarts. It also detects the environment and configures distributed data-parallel training automatically from SLURM, PBS, OpenMPI, MPICH/Intel MPI, or MVAPICH2 without any change to the script. The architectures --- PaiNN,\cite{Schutt2021} NequIP,\cite{Batzner2022} MACE,\cite{Batatia2022} Allegro,\cite{Musaelian2023} EquiformerV2,\cite{Liao2024} EquiformerV3,\cite{Liao2026} and UMA\cite{UMA2025} --- are available by their name in a configuration dictionary. Each is reimplemented against the shared \texttt{AtomsData} interface, so that one data object, one derivative route, and one exporter serve them all. The accuracies reported here are therefore those of the IANN implementations rather than of the original releases. The same trainer then supplies the objective and runs on a single GPU or in parallel across many\cite{Li2020} without a change to the script. The trainer configurations and the corresponding implementation details are described in full as Section~\ref{sec:si_impl} and Section~\ref{sec:si_config}.

\subsection{Calculators and plugins layer}

After training a model, a unified and user-friendly interface is needed to load the trained checkpoint for doing calculations, which is exactly what the third layer provides. The third layer provides unified calculators (\texttt{iann.calculators} and plugins \texttt{iann.plugins}), making the full ASE ecosystem directly accessible without requiring additional code. In addition, \texttt{EnsembleCalculator} can utilise ensemble checkpoints to do prediction with variance of total energy, forces, and atomic properties (like atomic energy) directly as an acquisition signal for active learning.\cite{Peterson2017,Schran2020} The LAMMPS plugin can use the compiled TorchScript single model or ensemble model to do inference in serial or in parallel, which does not depend on the specific architectures.

\subsection{Calculations layer and applications layer}

Building directly on the third layer, the calculations layer invokes these calculators and plugins to drive various atomistic simulations: single-point evaluation, geometry relaxation, and nudged-elastic-band searches through ASE as surrogates for the corresponding DFT calculations, molecular dynamics through either ASE or LAMMPS, and free-energy methods such as metadynamics through ASE-PLUMED calculator. These calculations serve the application domains at the top of \autoref{fig:framework}, which include catalysis, battery materials, and drug discovery.
\subsection{Foundation models}

To accelerate project workflows and avoid training models from scratch, IANN distributes pretrained foundation potentials (\texttt{iann.foundations}). The foundation models are released,\cite{IANNmodels} trained on 41.9 million curated DFT structures and grouped by exchange-correlation functional. Each model can serve directly as a zero-shot calculator or as an initialization checkpoint for fine-tuning on targeted in-house data to boost predictive accuracy (see Section~\ref{sec:fine_tuning}).

\subsection{Deployment to LAMMPS}

To bypass the size and timescale bottlenecks of Python-driven simulations, IANN deploys compiled models directly inside LAMMPS (\texttt{pair\_style iann}). Built on the LibTorch C\texttt{++} API, the pair styles run in serial or multi-GPU parallel configurations. Furthermore, the plugin natively supports ensemble predictions: loading an ensemble checkpoint automatically instantiates a companion compute that streams energy and force variances alongside standard thermodynamic variables, enabling real-time detection of out-of-distribution drift.

\subsection{Agent interface}

Beyond the core framework, an agent interface exposes these underlying operations programmatically, enabling fully autonomous model training, inspection, and deployment without human intervention. Instead of introducing a separate layer into \autoref{fig:framework}, the agent interface serves as a parallel entry point to the core stack through three complementary mechanisms: structured CLI output, an MCP server, and task-specific agent skills. Detailed descriptions of each interface and the constraints governing autonomous execution are provided in Section~\ref{sec:si_agent} of the Supporting Information.

\section{Underlying mechanism}
\label{sec:mechanism}

\begin{figure}[!ht]
    \centering
    \includegraphics[width=\textwidth]{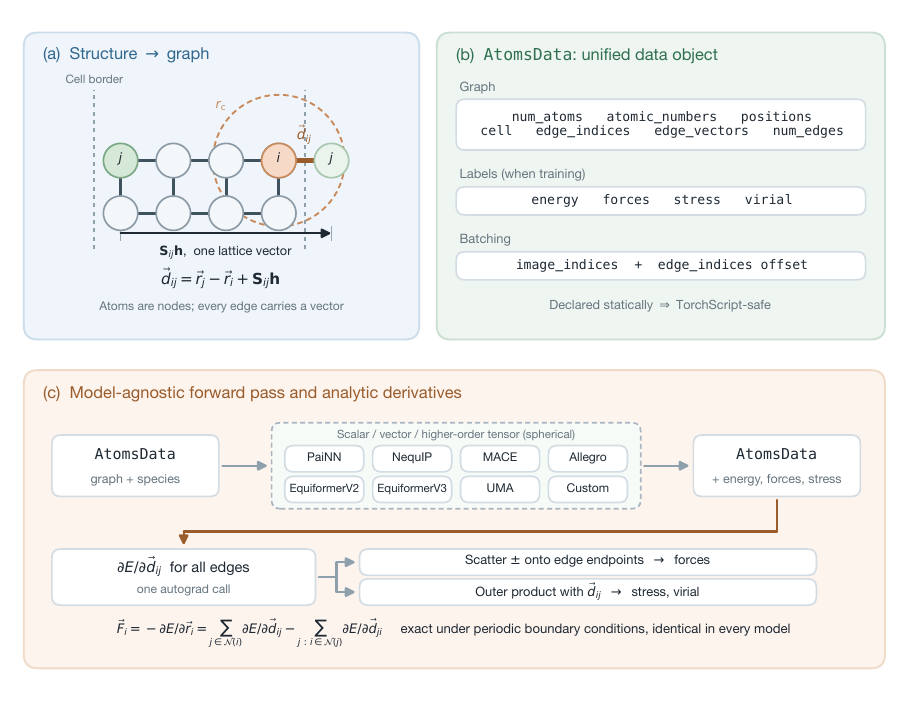}
        \caption{The underlying mechanism: a shared graph convention, a unified data object, and a shared derivative route. (a) A structure becomes a graph: atom $i$ bonds to atom $j$ within the cutoff $r_\mathrm{c}$. (b) The unified data object: \texttt{AtomsData}, which describes the graph and its associated data. (c) The architecture-independent forward pass that yields forces, stress and virial.}
    \label{fig:mechanism}
\end{figure}

The core architectural mechanism enabling the uniformity across \autoref{fig:framework} is illustrated in \autoref{fig:mechanism}. It comprises three key components: a shared graph convention, a unified data object consumed and produced by every model, and a single analytic derivative pathway. The details of each component are described below.

\FloatBarrier

\subsection{Graph convention}

Before processing by any architecture, an atomic structure is converted into a graph representation. This graph is materialized as node and edge tensors according to \eqlink{eqn:graph}. The nodes are represented by a single integer array of atomic numbers, and the edges by integer indices array and edge vectors $\vec{d}_{ij}$. The graph is directed, so each edge appears twice: once from $i$ to $j$ and once from $j$ to $i$. For periodic boundary conditions, the lattice shift vector $\mathbf{S}_{ij}\mathbf{h}$ is incorporated during graph construction. Crucially, the edge vectors are flagged to track gradients, enabling the unified, single-step differentiation detailed below.

\subsection{Unified data object}

The graph is packaged into a unified data structure (\texttt{AtomsData}) that every architecture consumes and returns, making the trainer and calculators architecture-agnostic. This data object also carries training labels, batching indices, optional intermediate representations, predictive variances, and ensemble metadata. The data object is also compatible with TorchScript, which allows the same model object to be compiled for LAMMPS without an architecture-specific adapter. 

\subsection{Derivative route}

The architecture-independent forward pass in \autoref{fig:mechanism}c closes the loop. Every architecture implements an identical \texttt{forward} signature—mapping \texttt{AtomsData} to \texttt{AtomsData}—and extracts atomic forces through a single differentiation with respect to edge vectors (\eqlink{eqn:forces}). A subsequent contraction of these same per-edge gradients against the edge vectors yields the virial and Cauchy stress (\eqlink{eqn:virial}), allowing all three derived physical quantities to be obtained in a single backward pass. The loss functions, trainer, calculators, and TorchScript exporter are implemented against that signature alone and contain no architecture-specific logic. This design offers two key advantages. First, selecting a network architecture is reduced to a simple configuration parameter, allowing different models to be trained and evaluated using a single unified script. Second, a universal wrapper exports any supported architecture to LAMMPS, ensuring that adding a new model leaves the downstream deployment pipeline untouched.

\subsection{A complete workflow}

A primary design objective of IANN is usability, enabling practitioners to train and deploy models with minimal code and without low-level framework knowledge. As shown below, \autoref{lst:train} demonstrates how to train a model directly from an ASE trajectory. Switching between supported architectures—including \texttt{nequip}, \texttt{mace}, \texttt{allegro}, \texttt{equiformerV2}, \texttt{equiformerV3}, and \texttt{UMA}—requires only modifying the \texttt{model} name and corresponding hyperparameters in the \texttt{config} dictionary. Furthermore, custom architectures can be integrated into the framework with minimal boilerplate, as illustrated below.

\begin{lstlisting}[caption={Training a model with a registered architecture.},label={lst:train}]
from iann.trainer import Trainer

trainer = Trainer(
    model="painn",
    config={"num_channels": 128,
            "num_layers": 3,
            "cutoff": 5.5,
            "batch_size": 16,
            "learning_rate": 1e-4,
            "forces_weight": 0.9,
            "val_ratio": 0.1,
            "device": "cuda",
            "output_dir": "output"},
    distributed=True,
)
trainer.train("dataset.traj")
\end{lstlisting}

The resulting checkpoint is loaded into an ASE calculator in \autoref{lst:predict} to drive downstream calculations. Because each checkpoint encapsulates its network architecture and hyperparameters, the calculator automatically reconstructs the underlying model without manual specification. Uncertainty quantification is equally seamless: swapping \texttt{MLCalculator} for \texttt{EnsembleCalculator} with a list of checkpoints yields both mean predictions and calibrated variances. Finally, deployment to LAMMPS requires only a single export step to generate a TorchScript model compatible with \texttt{pair\_style iann}.

\begin{lstlisting}[caption={Running inference with a model checkpoint with its own architecture and parameters.},label={lst:predict}]
from ase.io import read
from ase.optimize import BFGS
from iann.calculators import MLCalculator

atoms = read("start.traj")
atoms.calc = MLCalculator("output/model.pt", device="cuda")
BFGS(atoms, trajectory="relax.traj").run(fmax=0.03)
\end{lstlisting}

\section{Results and discussion}
\label{sec:results}

\subsection{Curated DFT databases}

\begin{table}[!ht]
    \centering
    \caption{Curated DFT databases grouped by exchange-correlation functional.}
    \label{tab:databases}
    \small
    \setlength{\tabcolsep}{3.5pt}
    \begin{tabular}{lllrrrc}
    \toprule
    Functional & Source dataset & System type & Structures & Atoms (M) & Elem. & Ref. \\
    \midrule
    \multirow{6}{*}{PBE}
      & MPtrj             & Bulk, relaxations    & 1,579,091  & 49.2    & 89 & \cite{Deng2023,MPtrjData} \\
      & sAlex             & Bulk, relaxations    & 10,423,284 & 107.5   & 89 & \cite{Alexandria,Barroso2024} \\
      & OMat24            & Bulk, off-equilibrium & 11,274,650 & 157.1   & 89 & \cite{Barroso2024} \\
      & MatPES            & Bulk, MD snapshots   & 434,092     & 3.9     & 89 & \cite{MatPES2025} \\
      & OC22              & Surfaces, oxides   & 8,199,197  & 654.3   & 57 & \cite{Tran2023} \\
      & \emph{PBE total}   &                                 & \textbf{31,910,314} & \textbf{972.0} & 89 & \\
    \midrule
    \multirow{3}{*}{RPBE}
      & OC20 (2M)  & Surfaces, relaxations   & 1,999,572  & 146.5   & 56 & \cite{Chanussot2021} \\
      & OC25              & Solid--liquid interf. & 7,370,461  & 1064.3 & 73 & \cite{OC25} \\
      & \emph{RPBE total}   &                                 & \textbf{9,370,033} & \textbf{1210.8} & 73 & \\
    \midrule
    \multirow{3}{*}{r$^2$SCAN}
      & MP r$^2$SCAN      & Bulk, relaxations    & 238,241     & 5.0     & 88 & \cite{MPr2SCAN} \\
      & MatPES r$^2$SCAN  & Bulk, MD snapshots   & 387,306     & 3.0     & 89 & \cite{MatPES2025} \\
      & \emph{r$^2$SCAN total}     &                          & \textbf{625,547} & \textbf{8.1} & 89 & \\
    \bottomrule
    \end{tabular}
\end{table}

Training a robust foundation potential requires DFT data generated at a consistent level of theory. However, existing public DFT repositories use a range of exchange-correlation functionals and vary in structural quality. To address this, we systematically screened the source databases for conspicuous energy and force outliers and removed them prior to training. After filtering, we categorized the databases by exchange-correlation functionals and converted them into a standardized ASE trajectory format (\autoref{tab:databases}), which the framework ingests directly without requiring auxiliary preprocessing or database conversion. The PBE\cite{Perdew1996} database includes MPtrj,\cite{Deng2023,MPtrjData}, sAlex (subsample of Alexandria),\cite{Alexandria,Barroso2024} OMat24,\cite{Barroso2024} MatPES,\cite{MatPES2025} and OC22, which cover 89 elements for bulk and surface chemistry;\cite{Tran2023} the RPBE\cite{Hammer1999} database includes OC20\cite{Chanussot2021} and OC25,\cite{OC25} which together cover the catalysis-relevant surface and solid--liquid-interface chemistry with 73 elements; and the r$^2$SCAN\cite{Furness2020} database includes the meta-GGA subsets of the Materials Project\cite{MPr2SCAN} and MatPES, which cover 89 elements.\cite{MatPES2025} The detailed system information, total structure counts, atom counts (in millions), and element coverages for each database are provided in \autoref{tab:databases}. It can be noted that the PBE database is the largest by structure count, with 31.9 million structures, followed by the RPBE database with 9.37 million, and the r$^2$SCAN database with 0.63 million. When measured by atom count, the order of the first two is inverted: the RPBE dataset comprises 1,210.8 million atoms compared to 972.0 million for PBE and 8.1 million for r$^2$SCAN. This inversion arises because the surface and interface supercells in the RPBE sources are substantially larger than the bulk unit cells that dominate the PBE collection.

\FloatBarrier
\begin{figure}[!ht]
    \centering
    \includegraphics[width=\textwidth]{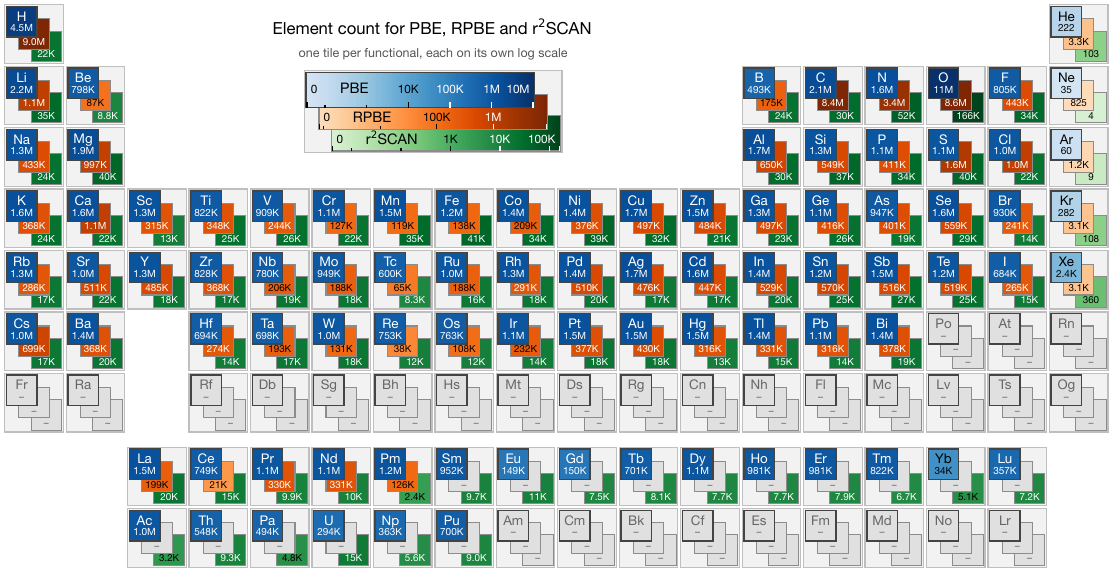}
    \caption{Elemental composition of three levels of curated DFT databases: number of structures containing each element on a logarithmic colour scale. Each element tile is divided into three stepped sub-cells from top left to bottom right, corresponding to PBE, RPBE, and r$^2$SCAN.}
    \label{fig:dft_db}
\end{figure}

\FloatBarrier

These three DFT levels databases also differ markedly in chemical diversity and dataset scale as shown in \autoref{fig:dft_db}. Each periodic table tile indicates the element count and is shaded on an independent logarithmic scale according to its functional-specific colormap: blue for PBE, orange for RPBE, and green for r$^2$SCAN. Darker tones correspond to higher counts, while dashed gray cells designate elements absent from that dataset. The PBE and r$^2$SCAN databases span nearly the entire periodic table—including the lanthanides and actinides up to plutonium (Pu)—yet differ in depth by more than an order of magnitude. Conversely, the RPBE database exhibits a narrower elemental breadth, containing no actinides and only five lanthanides, but compensates with massive statistical depth: it encompasses 9.0~million structures containing hydrogen and 8.6~million containing oxygen, directly reflecting its provenance in surface science and solid--liquid interface catalysis. In addition, complete per-element, per-atom, and structure counts for each database are mapped onto periodic tables in Section~\ref{sec:si_ptables} in the Supporting Information. Altogether, the curated collection contains 41.9 million labeled configurations, accessible through Hugging Face Datasets.\cite{IANNdata}

\subsection{Foundation models across architectures}

Seamless cross-architecture training is a cornerstone of IANN, enabling users to train diverse model architectures on an identical dataset with minimal modifications to the configuration script. Here, we present a systematic benchmark comparing the energy and force prediction accuracy across multiple architectures trained on the same database. While this benchmark is not designed as a controlled ablation study, it provides a realistic assessment of empirical performance in practical deployments. Model parameter count serves as the common metric of comparison. Consequently, aside from the distinct network architectures and their parameterizations, all models share identical training splits, graph conventions, loss formulations, and evaluation routines. We trained seven architectures on the curated MPtrj database (1,579,091 PBE structures) under uniform training settings: a \qty{5.5}{\angstrom} interaction cutoff, the Adam optimizer, identical gradient clipping, and standardized early-stopping patience. For the loss function in \eqlink{eqn:loss_energy}, we adopted a staged weighting scheme, transitioning from $w_F = 0.9$ to emphasize force learning early on to $w_F = 0.3$ to optimize asymptotic energy accuracy. For each architecture, channel widths, layer depths, and angular resolutions ($\ell_\mathrm{max}$, $m_\mathrm{max}$) were selected to standardize model capacity around 1~million parameters (spanning 0.57 to 1.43~M). Learning rates were tuned independently for each network to match its characteristic optimization dynamics.

\begin{table}[!ht]
    \centering
    \caption{Validation accuracy and parameter counts for seven representative architectures trained via IANN on the curated MPtrj dataset.~\cite{IANNmodels} Parameter count serves as the common metric of model capacity across disparate architectures; network depth, channel dimensions, and angular resolutions for each model are detailed in Table~\ref{tab:hparams}.}
    \label{tab:accuracy}
    \small
    \setlength{\tabcolsep}{4.5pt}
    \begin{tabular}{lrrr}
    \toprule
    Model         & Parameters & $E_\mathrm{MAE}$ (eV/atom) & $F_\mathrm{MAE}$ (eV/\AA) \\
    \midrule
    PaiNN         &   600,577 &           0.041 & 0.050 \\
    NequIP        & 1,379,513 &           0.050 & 0.069 \\
    Allegro       &   567,624 &           0.040 & 0.066 \\
    MACE          & 1,429,376 &           0.054 & 0.069 \\
    EquiformerV2  &   773,089 &           0.040 & 0.059 \\
    EquiformerV3  &   943,249 &           0.040 & 0.046 \\
    UMA           & 1,173,698 &           0.029 & 0.040 \\
    \bottomrule
    \end{tabular}
\end{table}

The energy and forces accuracies of the seven architectures are summarized in \autoref{tab:accuracy}. Validation errors are evaluated on a held-out 5\% split of the curated MPtrj dataset using the mean absolute error (MAE). Here, models are deliberately constrained to approximately 1~million parameters to ensure rapid inference and facilitate lightweight fine-tuning. Two key trends emerge from these results. First, when evaluated on the same benchmark at comparable model capacities, energy errors vary relatively little across architectures, from 0.029 to \qty{0.054}{eV/atom}, whereas force errors show a wider spread, from 0.040 to \qty{0.069}{eV/\angstrom}. Second, attention-based architectures achieve the strongest overall accuracy on this dataset: EquiformerV3 delivers competitive energy and force errors of \qty{0.040}{eV/atom} and \qty{0.046}{eV/\angstrom} with 0.94~M parameters, while UMA achieves the lowest errors overall, recording an energy error of \qty{0.029}{eV/atom} and a force error of \qty{0.040}{eV/\angstrom} but uses more parameters, at 1.17~M. PaiNN is also highly competitive: with just 0.60~M parameters, it achieves errors of \qty{0.041}{eV/atom} for energy and \qty{0.050}{eV/\angstrom} for forces, while delivering the fastest inference throughput (\autoref{tab:si_cost}). This balance of accuracy, efficiency, and speed motivates our choice of PaiNN as the core architecture for the foundation potentials at all three DFT levels.

\subsection{Foundation models at three DFT levels}

\FloatBarrier

\begin{table}[!ht]
    \centering
    \caption{PaiNN foundation models within IANN grouped by three levels of DFT exchange-correlation functional.\cite{IANNmodels} Each model is trained on a single source database. Errors here are evaluated across entire source databases rather than on validation sets.}
    \label{tab:foundation}
    \small
    \setlength{\tabcolsep}{4pt}
    \begin{tabular}{lllrrr}
    \toprule
    Functional & Model Name & Database & Structure Count & $E_\mathrm{MAE}$ (eV/atom) & $F_\mathrm{MAE}$ (eV/\AA) \\
    \midrule
    \multirow{5}{*}{PBE}
      & PBE-MPtrj              & MPtrj   & 1,579,058  & 0.0517 & 0.0498 \\
      & PBE-sAlex              & sAlex   & 10,422,893 & 0.0408 & 0.0513 \\
      & PBE-OMat24             & OMat24  & 11,261,400 & 0.0261 & 0.1382 \\
      & PBE-MatPES             & MatPES  & 434,083    & 0.0463 & 0.0930 \\
      & PBE-OC22               & OC22    & 8,198,695  & 0.0381 & 0.0528 \\
    \midrule
    \multirow{2}{*}{RPBE}
      & RPBE-OC20              & OC20    & 1,999,216  & 0.0263 & 0.0630 \\
      & RPBE-OC25              & OC25    & 7,369,601  & 0.0116 & 0.0753 \\
    \midrule
    \multirow{2}{*}{r$^2$SCAN}
      & r$^2$SCAN-MPtrj        & MPtrj   & 238,241    & 0.0206 & 0.0117 \\
      & r$^2$SCAN-MatPES       & MatPES  & 387,285    & 0.0446 & 0.1127 \\
    \bottomrule
    \end{tabular}
\end{table}

\begin{figure}[!tbp]
    \centering
    \includegraphics[width=0.85\textwidth]{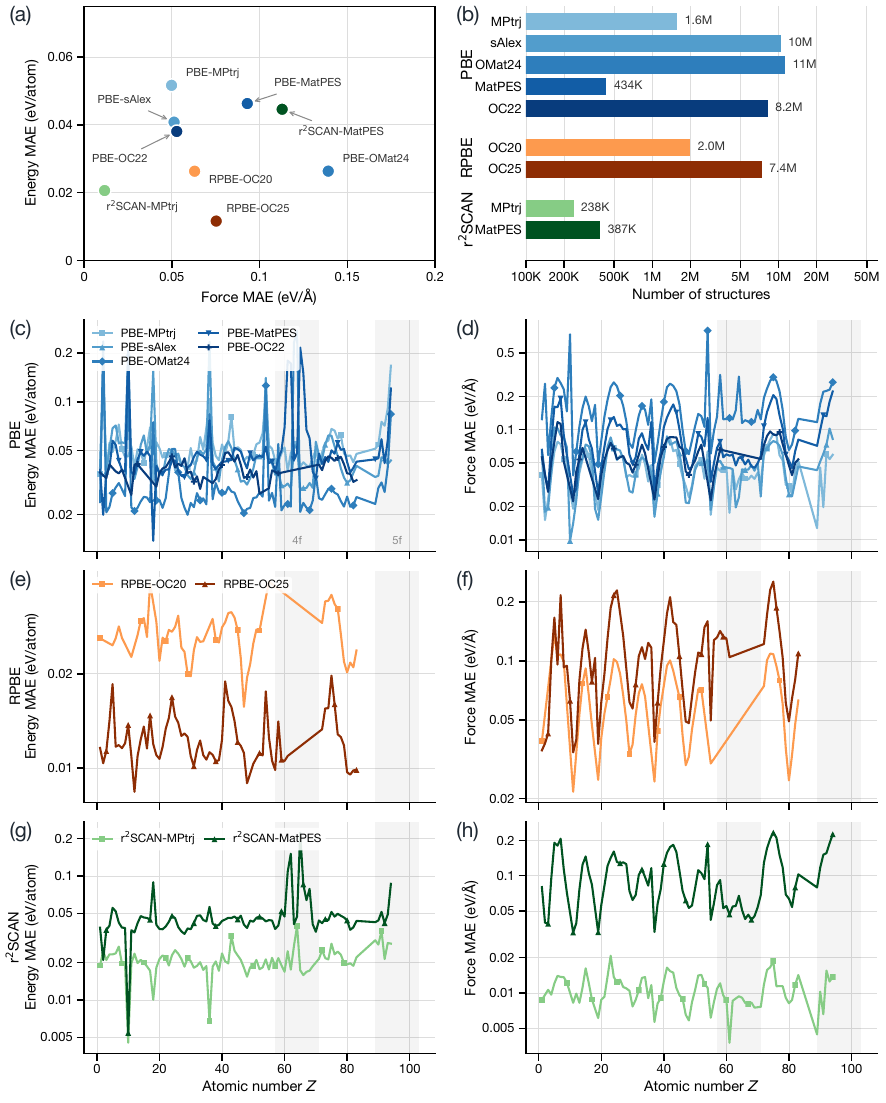}
    \caption{The PaiNN foundation models trained on each DFT database. (a) Energy MAE against force MAE. (b) Number of structures that each was trained on. (c, e, g) Per-element energy MAE against atomic number. (d, f, h) Per-element force MAE against atomic number for the PBE, RPBE and r$^2$SCAN functionals, respectively.}
    \label{fig:foundation_models}
\end{figure}

Leveraging the curated datasets in \autoref{tab:databases}, we trained foundation potentials across all three DFT levels, using PaiNN as the underlying architecture. Training was conducted under identical architectural hyperparameters—128 channels, 3 interaction layers, a \qty{5.5}{\angstrom} cutoff, and 600,577 total parameters—ensuring that performance variations across rows reflect dataset composition rather than model capacity. Each model was optimized using the per-atom energy loss, a batch size of 16, and an initial learning rate of $10^{-4}$. In the final training phase, each run was restarted at a reduced learning rate directly through the trainer, facilitating asymptotic convergence without resetting the optimizer state from scratch. As summarized in \autoref{tab:foundation}, we provide five foundation potentials at the PBE level, two at the RPBE level, and two at the r$^2$SCAN level, each trained on an individual database. We deliberately avoid merging datasets within a given functional because underlying DFT parameters, such as Hubbard $U$ corrections and D3 dispersion schemes, differ across source repositories. Distributing separate models per source preserves internal thermodynamic consistency and allows practitioners to select the checkpoint whose reference settings most closely match their target application. Across all models, training dataset size does not strictly correlate with final validation error. For PBE, per-database energy MAEs range from 0.0261 to \qty{0.0517}{eV/atom}, higher than those for RPBE (0.0116 to \qty{0.0263}{eV/atom}) and comparable to r$^2$SCAN (0.0206 to \qty{0.0446}{eV/atom}). Force MAEs for PBE span 0.0498 to \qty{0.1382}{eV/\angstrom}, exhibiting wider variance than RPBE (0.0630 to \qty{0.0753}{eV/\angstrom}) and r$^2$SCAN (0.0117 to \qty{0.1127}{eV/\angstrom}). The elevated force errors observed for PBE-OMat24 and r$^2$SCAN-MatPES reflect their strong off-equilibrium sampling, which includes heavily rattled geometries in OMat24 and high-temperature dynamic configurations in MatPES.

To enable direct comparison, the validation energy and force MAEs from \autoref{tab:foundation} are plotted against one another in \autoref{fig:foundation_models}a, with the corresponding training dataset sizes shown in \autoref{fig:foundation_models}b. We can note that r$^2$SCAN-MPtrj achieves the lowest force MAE (\qty{0.0117}{eV/\angstrom}), while RPBE-OC25 attains the lowest energy MAE (\qty{0.0116}{eV/atom}). Notably, predictive accuracy does not correlate monotonically with dataset scale: RPBE-OC25 is trained on 7.4~million configurations yet records the lowest energy error, whereas r$^2$SCAN-MatPES is trained on 387,285 structures but yields an elevated force MAE of \qty{0.1127}{eV/\angstrom}. This demonstrates that foundation model fidelity is governed not merely by dataset volume, but critically by structural diversity and configuration space sampling. While \autoref{fig:foundation_models}a summarizes global accuracies, elemental transferability is often more decisive in practice. To this end, \autoref{fig:foundation_models}c--h resolve energy and force errors as a function of atomic number across all PBE, RPBE, and r$^2$SCAN functionals. Although per-element accuracy varies markedly across the periodic table, the elemental profiles within a single functional remain strikingly congruent across different source databases. This consistency demonstrates that elemental error trends are primarily dictated by the underlying level of theory rather than dataset provenance. Detailed per-element MAE values for each model are summarized in Section~\ref{sec:si_ptables} in the Supporting Information.

\FloatBarrier

\subsection{Fine-tuning foundation models}
\label{sec:fine_tuning}

\FloatBarrier
\begin{figure}[!ht]
  \centering
  \includegraphics[width=\textwidth]{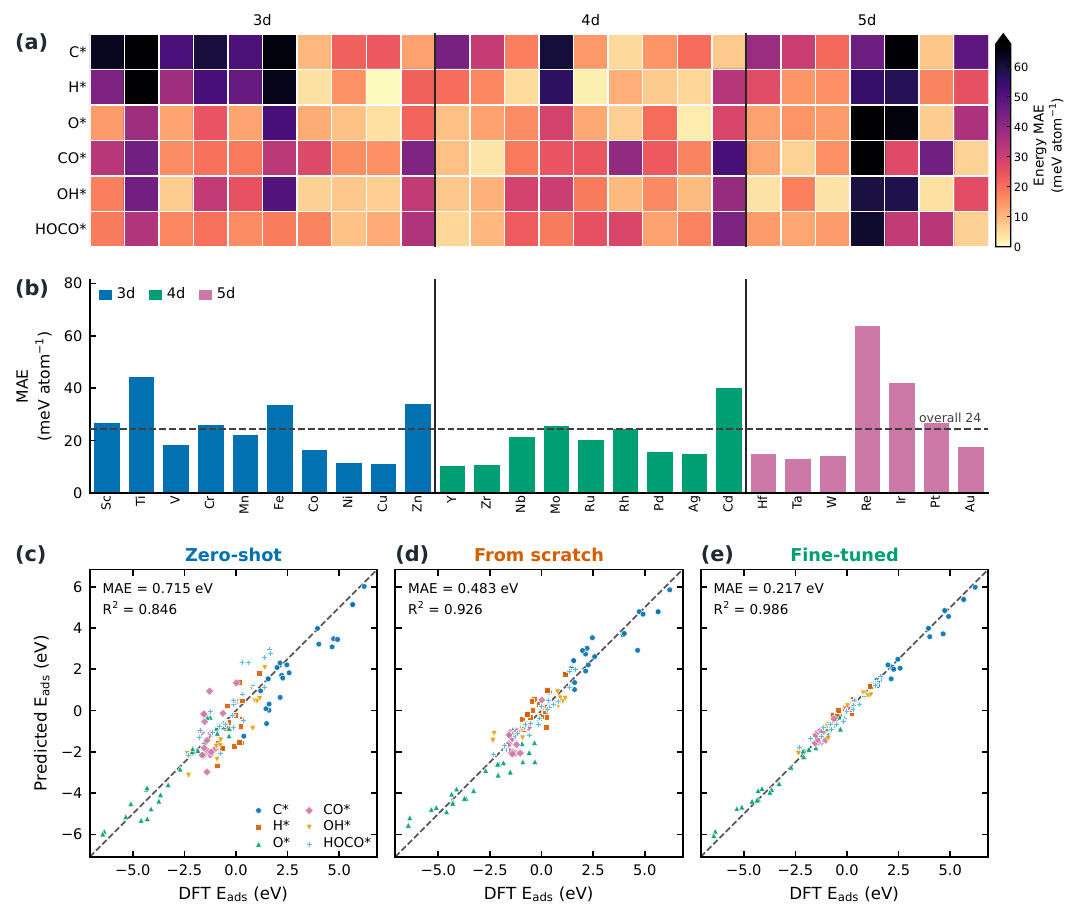}
  \caption{Benchmark performance of the RPBE-OC20 PaiNN foundation model on the AdsTraj. (a) Heatmap of energy MAE across transition metal surfaces (Sc to Au) and key catalytic adsorbates (C*, H*, O*, CO*, OH*, and HOCO*). (b) Energy MAE resolved across metals. Parity plots of adsorption energy under (c) zero-shot, (d) from scratch, and (e) fine-tuning.}
  \label{fig:combined}
\end{figure}
The foundation models provide the intended starting baselines for downstream fine-tuning where high task-specific precision is required. In typical computational catalysis workflows, the DFT calculations for a specific reaction coordinate are limited to a few hundred configurations, well below the sample volume required for training from scratch. Fine-tuning a pre-trained foundation model on this sparse dataset readily adapts the potential to the target chemistry. Within IANN, fine-tuning is executed simply by loading the pre-trained checkpoint and resuming optimization on the specialized in-house dataset. The training routine remains identical to \autoref{lst:train}, requiring only that the \texttt{model} arguments point to the desired foundation checkpoint. To evaluate the fine-tuning performance, we validated the model on AdsTraj, an independent RPBE dataset comprising 33,920 configurations across 26 transition metals, six representative adsorbates (H*, C*, O*, CO*, OH*, and HOCO*), and diverse adsorption sites.~\cite{AdsTraj} The dataset was split into 90\% training and 10\% test partitions. As shown in \autoref{fig:combined}a, the zero-shot energy MAE of the RPBE-OC20 PaiNN foundation model fluctuates across metals and adsorbates, highlighting the inherent non-uniformity of generalized priors across chemical space. The aligned column bars in \autoref{fig:combined}b compute the ensemble MAE per element across all sampled frames, showing an overall baseline error of \qty{0.024}{eV/atom} with particularly high fidelity on metals including V, Co, Ni, Cu, Y, Zr, Pd, Ag, Hf, Ta, W, and Au. Parity plots comparing adsorption energies under zero-shot, trained from scratch, and fine-tuned are illustrated in \autoref{fig:combined}c--e, with quantitative benchmarks summarized in \autoref{tab:finetune}. It can be observed that zero-shot evaluation shows an MAE of \qty{0.715}{eV} with an $R^2$ of 0.846, which has lower accuracy than the model trained from scratch on AdsTraj alone (\qty{0.483}{eV} MAE, $R^2 = 0.926$). Notably, the fine-tuned RPBE-OC20 checkpoint dramatically outperforms both baselines, reducing the MAE to \qty{0.217}{eV} and achieving an $R^2$ of 0.986. These findings demonstrate that pre-trained foundation models provide superior inductive priors, enabling fine-tuned potentials to decisively surpass models trained from scratch within identical, data-scarce regimes.

Per-atom energy and force errors across varying training-set sizes are summarized in Table~\ref{tab:si_finetune} in the Supporting Information. Notably, when trained on the full dataset, the fine-tuned model attains the lowest per-atom energy error (\qty{7.1}{meV/atom}) and force error (\qty{36.7}{meV/\angstrom}), substantially outperforming both the zero-shot foundation model (\qty{24.0}{meV/atom} and \qty{67.2}{meV/\angstrom}) and the model trained from scratch (\qty{14.1}{meV/atom} and \qty{45.4}{meV/\angstrom}). Throughout the data-scaling series, the fine-tuned model consistently surpasses the model trained from scratch at every training-set size. This demonstrates that pre-trained representations accelerate convergence and enhance predictive accuracy in data-limited regimes, confirming that the pre-trained weights supply superior inductive priors relative to random initialization across all stages of fine-tuning.

\begin{table}[!ht]
  \centering
  \caption{Performance comparison across zero-shot, training from scratch, and fine-tuning.}
  \label{tab:finetune}
  \begin{tabular}{lrrr}
    \toprule
    Strategy & Training structures & $E_\mathrm{ads}$ MAE (eV) & $R^2$ \\
    \midrule
    Zero-shot (foundation model)     & 0          & 0.715 & 0.846 \\
    From scratch (train from scratch)                    & 24,154    & 0.483 & 0.926 \\
    Fine-tuned (train on foundation model)    & 24,154    & \textbf{0.217} & \textbf{0.986} \\
    \bottomrule
  \end{tabular}
\end{table}

\subsection{Inference cost across architectures}

\begin{figure}[!ht]
    \centering
    \includegraphics[width=\textwidth]{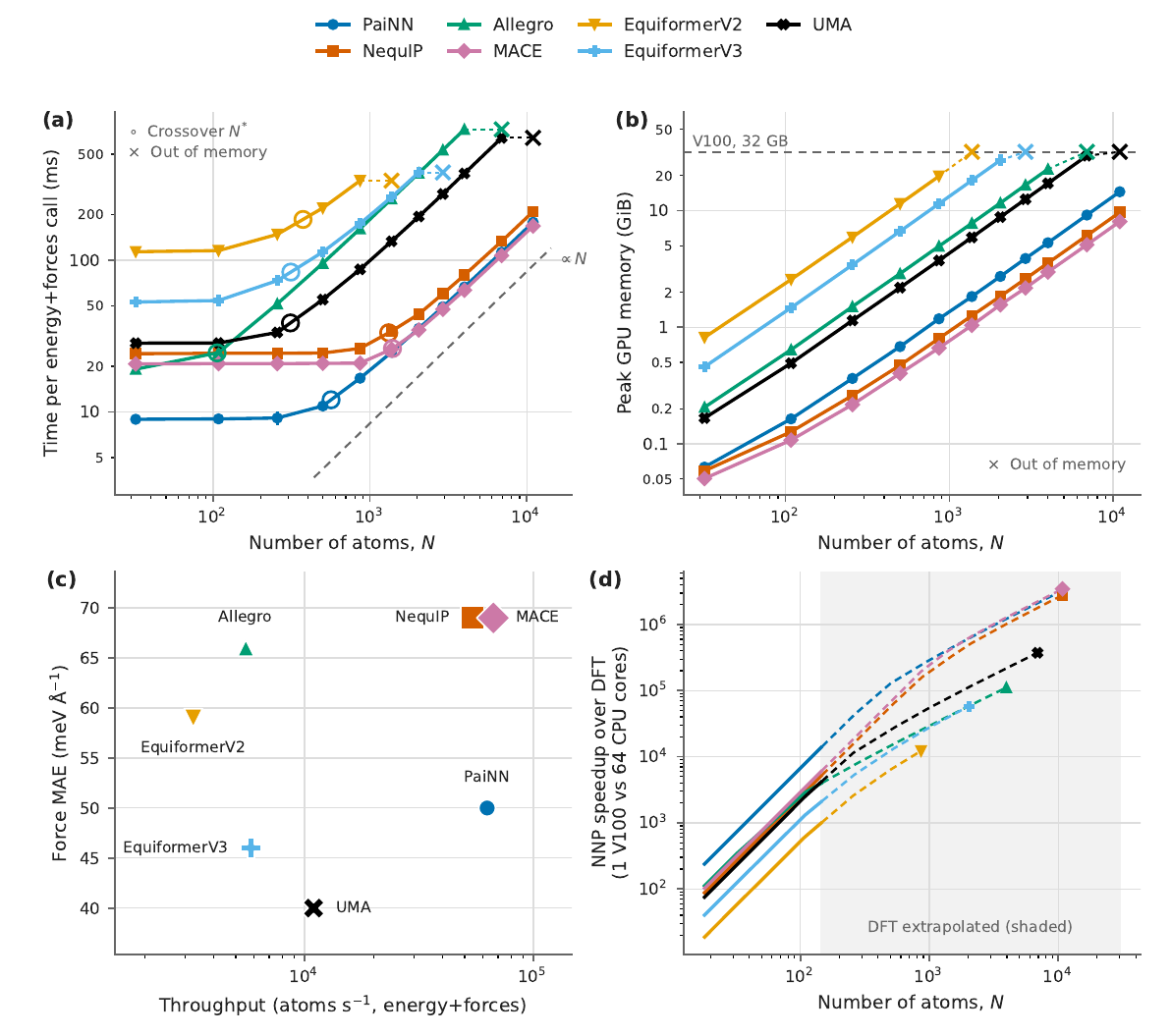}
    \caption{Inference cost of the architectures via IANN. (a) Wall time per energy-and-force call against system size. (b) Peak GPU memory versus the number of atoms. (c) Force MAE against sustained throughput for each model. (d) NNP Speedups over DFT (VASP, RPBE). The shaded region is where the DFT reference is extrapolated.}
    \label{fig:models_cost}
\end{figure}

While model accuracy is a primary criterion of merit, computational efficiency is an equally vital consideration, especially for long-timescale and large-scale applications such as molecular dynamics and high-throughput screening.\cite{Schran2021} To provide a comprehensive assessment beyond static error metrics, we benchmark the inference cost across all seven architectures trained on the MPtrj database. The wall-clock time per evaluation against the number of atoms, peak GPU memory versus the number of atoms, force MAE versus sustained throughput, and NNP speedup over DFT (VASP, RPBE functional) are considered to evaluate the cost for each model (Figure~\ref{fig:models_cost}). All benchmarks were conducted in single precision (\texttt{float32}) on a single NVIDIA Tesla V100 GPU (32~GB), evaluating bulk supercells ranging from 32 to 10,976 atoms under the standardized \qty{5.5}{\angstrom} cutoff.

\FloatBarrier

Figure~\ref{fig:models_cost}a plots the wall-clock execution time per energy-and-force evaluation as a function of system size across all architectures. Open circles identify the transition threshold $N^*$, marking the crossover from the latency-bound regime to the throughput-bound scaling regime, while crosses denote the maximum system size reached before encountering out-of-memory (OOM). We can see that each architecture exhibits a distinct latency floor below which evaluation time is invariant to system size, reflecting a regime where kernel launch overhead dominates and GPU compute capacity remains underutilized. These latency floors range from \qty{9.0}{ms} for PaiNN, to \qty{19}{}--\qty{28}{ms} for NequIP, MACE, Allegro and UMA, reaching \qty{54}{ms} for EquiformerV3 and \qty{115}{ms} for EquiformerV2. This indicates that PaiNN requires the shortest time for GPU kernel launches and data transfers, whereas EquiformerV2 takes the longest. Meanwhile, NequIP, MACE, Allegro, and UMA exhibit similar latency floor times, and EquiformerV3 lies intermediate between this group and EquiformerV2. The crossover $N^*$ at which the per-atom cost begins to dominate ranges from about 100 atoms for Allegro to roughly 1400 for MACE and NequIP. Above $N^*$ the cost is linear in $N$ as the locality of \eqlink{eqn:energy_decomposition} requires. For the few-hundred-atom slabs typical of surface catalysis, architectural selection is effectively governed by the baseline latency floor rather than marginal scaling, as per-atom overhead remains negligible until systems reach significantly larger scales. In this regime, MACE, PaiNN, and NequIP demonstrate the lowest marginal per-atom cost, while UMA, EquiformerV3, Allegro, and EquiformerV2 exhibit steeper scaling slopes. \autoref{fig:models_cost}b displays peak GPU memory allocation versus system size, revealing strict linear scaling with atom count across all models. Consistent with the computational latency trends in \autoref{fig:models_cost}a, PaiNN, NequIP, and MACE maintain the lowest memory footprints, allowing them to simulate substantially larger systems before encountering GPU memory ceilings compared to the heavier architectures.

Force MAE against sustained throughput for each model is shown in \autoref{fig:models_cost}c. Sustained throughput is defined as the number of atoms normalized by wall-clock time per energy-and-force call. In this metric, PaiNN, NequIP, and MACE exhibit higher throughput than UMA, EquiformerV3, Allegro, and EquiformerV2, reflecting their lower per-call computational cost. However, UMA, EquiformerV3, and PaiNN attain the lower force MAEs than NequIP, MACE, Allegro, and EquiformerV2, which indicates that they are more accurate in terms of force prediction in current configurations. Balancing this trade-off between predictive error and inference latency is essential when tailoring architecture selection to target application scales.

\autoref{fig:models_cost}d compares the computational speedups of different models relative to plane-wave DFT (VASP, RPBE). Across all architectures, the neural network potentials achieve speedups ranging from three to six orders of magnitude over DFT. In the extrapolated DFT regime (shaded region in \autoref{fig:models_cost}d), this computational advantage widens progressively with system size, underscoring the benefit of neural network potentials for large-scale simulations. Furthermore, inter-model performance varies by up to 10$\times$ between the fastest and slowest architectures—a performance gap that translates into substantial wall-clock savings during high-throughput screening or long-timescale molecular dynamics runs.

\FloatBarrier
\subsection{Parallel performance}

Parallel performance divides into two parts: training parallelism and inference parallelism. Scalable multi-GPU training is essential for pre-training large models on massive structural datasets. To evaluate multi-GPU scaling efficiency and sustained throughput across architectures, we benchmarked training on the curated MPtrj dataset across 1 to 16 NVIDIA A100 GPUs (configured with one MPI rank per GPU) under a fixed per-device batch size, as shown in Figure~\ref{fig:si_train_a100}. It can be seen that training acceleration scales near-linearly as GPU counts increase, with individual architectures achieving speedups between 10.5$\times$ and 13.7$\times$ across 16 GPUs. Consequently, parallel training efficiency gradually attenuates with scale, spanning from 65.7\% to 86.0\% at 16 GPUs. Aggregate training throughput scales concurrently with GPU allocation, though absolute performance diverges markedly across architectures, ranging from 323 to 3,892 structures per second on 16 GPUs. Overall, while parallel scaling efficiency gradually declines with increasing GPU counts across all models, MACE, NequIP, and PaiNN sustain the highest multi-GPU training throughput, in close alignment with their favorable memory and latency profiles in \autoref{fig:models_cost}b. In addition, we profiled the computational budget of an individual training iteration across five key components: data loading with neighbor-list construction, host-to-device transfer, forward and backward passes, the gradient all-reduce, and the optimizer update with gradient clipping. It can be seen that graph evaluation, neighbor-list building, and gradient synchronization clearly dominate the training overhead. Complete scaling measurements and timing distributions are detailed in Figure~\ref{fig:si_train_a100} and Tables~\ref{tab:si_train_a100} and~\ref{tab:si_train_sweep} in the Supporting Information.

\FloatBarrier
\begin{figure}[!htbp]
    \centering
    \includegraphics[width=\textwidth]{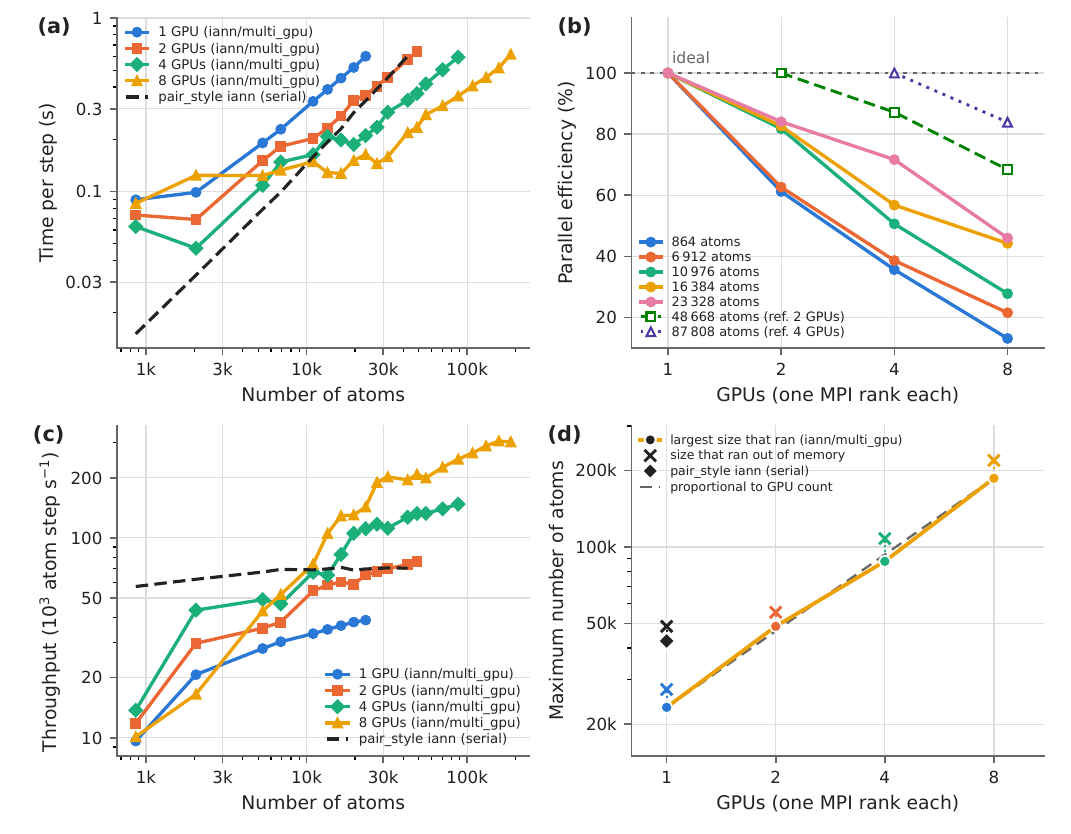}
    \caption{Performance of the IANN LAMMPS interface on A100-80GB GPUs. (a) Time per MD step against the number of atoms for 1--8 GPUs. (b) Parallel efficiency against GPU count. (c) Throughput against the number of atoms. (d) Maximum number of atoms versus GPU count.}
    \label{fig:lammps_scaling}
\end{figure}

Inference parallelism denotes distributing model evaluation across multiple GPUs, which is essential for large-scale simulations. We evaluated the parallel inference scalability of IANN using the PaiNN foundation potential deployed through our native LAMMPS interface (\texttt{pair\_style iann/multi\_gpu}). Benchmarks were conducted on thermally rattled FCC Cu supercells across 1 to 8 NVIDIA A100 GPUs (\qty{80}{GB}), assigning one MPI rank per GPU. \autoref{fig:lammps_scaling}a presents the wall-clock time per MD step across system sizes and GPU allocations. It can be seen that
the per-step wall-clock latency decreases systematically as additional GPUs are allocated, with multi-GPU acceleration becoming particularly pronounced for larger systems. Conversely, for smaller systems below approximately 5000 atoms, scaling up to 8 GPUs yields no performance gain due to device underutilization and communication bottlenecks. When compared against the serial baseline, execution on 1 or 2 GPUs using the parallel implementation incurs higher latency across all system sizes. This runtime penalty occurs because spatial domain decomposition in \texttt{pair\_style iann/multi\_gpu} requires each MPI rank to evaluate extended ghost-atom buffers across subdomain boundaries. Consequently, the aggregate atom count evaluated across ranks exceeds that of serial execution, increasing per-step computation. Additionally, inter-GPU and cross-node communication overhead adds to the total wall-clock time per MD step. Beyond two GPUs, however, per-step wall-clock latency drops below the serial baseline for sufficiently large systems, confirming that distributed compute gains outpace communication overhead as system scale expands. Parallel inference efficiency is illustrated in \autoref{fig:lammps_scaling}b: parallel efficiency diminishes with higher GPU counts, with smaller systems suffering greater efficiency losses. For instance, in an 864-atom supercell, 8-GPU efficiency drops below 20\%. However, when the number of atoms is larger than 23k, the parallel efficiency is about 50\%, which is fully consistent with \autoref{fig:lammps_scaling}a.

Sustained throughput, defined as atom count normalized by wall-clock time per MD step, is illustrated in \autoref{fig:lammps_scaling}c. Computational throughput increases steadily with GPU allocation across all scales, except for the 8-GPU configuration on small systems. In large-scale regimes (around 30k atoms), the 8-GPU throughput reaches approximately 5$\times$ that of single-GPU parallel execution (\texttt{iann/multi\_gpu}) and 3$\times$ that of the serial baseline. The maximum number of atoms that can be handled is also evaluated, and shown in \autoref{fig:lammps_scaling}d. Cross markers indicate configurations that encountered out-of-memory errors. It can be seen that the maximum atom capacity scales linearly with GPU count: only 25,000 atoms can be handled on 1 GPU, while 200,000 atoms can be handled on 8 GPUs, which is an exact eightfold increase. This indicates that the maximum number of atoms is limited by the available GPU memory. Furthermore, the serial implementation handles roughly twice the capacity of the single-GPU parallel configuration, as serial execution bypasses ghost-atom buffer allocation across subdomain boundaries. Overall, parallel inference achieves high scalability in the large-system regime, while total atom capacity remains strictly memory-bound. Comprehensive numerical records, including per-step execution times, parallel efficiencies, and memory thresholds across all configurations, are detailed in Table~\ref{tab:si_lmp_scaling} in the Supporting Information.

\subsection{Limitations}

Long-range electrostatics, charge transfer, and magnetic degrees of freedom are not captured by the strictly local decomposition of \eqlink{eqn:energy_decomposition} and remain currently unsupported.\cite{Schran2021} Furthermore, because the released foundation models inherit the DFT parameters of their underlying training datasets, absolute energies are not transferable across different exchange-correlation functionals---the primary motivation for organizing them by functional rather than pooling them into a single model. Regarding the curated databases in \autoref{tab:databases}, outlier filtering enhances training stability but renders them unsuitable for direct benchmarking against published literature, as the excluded structures often constitute the most physically challenging configurations; the uncurated source datasets should be used for such benchmarks. Finally, because architectures are natively reimplemented within our unified interface rather than wrapped as upstream external dependencies, checkpoint binaries from original third-party implementations cannot be loaded directly into IANN. Consequently, models within this framework must either be trained from scratch or fine-tuned from an existing IANN foundation checkpoint.

\section{Conclusions}
In this work, we presented IANN, an open-source framework for interatomic neural networks designed to bridge architectural development and production molecular modeling. By integrating diverse equivariant graph neural networks within a unified data structure, trainer, and deployment interface, IANN enables seamless architectural transitions via a single-word modification in the configuration dictionary. Across all supported architectures, atomic forces, virials, and stresses are derived through unified analytic differentiation with respect to graph edge vectors, ensuring exact evaluations under periodic boundary conditions. Furthermore, distributed data-parallel training automatically adapts to SLURM, PBS, or standard MPI schedulers, allowing a single execution script to scale without modification from multi-GPU workstations to multi-node supercomputers. Trained, self-describing checkpoints can be deployed directly as ASE calculators with optional uncertainty estimation, compiled to TorchScript predictors supporting on-the-fly variance evaluation during single- or multi-GPU LAMMPS simulations, or queried via an agent interface. This architectural uniformity facilitates rigorous, controlled benchmarking: across 1.6 million PBE-level structures, energy and force errors vary only modestly among architectures—spanning \qty{0.029}{} to \qty{0.054}{eV/atom} and \qty{0.040}{} to \qty{0.069}{eV/\angstrom}, respectively. In contrast, computational latency and memory footprints diverge substantially, demonstrating that architecture selection for production simulations must be governed by accuracy–efficiency trade-offs rather than error metrics alone. Finally, our released foundation models—systematically organized by exchange-correlation functional—provide reusable inductive priors that close the loop on active-learning workflows natively backed by uncertainty quantification.

\section*{Code availability}
IANN is released under the MIT license and is available at \url{https://github.com/changzhiai/IANN}, with documentation at \url{https://iann.readthedocs.io}. The version described in this paper is archived at \url{https://doi.org/10.5281/zenodo.17809948}.

\section*{Data availability}
The curated reference databases of \autoref{tab:databases} are available at \url{https://huggingface.co/datasets/changzhiai/dft-databases}. The foundation models of \autoref{tab:foundation} and the architecture comparison of \autoref{tab:accuracy}, together with the training script, are available at \url{https://huggingface.co/changzhiai/iann-foundation-models}. The underlying published datasets are cited in \autoref{tab:databases} and remain the appropriate reference for benchmarking.

\section*{Acknowledgments}
This research was supported by the U.S. Department of Energy, Office of Science, Office of Basic Energy Sciences, Chemical Sciences, Geosciences, and Biosciences Division, Catalysis Science Program to the SUNCAT Center for Interface Science and Catalysis. This research used high-performance computing (HPC) resources of the National Energy Research Scientific Computing Center (NERSC), a Department of Energy User Facility using NERSC award BES-ERCAP0032527. This work also used the HPC resources of the SLAC Shared Science Data Facility (S3DF) at SLAC National Accelerator Laboratory operated by Stanford University for the U.S. Department of Energy’s Office of Science. Work was also performed on the Carbon HPC cluster at the Center for Nanoscale Materials, a U.S. Department of Energy Office of Science User Facility, and was supported by the U.S. DOE, Office of Basic Energy Sciences, under Contract No. DE-AC02-06CH11357.

\section*{Keywords}
Machine-learning interatomic potentials; Equivariant graph neural networks; Molecular dynamics; Heterogeneous catalysis; Scientific software; Uncertainty quantification

\newpage
\printbibliography 



\end{document}


\maketitle
\clearpage

\setcounter{tocdepth}{1}
\tableofcontents
\clearpage

\section{Installation}

IANN requires Python 3.8 or newer and PyTorch 1.9 or newer. The package and its dependencies (\texttt{numpy}, \texttt{ase}, \texttt{scipy}, \texttt{asap3}, \texttt{e3nn}, \texttt{toml}, \texttt{py-cpuinfo}) are installed with

\begin{lstlisting}[language=bash]
git clone https://github.com/changzhiai/IANN.git
cd IANN
pip install -e .
\end{lstlisting}

\noindent GPU support follows the standard PyTorch CUDA installation. \texttt{e3nn} is used for equivariant tensor algebra by default. The optional NVIDIA \texttt{cuEquivariance}\cite{cuEquivariance} library can also be used to accelerate the equivariant tensor products, especially for NequIP and MACE. The LAMMPS interface additionally requires LibTorch (the PyTorch C++ API), CMake $\ge$ 3.18 and a C++11 compiler; build instructions are given in the \href{https://iann.readthedocs.io}{Documentation}.

\section{Framework implementation details}
\label{sec:si_impl}

This section collects the implementation detail behind the framework overview of the manuscript, in the order the modules are described there. It documents the choices a reader may need in order to reproduce or extend it.

\subsection{Graph construction and batching}

Graph construction is handled by \texttt{AseDataReader}. Periodic systems use the linear-scaling cell-list neighbour list from \texttt{asap3}, non-periodic systems a direct distance-matrix cutoff; in both cases the reader returns edge indices together with the minimum-image edge vectors, and flags the edge vectors as requiring gradients when forces are needed. \texttt{AseDataset} wraps an ASE trajectory or database and reads structures lazily, so that datasets larger than memory are handled without a preprocessing step or a database conversion. Batching is performed by a collate function that concatenates the per-structure tensors, builds an image index that assigns every atom to its structure, and offsets the edge indices of each structure by the cumulative atom count, so that a batch is itself a single valid \texttt{AtomsData} object describing one disconnected graph.

\subsection{Shared building blocks}

Equivariant tensor algebra is provided by \texttt{e3nn},\cite{Geiger2022} and the NequIP and MACE implementations transparently use NVIDIA's \texttt{cuEquivariance}\cite{cuEquivariance} kernels when the optional dependency is installed and chosen. Building blocks that are architecture-independent --- the Bessel radial basis of the manuscript, the polynomial cutoff envelope, gated nonlinearities and activation functions --- live in \texttt{iann.tools} and are shared.

\subsection{Adding an architecture}

Adding an architecture requires implementing a \texttt{torch.nn.Module} whose \texttt{forward} accepts and returns \texttt{AtomsData} and registering it in the model registry; the data pipeline, trainer, calculators and LAMMPS export are then inherited unchanged.

\subsection{Trainer configuration}

\begin{sloppypar}
\texttt{Trainer} is configured by a single dictionary that is merged into a documented set of defaults (reproduced in full in the next section). It covers the architecture (\texttt{num\_channels}, \texttt{num\_layers}, \texttt{cutoff}, and architecture-specific keys such as $\ell_\mathrm{max}$, $m_\mathrm{max}$), the objective (\texttt{forces\_weight}, \texttt{virial\_weight}, \texttt{stress\_weight}, \texttt{loss\_per\_atom}), optimisation (\texttt{optimizer\_type}, \texttt{learning\_rate}, \texttt{scheduler\_type}, \texttt{max\_grad\_norm}, \texttt{max\_steps} or \texttt{max\_epochs}, \texttt{stop\_patience}), data handling (\texttt{val\_ratio}, \texttt{random\_seed}, \texttt{save\_split}, \texttt{load\_split}, \texttt{norm\_data}, \texttt{norm\_per\_atom}), distributed execution and output. Adam, AdamW\cite{Kingma2015,Loshchilov2019} and SGD optimisers are available, together with seven learning-rate schedules selected by \texttt{scheduler\_type}: \texttt{LambdaLR} (the default), \texttt{ReduceLROnPlateau}, \texttt{CosineAnnealingLR}, \texttt{CosineAnnealingWarmRestarts}, \texttt{StepLR}, \texttt{MultiStepLR} and \texttt{ExponentialLR}.\cite{Loshchilov2017}
\end{sloppypar}

\subsection{Reproducibility safeguards in the trainer}

Several decisions in the trainer exist specifically to remove recurring sources of error. Dataset splits can be written to and reloaded from a JSON file, so that a restarted or fine-tuned run provably reuses the original validation set. Energy statistics for normalisation are computed over the training split only, optionally from a random subsample and with threaded file I/O, which keeps the start-up cost of a multi-million-structure dataset bounded. For architectures that normalise messages by the mean coordination number, the average number of neighbours is measured from the training data automatically rather than being left as a user-supplied constant. Early stopping monitors an exponentially smoothed validation loss, which avoids terminating on a single noisy evaluation. Checkpoints store the architecture name and the complete set of architecture keyword arguments alongside the parameters, so that a model can be reconstructed exactly from the file alone; for older checkpoints that predate this convention, the loader infers the architecture from signatures in the parameter names. Restarts distinguish two intentions explicitly: resuming continues the step counter, optimiser state and schedule exactly, whereas fine-tuning (\texttt{reset\_lr}) reloads the parameters but starts a fresh schedule at the requested learning rate.

\subsection{Detecting the distributed environment}

Distributed data-parallel training is enabled with \texttt{distributed=True} and configured by inspecting the environment. The trainer recognises SLURM (\texttt{SLURM\_PROCID}, \texttt{SLURM\_NTASKS},
\texttt{SLURM\_LOCALID}) and the rank variables of OpenMPI, MPICH/Intel MPI and
MVAPICH2, and from these derives the global rank, world size, per-node local rank and
rendezvous address. PBS/Torque allocations are supported through the same path: PBS
exports no rank variables of its own, so the ranks are read from whichever MPI launcher
started the processes, while \texttt{PBS\_NODEFILE} supplies the rendezvous host and the
node list recorded in the log, deduplicated because PBS repeats each host once per
requested core. The NCCL backend is selected on GPUs and Gloo on CPUs. On a bare multi-GPU node with no external launcher, the trainer spawns one worker process per visible GPU itself; under \texttt{srun} or \texttt{mpirun} it detects that the launcher has already created the processes and only reads its own coordinates. The practical consequence is that the training script of the manuscript is submitted unchanged on a laptop, a workstation and a multi-node allocation, with the job script alone determining the degree of parallelism.

\subsection{Resolving and fine-tuning foundation models}

A small selection ships inside the package and is resolved by name through \texttt{foundation\_model()} irrespective of the working directory; the complete set is hosted separately. Any of them can be used as a zero-shot calculator or as the initialisation for fine-tuning, by passing the checkpoint path as \texttt{load\_model} in the trainer configuration together with \texttt{reset\_lr} to start a fresh learning-rate schedule.

\subsection{Exporting a model to LAMMPS}

For production MD, a trained model is compiled to TorchScript by \texttt{convert\_model\_for\_lammps}, which wraps the model in a module whose \texttt{forward} takes only plain tensors --- the same seven fields that define the graph --- and returns a dictionary of energy, forces and atomic energies. Because that signature is architecture-independent, one wrapper and one exporter cover all models, and a companion function combines several checkpoints into a single ensemble module.

\subsection{Agent interface}
\label{sec:si_agent}

\textbf{Command line.} The package installs an \texttt{iann} entry point with nine subcommands --- \texttt{doctor}, \texttt{models}, \texttt{foundation}, \texttt{inspect}, \texttt{predict}, \texttt{train}, \texttt{status}, \texttt{export} and \texttt{agent} --- covering the operations of the layers above; \texttt{doctor} reports whether an environment can run IANN, and \texttt{inspect} which structural parameters a checkpoint does \emph{not} record --- the usual cause of a parameter-shape mismatch when a model is rebuilt from its weights. Every subcommand accepts \texttt{--json}, under which standard output carries nothing but the JSON document, and exit codes separate success from a failed operation and from a malformed command line, so a failure is never a traceback nor a silent empty result. Training requires an explicit step budget, because \texttt{max\_steps} defaults to $10^{6}$ and a call that omitted one would begin a run lasting days.

\textbf{MCP server.} The same operations are served as nine tools over stdio by \texttt{mcp\_server}, so that a client implementing the Model Context Protocol, and therefore a large-language-model agent, can call them without knowing how the framework is invoked. Both front ends call one implementation --- a single function per operation, returning a plain dictionary --- so the two cannot drift apart; the protocol SDK is an optional dependency, and the command line needs nothing beyond the base requirements.

\textbf{Agent skills.} \texttt{iann agent install} writes a set of repository notes and five skills into a working copy, recording what cannot be inferred from the source: which environment to use, that scripts are run from the repository root, and that a training run must be given a step budget. The notes are not specific to any one assistant. Job submission to SLURM or PBS, uploads to the model hub and deletion of files are not exposed at all, which confines an automated caller to reads and a bounded amount of local computation; the whole interface is contained in the \texttt{iann.agent} subpackage and documented in the \href{https://iann.readthedocs.io/en/latest/agents.html}{Documentation}.

\section{Default trainer configuration}
\label{sec:si_config}

Table~\ref{tab:config} lists the configuration keys recognised by
\texttt{iann.trainer.Trainer} together with their defaults. Any subset can be
overridden through the \texttt{config} dictionary passed to the constructor;
keys that are not recognised by the trainer are forwarded to the model
constructor, which is how architecture-specific parameters such as
\texttt{lmax}, \texttt{num\_basis} or \texttt{parity} are set.

\begingroup
\small
\setlength{\tabcolsep}{4pt}
\begin{longtable}{p{0.30\textwidth} p{0.16\textwidth} p{0.46\textwidth}}
\caption{Configuration keys of \texttt{iann.trainer.Trainer} and their default
values.}\label{tab:config}\\
\toprule
Key & Default & Meaning \\
\midrule
\endfirsthead
\toprule
Key & Default & Meaning \\
\midrule
\endhead
\multicolumn{3}{l}{\emph{Model}}\\
\texttt{num\_channels}      & 128    & Feature channels per node \\
\texttt{num\_layers}        & 3      & Message-passing layers \\
\texttt{cutoff}             & 5.5    & Cutoff radius (\AA) \\
\midrule
\multicolumn{3}{l}{\emph{Objective}}\\
\texttt{forces\_weight}     & 0.9    & Weight $w_F$ of the force term \\
\texttt{virial\_weight}     & 0.0    & Weight $w_W$ of the virial term \\
\texttt{stress\_weight}     & 0.0    & Weight $w_\sigma$ of the stress term \\
\texttt{loss\_per\_atom}    & True   & Divide the energy residual by the number of atoms \\
\midrule
\multicolumn{3}{l}{\emph{Optimisation}}\\
\texttt{optimizer\_type}    & adam   & \texttt{adam}, \texttt{adamw} or \texttt{sgd} \\
\texttt{learning\_rate}     & 1e-4   & Initial learning rate \\
\texttt{scheduler\_type}    & LambdaLR & One of \texttt{ReduceLROnPlateau}, \texttt{LambdaLR}, \texttt{CosineAnnealingLR}, \texttt{CosineAnnealingWarmRestarts}, \texttt{StepLR}, \texttt{MultiStepLR}, \texttt{ExponentialLR} \\
\texttt{max\_grad\_norm}    & 10.0   & Gradient-norm clipping threshold \\
\texttt{batch\_size}        & 12     & Structures per batch and per rank \\
\texttt{max\_steps}         & 1000000 & Optimiser steps before stopping \\
\texttt{max\_epochs}        & None   & Alternative stopping criterion \\
\texttt{stop\_patience}     & 200    & Evaluations without improvement before early stopping \\
\texttt{log\_interval}      & 2000   & Steps between validation and logging \\
\midrule
\multicolumn{3}{l}{\emph{Data}}\\
\texttt{val\_ratio}         & 0.1    & Fraction held out for validation \\
\texttt{random\_seed}       & 666    & Seed for splitting and initialisation \\
\texttt{save\_split}        & False  & File name for storing the split \\
\texttt{load\_split}        & False  & File name of a split to reuse \\
\texttt{norm\_data}         & False  & Standardise energy targets \\
\texttt{norm\_per\_atom}    & False  & Standardise per-atom energies \\
\texttt{norm\_sample\_size} & None   & Subsample size for the normalisation statistics \\
\texttt{norm\_num\_workers} & 0      & Threads used to read the statistics \\
\midrule
\multicolumn{3}{l}{\emph{Restart}}\\
\texttt{load\_model}        & False  & Checkpoint path to resume or fine-tune from \\
\texttt{reset\_lr}          & False  & \texttt{True}: fresh schedule and step counter (fine-tuning); \texttt{False}: resume exactly \\
\midrule
\multicolumn{3}{l}{\emph{Distributed}}\\
\texttt{dist\_timeout}      & 600    & Timeout of collective operations (s) \\
\texttt{master\_port}       & 12356  & Rendezvous port \\
\midrule
\multicolumn{3}{l}{\emph{Output}}\\
\texttt{device}             & None   & \texttt{cpu} or \texttt{cuda}; auto-detected if unset \\
\texttt{output\_dir}        & output & Output directory \\
\texttt{output\_log}        & output.log & Log file \\
\texttt{output\_model}      & model.pt & Checkpoint file \\
\texttt{log\_input}         & False  & Echo the user configuration to the log \\
\texttt{debug}              & False  & Verbose diagnostics \\
\bottomrule
\end{longtable}
\endgroup

\section{DFT databases}
\label{sec:si_refdata}

The DFT databases are collected and curated as shown in Table~1 in the manuscript, regrouped by exchange-correlation functional and converted to ASE trajectory format. The sources span three functionals: PBE for MPtrj, sAlex, OMat24, MatPES and OC22; RPBE for OC20 and OC25; and r$^2$SCAN for the meta-GGA subsets of the Materials Project and MatPES. Note that a shared functional does not imply identical settings. For example, MPtrj, OMat24 and OC22 apply a Hubbard correction, the remaining PBE sources do not. OC25 has D3 dispersion corrections, OC20 does not. Outlier structures were removed from every source dataset based on abnormally high energy or force components, abnormally short interatomic distances, or the presence of overlapping atoms.

\section{Foundation models details across architectures}

Every architecture compared here is reimplemented within IANN against the shared \texttt{AtomsData} interface rather than taken from the authors' released code, so the settings below are those of the IANN implementations and need not coincide with the ones published with the original models. 

The configuration details of each architecture in the accuracy comparison of the manuscript are given in Table~\ref{tab:hparams}. The table lists the number of channels, the highest spherical harmonic degree $\ell_\mathrm{max}$ and order $m_\mathrm{max}$, the number of message-passing layers, the total number of trainable parameters, the per-rank batch size, the initial learning rate and the gradient-norm clipping threshold. All runs used the curated MPtrj database (1,579,091 PBE structures), a cutoff of \qty{5.5}{\angstrom}, the Adam optimiser with the \texttt{LambdaLR} schedule, the per-atom energy loss and an early-stopping patience of 500 evaluations; all other keys retain the defaults of Table~\ref{tab:config}. 

\begin{table}[H]
\centering
\caption{Settings of the training of each architecture.
$\ell_\mathrm{max}$ is the highest spherical harmonic degree the features carry and 
$m_\mathrm{max}$ the highest order kept at each degree; NequIP, MACE and Allegro build 
full $O(3)$ irreps, so $m_\mathrm{max}=\ell_\mathrm{max}$, whereas the Equiformers 
truncate below $\ell_\mathrm{max}$ to cut the cost of the SO(2) convolution. 
$^\ast$Equivalent values: PaiNN exposes neither setting, but its Cartesian scalars and 
vectors are equivariant to $\ell\leq1$ with all orders present.}
\label{tab:hparams}
\small
\setlength{\tabcolsep}{4pt}
\begin{tabular}{lrrrrrrrr}
\toprule
Model & Channels & $\ell_\mathrm{max}$ & $m_\mathrm{max}$ & Layers & Parameters & Batch & LR & Clip \\
\midrule
PaiNN$^\ast$ & 128 & 1 & 1 & 3 & 600,577 & 16 & $1\times10^{-4}$ & 10.0 \\
NequIP & 32 & 1 & 1 & 3 & 1,379,513 & 16 & $1\times10^{-4}$ & 10.0 \\
Allegro & 128 & 1 & 1 & 4 & 567,624 & 16 & $1\times10^{-4}$ & 10.0 \\
MACE & 48 & 1 & 1 & 3 & 1,429,376 & 16 & $1\times10^{-4}$ & 10.0 \\
EquiformerV2 & 16 & 4 & 2 & 3 & 773,089 & 8 & $1\times10^{-4}$ & 10.0 \\
EquiformerV3 & 32 & 3 & 2 & 3 & 943,249 & 16 & $1\times10^{-4}$ & 10.0 \\
UMA & 64 & 2 & 2 & 3 & 1,173,698 & 16 & $1\times10^{-4}$ & 10.0 \\
\bottomrule
\end{tabular}
\end{table}

\begin{sloppypar}
EquiformerV2 additionally used \texttt{grid\_resolution=12} and
\texttt{avg\_degree=43.0}; UMA used \texttt{hidden\_channels=64},
\texttt{edge\_channels=64} and \texttt{num\_distance\_basis=128}. All
architectures except PaiNN were trained with per-atom energy standardisation
(\texttt{norm\_per\_atom=True}), with the statistics estimated from a random
subsample of 5000 training frames (\texttt{norm\_sample\_size}). For MACE, NequIP
and Allegro the trainer measured the average number of neighbours from the
training data automatically.
\end{sloppypar}

\section{Foundation model details across datasets}

\begin{table}[H]
\centering
\caption{Accuracy of the released PaiNN foundation models. Energies are per
atom; forces are per Angstrom. $N_\mathrm{struct}$ and
$N_\mathrm{atom}$ count the structures and atoms entering each fit.}
\label{tab:si_parity}
\small
\setlength{\tabcolsep}{10pt}
\begin{tabular}{lrrrrrr}
\toprule
\multirow{2}{*}{Dataset} & \multirow{2}{*}{$N_\mathrm{struct}$}
  & \multirow{2}{*}{$N_\mathrm{atom}$} & \multicolumn{2}{c}{Energy (eV\,atom$^{-1}$)}
  & \multicolumn{2}{c}{Force (eV\,\AA$^{-1}$)} \\
\cmidrule(lr){4-5} \cmidrule(lr){6-7}
& & & MAE & $R^2$ & MAE & $R^2$ \\
\midrule
\multicolumn{7}{l}{\emph{PBE}} \\
\quad Alexandria & 10,422,893 & 107,451,444 & 0.0408 & 0.9988 & 0.0513 & 0.9512 \\
\quad MatPES & 434,083 & 3,865,201 & 0.0463 & 0.9984 & 0.0930 & 0.9583 \\
\quad MPtrj & 1,579,058 & 49,229,839 & 0.0517 & 0.9980 & 0.0498 & 0.9767 \\
\quad OC22 & 8,198,695 & 654,300,266 & 0.0381 & 0.9986 & 0.0528 & 0.8373 \\
\quad OMat24 & 11,261,400 & 156,856,550 & 0.0261 & 0.9991 & 0.1382 & 0.9961 \\
\addlinespace
\multicolumn{7}{l}{\emph{RPBE}} \\
\quad OC20 & 1,999,216 & 146,438,327 & 0.0263 & 0.9997 & 0.0630 & 0.9422 \\
\quad OC25 & 7,369,601 & 1,064,203,162 & 0.0116 & 0.9998 & 0.0753 & 0.9864 \\
\addlinespace
\multicolumn{7}{l}{\emph{r$^2$SCAN}} \\
\quad MatPES & 387,285 & 3,039,030 & 0.0446 & 0.9999 & 0.1127 & 0.9507 \\
\quad MPtrj & 238,241 & 5,022,951 & 0.0206 & 0.9999 & 0.0117 & 0.9710 \\
\bottomrule
\end{tabular}
\end{table}

\section{Parity plots for the foundation models}

Figures~\ref{fig:si_parity_pbe_mptrj}--\ref{fig:si_parity_r2scan_matpes} give fitting parity plots of energy and
forces between each foundation model and its DFT reference data, one figure per
released model in the order of Table~3 in the manuscript. The colorbar shows the
number of structures in each bin, and the dashed line is the $y=x$ diagonal. The
details of the training databases and the full error statistics are given in
Table~\ref{tab:si_parity} of the next section.

\begingroup
\setlength{\intextsep}{4pt plus 2pt minus 2pt}
\setlength{\abovecaptionskip}{4pt}
\begin{figure}[H]
\centering
\includegraphics[width=1\textwidth]{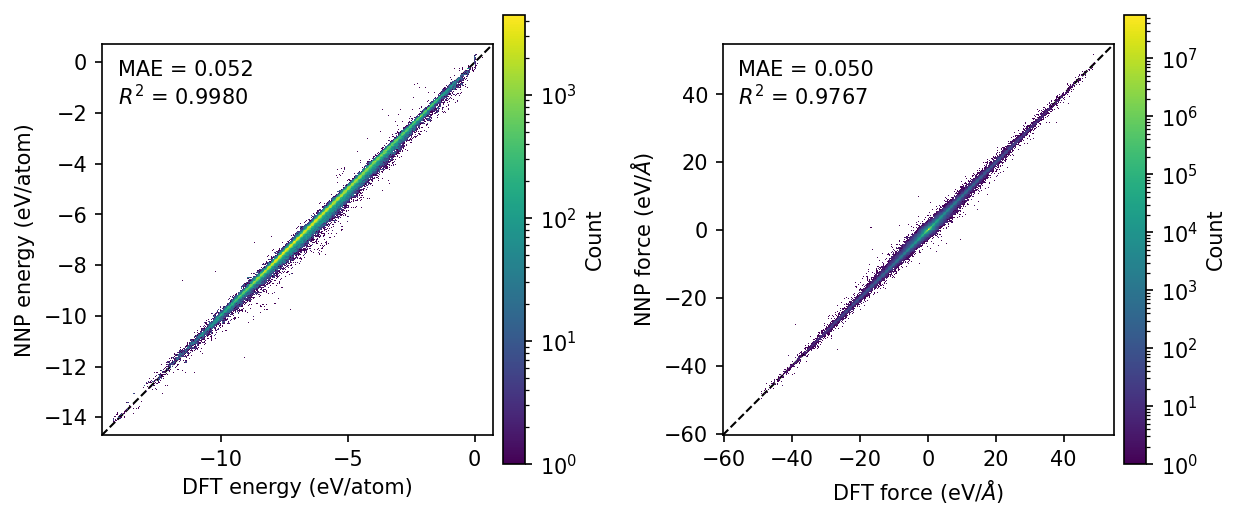}
\caption{Energy and force parity plots between the prediction of the PBE-MPtrj foundation model and the DFT reference.}
\label{fig:si_parity_pbe_mptrj}
\end{figure}

\begin{figure}[H]
\centering
\includegraphics[width=1\textwidth]{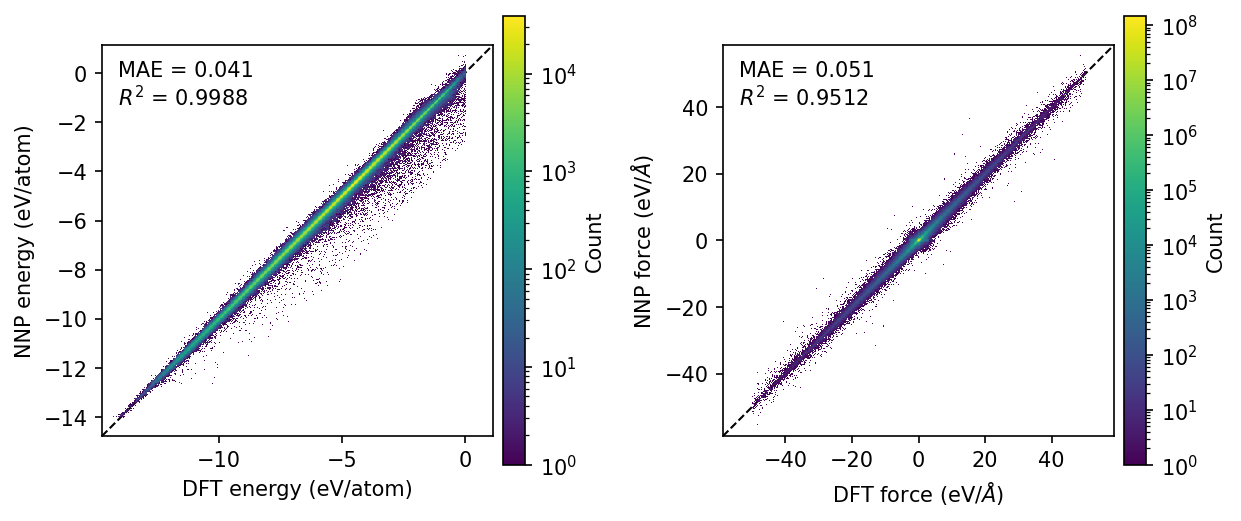}
\caption{Energy and force parity plots between the prediction of the PBE-sAlex foundation model and the DFT reference.}
\label{fig:si_parity_pbe_salex}
\end{figure}

\begin{figure}[H]
\centering
\includegraphics[width=1\textwidth]{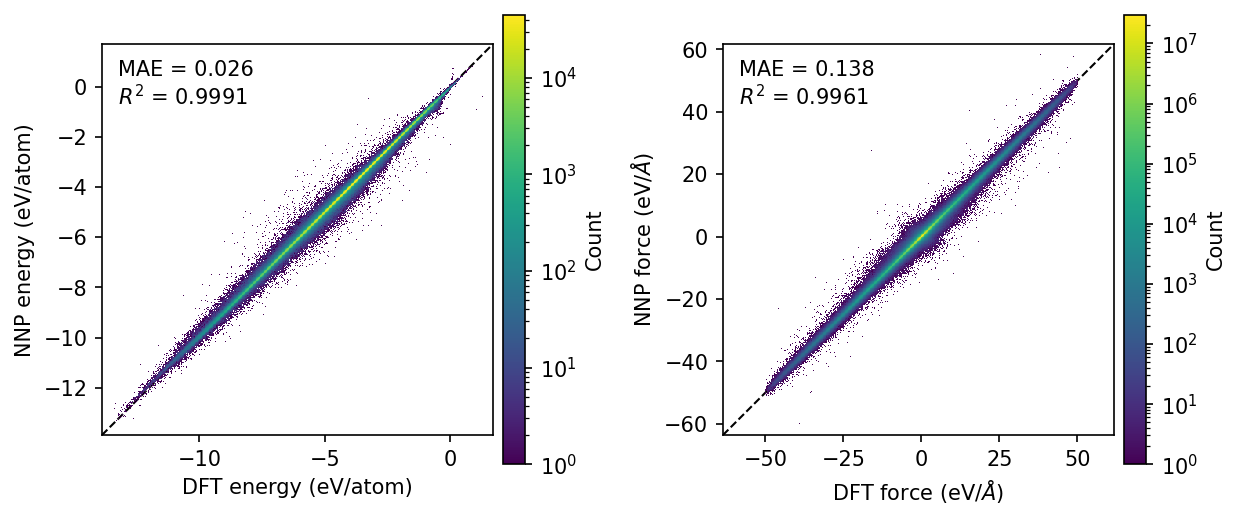}
\caption{Energy and force parity plots between the prediction of the PBE-OMat24 foundation model and the DFT reference.}
\label{fig:si_parity_pbe_omat24}
\end{figure}

\begin{figure}[H]
\centering
\includegraphics[width=1\textwidth]{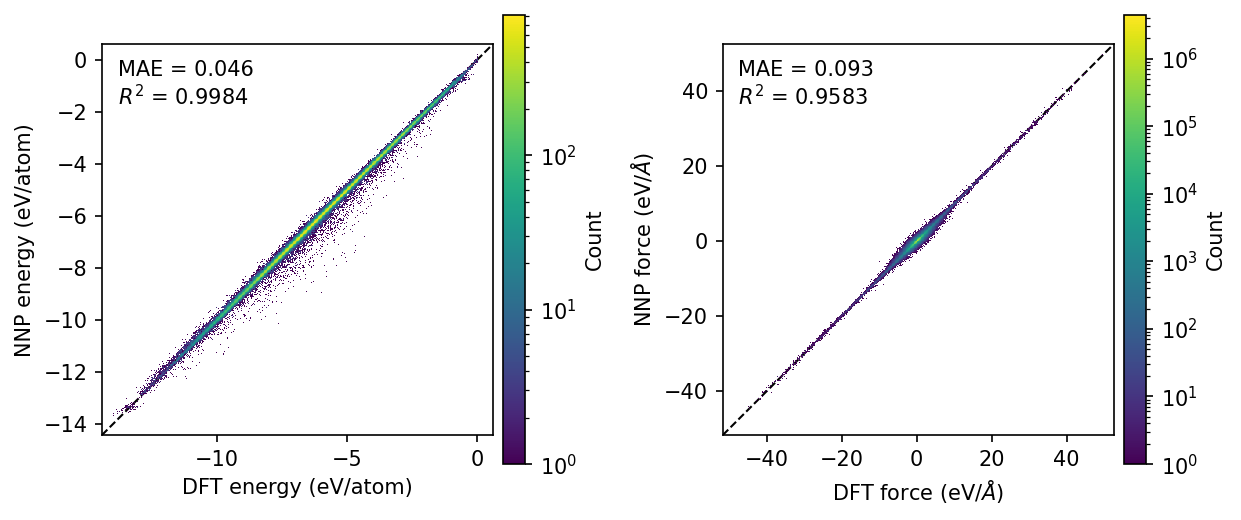}
\caption{Energy and force parity plots between the prediction of the PBE-MatPES foundation model and the DFT reference.}
\label{fig:si_parity_pbe_matpes}
\end{figure}

\begin{figure}[H]
\centering
\includegraphics[width=1\textwidth]{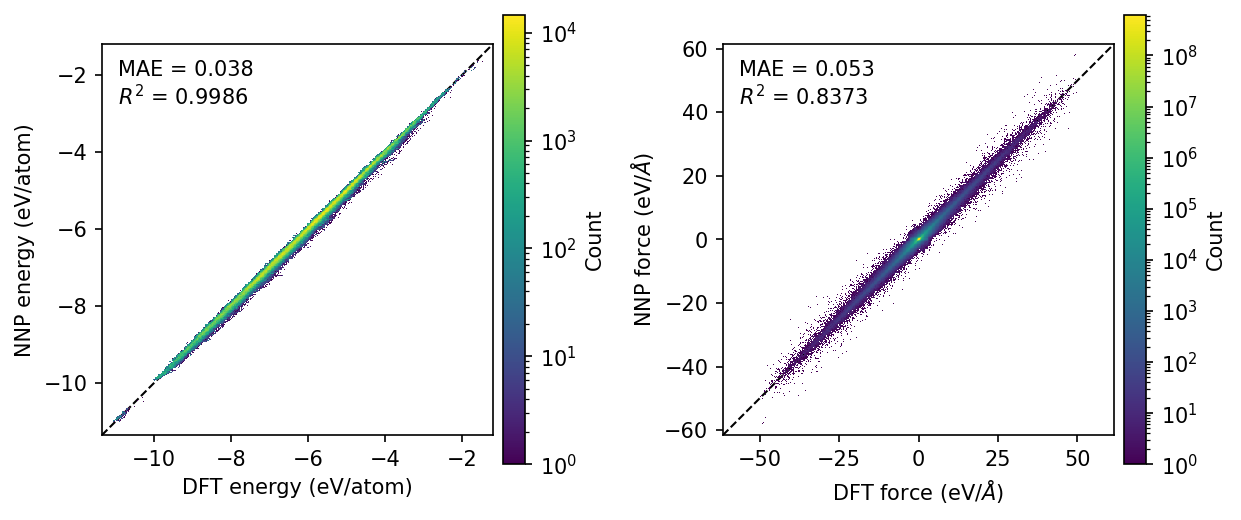}
\caption{Energy and force parity plots between the prediction of the PBE-OC22 foundation model and the DFT reference.}
\label{fig:si_parity_pbe_oc22}
\end{figure}

\begin{figure}[H]
\centering
\includegraphics[width=1\textwidth]{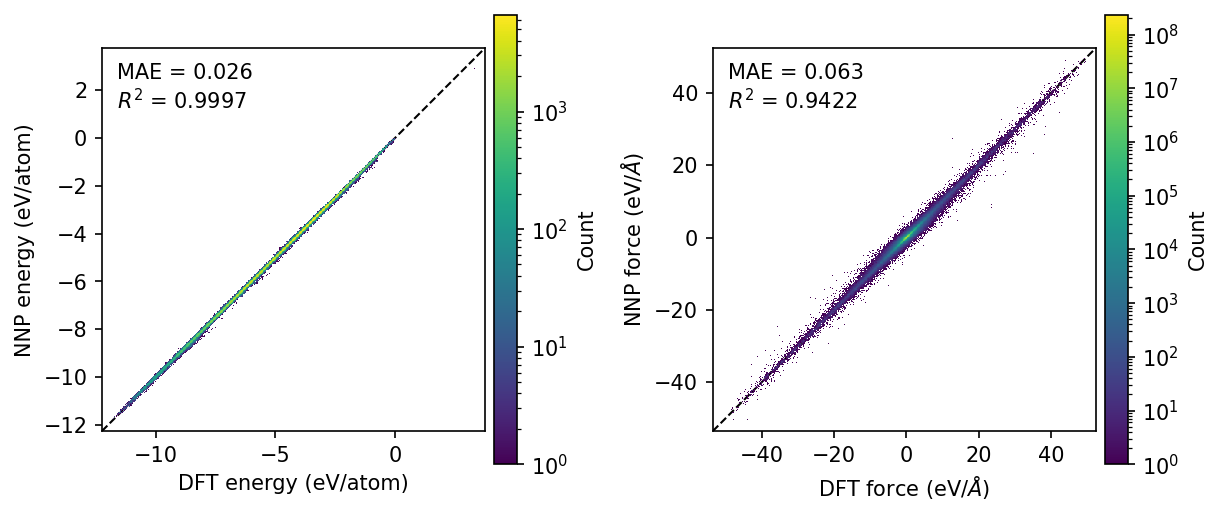}
\caption{Energy and force parity plots between the prediction of the RPBE-OC20 foundation model and the DFT reference.}
\label{fig:si_parity_rpbe_oc20}
\end{figure}

\begin{figure}[H]
\centering
\includegraphics[width=1\textwidth]{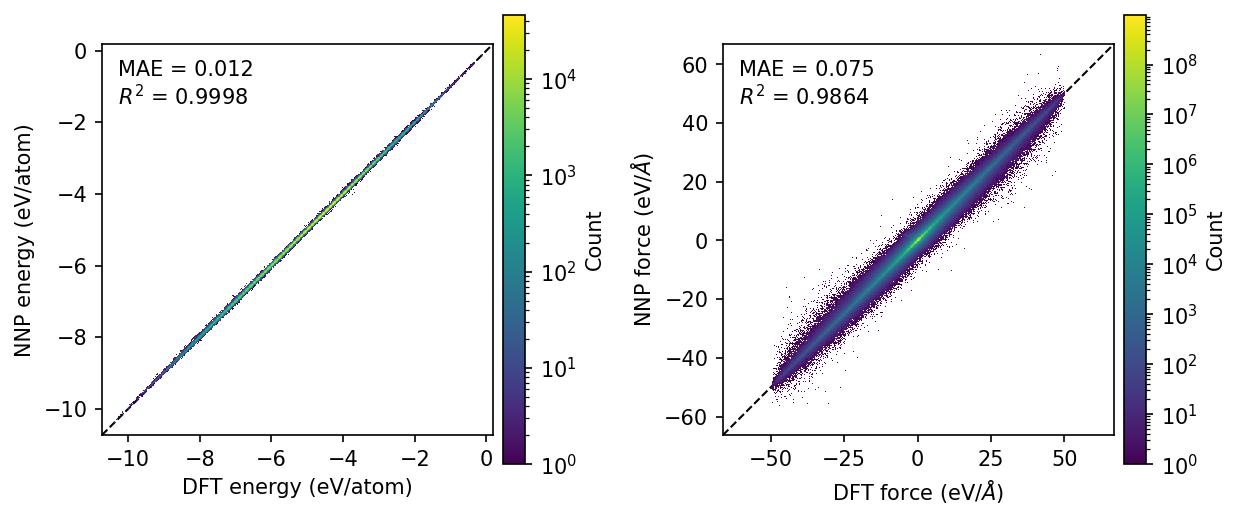}
\caption{Energy and force parity plots between the prediction of the RPBE-OC25 foundation model and the DFT reference.}
\label{fig:si_parity_rpbe_oc25}
\end{figure}

\begin{figure}[H]
\centering
\includegraphics[width=1\textwidth]{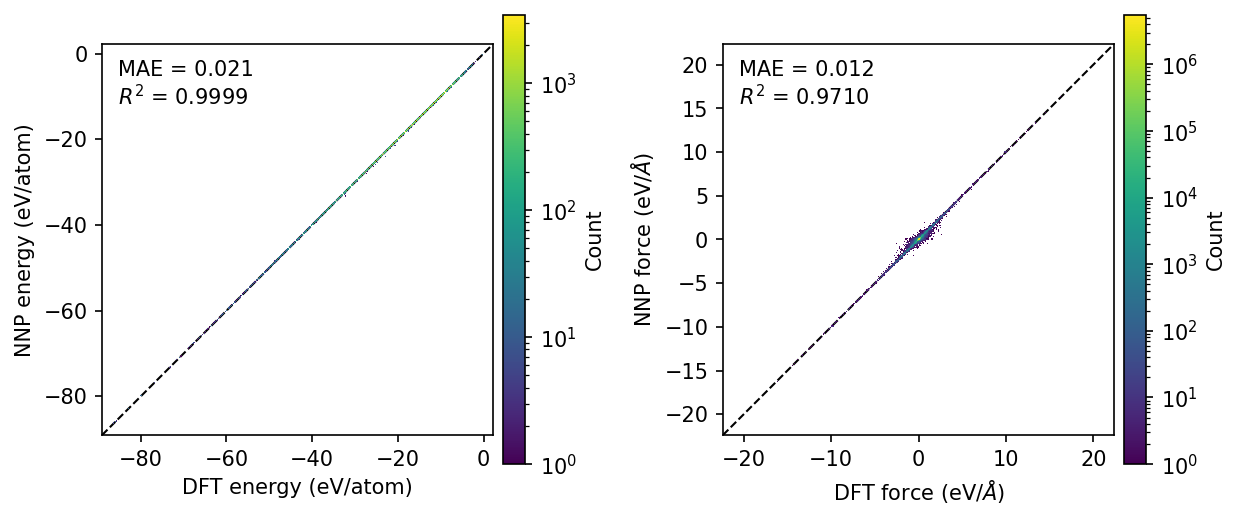}
\caption{Energy and force parity plots between the prediction of the r$^2$SCAN-MPtrj foundation model and the DFT reference.}
\label{fig:si_parity_r2scan_mptrj}
\end{figure}

\begin{figure}[H]
\centering
\includegraphics[width=1\textwidth]{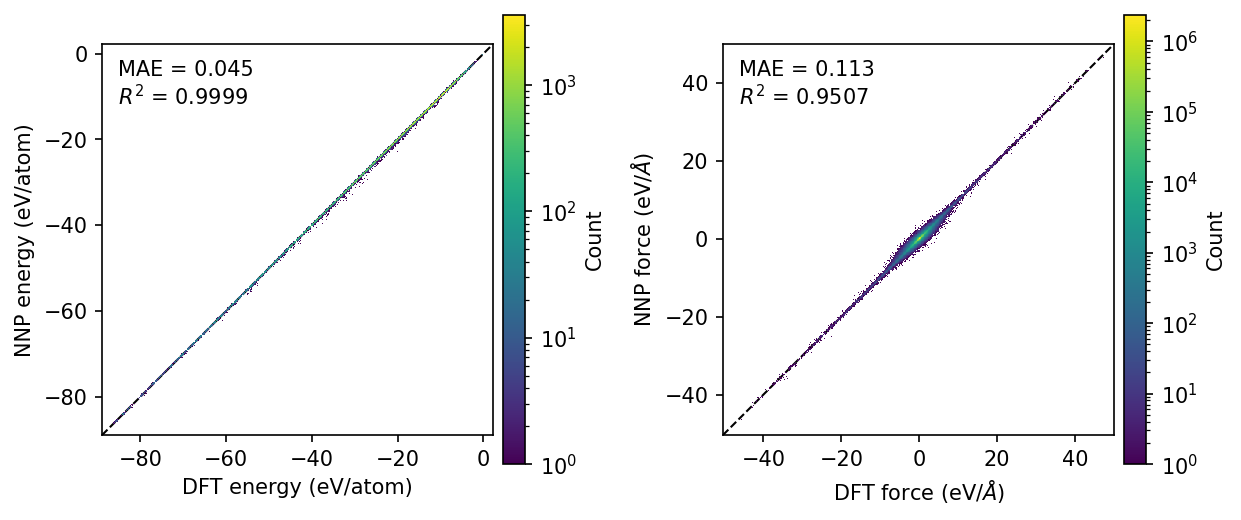}
\caption{Energy and force parity plots between the prediction of the r$^2$SCAN-MatPES foundation model and the DFT reference.}
\label{fig:si_parity_r2scan_matpes}
\end{figure}

\endgroup

\section{Fine-tuning details}

Table~\ref{tab:si_finetune} gives the full series behind the full-budget comparison. Training was performed on the same train set of 24,154 structures from the AdsTraj training set, evaluated throughout on the same test set of 7,296 structures. The zero-shot row is the error of the \emph{RPBE-OC20} foundation model on the test set, without any fine-tuning. The ``from scratch'' rows are the errors of models trained from random initialisation on the indicated number of training structures, and the ``fine-tuned'' rows are the errors of the same model after fine-tuning from \emph{RPBE-OC20} on the same number of training structures. The final row is the same as Table~4 in the manuscript.

\begin{table}[H]
\centering
\caption{Data efficiency of fine-tuning on the AdsTraj dataset.}
\label{tab:si_finetune}
\small
\setlength{\tabcolsep}{10pt}
\begin{tabular}{lrrrr}
\toprule
& & \multicolumn{3}{c}{Test error} \\
\cmidrule(lr){3-5}
Strategy & Structures & $E_\mathrm{MAE}$ & $F_\mathrm{MAE}$ & $E_\mathrm{ads, MAE}$ \\
  &  & (meV~atom$^{-1}$) & (meV~\AA$^{-1}$) & (eV) \\
\midrule
Zero-shot & 0 & 25.0 & 67.2 & 0.715 \\
\midrule
From scratch & 86 & 2283.6 & 565.7 & 71.1 \\
  & 1,023 & 794.0 & 181.4 & 28.2 \\
  & 2,119 & 524.6 & 223.7 & 18.1 \\
  & 5,424 & 32.2 & 91.5 & 1.033 \\
  & 11,991 & 19.7 & 65.5 & 0.689 \\
  & 24,154 & 14.1 & 45.4 & 0.483 \\
\midrule
Fine-tuned & 86 & 89.6 & 79.2 & 3.058 \\
  & 1,023 & 42.9 & 77.7 & 1.669 \\
  & 2,119 & 26.0 & 64.0 & 0.969 \\
  & 5,424 & 16.6 & 53.6 & 0.513 \\
  & 11,991 & 9.1 & 49.4 & 0.324 \\
  & 24,154 & \textbf{7.1} & \textbf{36.7} & \textbf{0.217} \\
\bottomrule
\end{tabular}
\end{table}

\section{Training cost}

\begin{figure}[H]
    \centering
    \includegraphics[width=\textwidth]{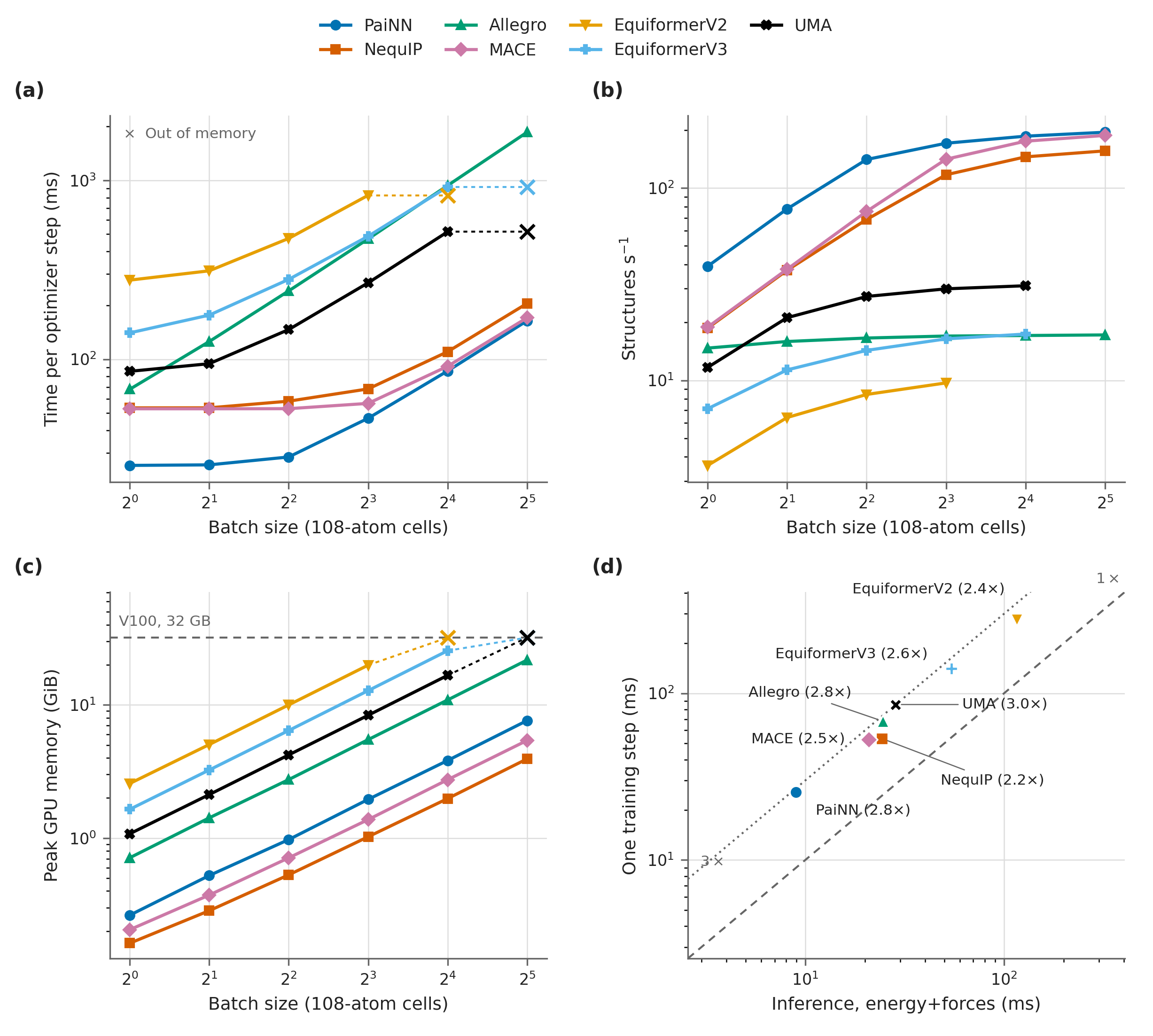}
    \caption{Cost of training measured on the same V100 as inference. (a) Wall time per optimizer step against batch size, in 108-atom cells. (b) Training throughput in structures per second, and where it saturates. (c) Peak GPU memory versus batch size. (d) One training step against one inference call on the same cell. The corresponding numbers are in Table~\ref{tab:si_train}.}
    \label{fig:si_cost_training_step}
\end{figure}

\begin{table}[H]
  \centering
  \caption{Cost and corresponding details of all architectures trained on Tesla V100 GPU.}
  \label{tab:si_train}
  \small
  \setlength{\tabcolsep}{2.5pt}
  \begin{tabular}{lrrrrrrr}
    \toprule
    Architecture & $B^{a}$ & step$^{b}$ & struct./s & GPU-h per & mem & step at $B{=}16^{c}$ & vs \\
     &  & (ms) &  & $10^{6}$ steps & (GiB) & (ms) & inference$^{d}$ \\
    \midrule
    PaiNN & 16 & 86.0 & 186 & 23.9 & 3.8 & 86.0 & 2.8$\times$ \\
    NequIP & 16 & 110.1 & 145 & 30.6 & 2.0 & 110.1 & 2.2$\times$ \\
    Allegro & 16 & 935.4 & 17 & 259.8 & 10.9 & 935.4 & 2.8$\times$ \\
    MACE & 16 & 91.3 & 175 & 25.4 & 2.7 & 91.3 & 2.5$\times$ \\
    EquiformerV2 & 8 & 824.9 & 10 & 229.1 & 19.9 & 1649.8 & 2.4$\times$ \\
    EquiformerV3 & 16 & 919.1 & 17 & 255.3 & 25.5 & 919.1 & 2.6$\times$ \\
    UMA & 16 & 515.4 & 31 & 143.2 & 16.7 & 515.4 & 3.0$\times$ \\
    \bottomrule
  \end{tabular}
  \par\vspace{3pt}
  \begin{minipage}{\linewidth}\footnotesize\raggedright
    $^{a}$~Each run's own \texttt{batch\_size}, and columns 3--6 use it. EquiformerV2 specifies 8, and 16 does not fit in 32\,GB, so that choice has to be forced by memory. Optimizer is Adam and \texttt{loss\_per\_atom} is true for all seven.\\
    $^{b}$~Cells are 108-atom bulk fcc Pt. \\
    $^{c}$~The same measurement at a common batch of 16, as a controlled comparison.\\
    $^{d}$~Ratio to one energy+forces inference call on the same 108-atom cell. \\
  \end{minipage}
\end{table}

\section{Inference cost}

\begin{table}[H]
  \centering
  \caption{Inference cost and accuracy of different architectures on one Tesla V100-PCIE-32GB, float32, 5.5\,\AA{} cutoff for energy+forces on bulk fcc Pt supercells.}
  \label{tab:si_cost}
  \small
  \setlength{\tabcolsep}{2pt}
  \begin{tabular}{lrrrrrrrrr}
    \toprule
    Architecture & Params & Floor$^{a}$ & Crossover$^{b}$ & $N^{*}{}^{c}$ & Max $N^{d}$ & Mem @ $10^{3}$ & MD$^{e}$ & $E_\mathrm{MAE}$$^{f}$ & $F_\mathrm{MAE}$$^{f}$ \\
    & & (ms) & ($\mu$s/atom) & & & (GiB) & (ns/day) & (meV/at.) & (meV/\AA) \\
    \midrule
    PaiNN & 600,581 & 9.0 & 15.9 & 564 & 10,976 & 1.36 & 3.00 & 41 & 50 \\
    NequIP & 1,383,079 & 24.3 & 18.5 & 1,316 & 10,976 & 0.92 & 2.03 & 50 & 69 \\
    Allegro & 570,062 & 19.1 & 180.7 & 106 & 4,000 & 5.72 & 0.76 & 40 & 66 \\
    MACE & 1,435,686 & 20.8 & 15.0 & 1,391 & 10,976 & 0.76 & 2.23 & 54 & 69 \\
    EquiformerV2 & 773,093 & 114.6 & 306.7 & 374 & 864 & 19.77 & 0.36 & 40 & 59 \\
    EquiformerV3 & 966,081 & 53.6 & 171.6 & 312 & 2,048 & 13.28 & 0.66 & 40 & 46 \\
    UMA & 1,174,121 & 28.4 & 91.4 & 311 & 6,912 & 4.32 & 1.19 & 29 & 40 \\
    \bottomrule
  \end{tabular}
  \par\vspace{3pt}
  \begin{minipage}{\linewidth}\footnotesize\raggedright
    $^{a}$~Measured plateau, the median of points within 15\% of the minimum -- what a call on a small cell actually costs.\\
    $^{b}$~Marginal cost per atom, fitted past the knee where the curve is linear.\\
    $^{c}$~Crossover below which the fixed per-call cost dominates.\\
    $^{d}$~Largest cell that fits in 32\,GB; the next size in the sweep raised OOM.\\
    $^{e}$~1\,fs timestep at $N=500$, interpolated from the measured curve, including the neighbour-list rebuild that MLCalculator performs every call.\\
    $^{f}$~The released checkpoints' own validation scores at the step recorded in each checkpoint.
  \end{minipage}
\end{table}

\begin{table}[H]
  \centering
  \caption{Energy-only inference against energy+forces. Energy-only is timed under no autograd graph retained.}
  \label{tab:si_energy}
  \small
  \setlength{\tabcolsep}{2pt}
  \begin{tabular}{lrrrrrrrr}
    \toprule
    Architecture & Floor $E$ & Crossover$_{E}$ & Floor$_{E{+}F}$ & Crossover$_{E+F}$ & Force cost,$^{a}$ & Force cost,$^{a}$ & Mem$_E^{b}$ & Mem$_{E{+}F}^{b}$ \\
    & (ms) & ($\mu$s/atom) & (ms) & ($\mu$s/atom) & Floor & Crossover & (GiB) & (GiB) \\
    \midrule
    PaiNN & 3.1 & 6.9 & 9.0 & 15.9 & 2.90$\times$ & 2.30$\times$ & 0.33 & 1.36 \\
    NequIP & 6.7 & 5.5 & 24.3 & 18.5 & 3.61$\times$ & 3.33$\times$ & 0.29 & 0.92 \\
    Allegro & 4.5 & 54.3 & 19.1 & 180.7 & 4.28$\times$ & 3.33$\times$ & 1.07 & 5.72 \\
    MACE & 8.2 & 5.5 & 20.8 & 15.0 & 2.55$\times$ & 2.72$\times$ & 0.21 & 0.76 \\
    EquiformerV2 & 39.0 & 95.5 & 114.6 & 306.7 & 2.94$\times$ & 3.21$\times$ & 7.73 & 19.77 \\
    EquiformerV3 & 19.3 & 58.9 & 53.6 & 171.6 & 2.78$\times$ & 2.91$\times$ & 2.07 & 13.28 \\
    UMA & 11.1 & 39.3 & 28.4 & 91.4 & 2.57$\times$ & 2.33$\times$ & 0.70 & 4.32 \\
    \bottomrule
  \end{tabular}
  \par\vspace{3pt}
  \begin{minipage}{\linewidth}\footnotesize\raggedright
    $^{a}$~Forces cost 2.3--3.3$\times$ the energy: the gradient pass must traverse every intermediate saved by the forward.\\
    $^{b}$~Peak device memory at $N=10^{3}$; dropping forces cuts it 3--6$\times$, which raises the largest tractable cell accordingly.
  \end{minipage}
\end{table}

\begin{table}[H]
  \centering
  \caption{DFT calculations on CPUs: VASP 6.3.2 single points on the same Pt(111)+CO cells, RPBE, cutoff with 450\,eV, and 64 MPI ranks on one node.}
  \label{tab:si_dft}
  \small
  \setlength{\tabcolsep}{4pt}
  \begin{tabular}{rrrr}
    \toprule
    Atoms$^{b}$ & LOOP+ (s)$^{a}$ & SCF steps & s per SCF step \\
    \midrule
    18 & 2.5 & 35 & 0.07 \\
    38 & 8.3 & 39 & 0.21 \\
    66 & 21.7 & 38 & 0.57 \\
    102 & 63.3 & 45 & 1.41 \\
    146 & 149.7 & 43 & 3.48 \\
    \bottomrule
  \end{tabular}
  \par\vspace{3pt}
  \begin{minipage}{\linewidth}\footnotesize\raggedright
    $^{a}$~VASP's own LOOP+ time, excluding ASE setup and POTCAR assembly. A power-law fit gives $t \propto N^{1.95}$ over this range, not $N^{3}$: at these sizes the FFTs still dominate and the cubic term has not taken over.\\
    $^{b}$~The $\Gamma$-only mesh is held fixed as the cell grows, so that a shrinking $k$-point count is not folded into the exponent.
  \end{minipage}
\end{table}

\begin{figure}[H]
    \centering
    \includegraphics[width=\textwidth]{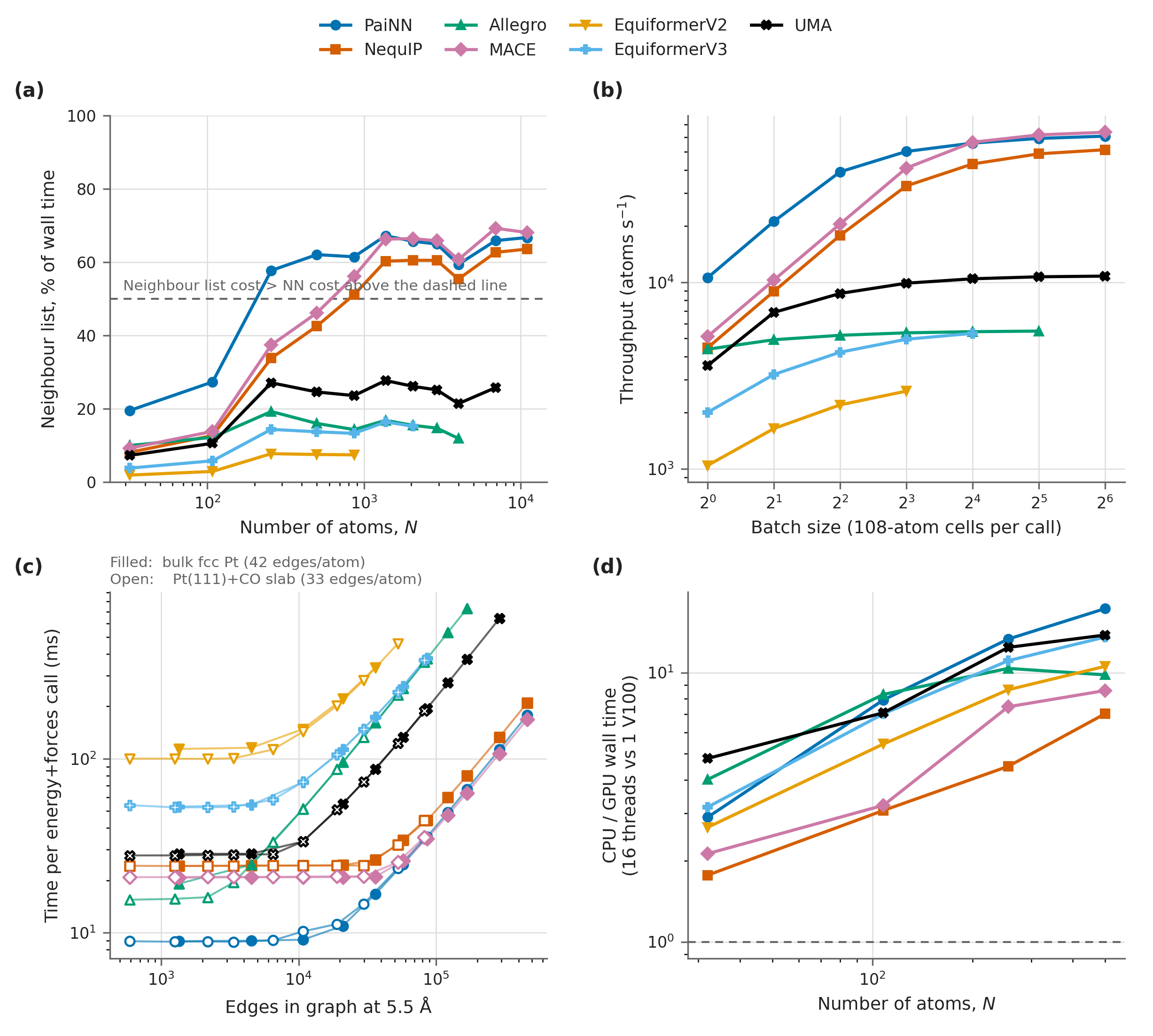}
    \caption{The wall time consumption in an energy+forces call with same models and hardware as Figure~6 in the manuscript. (a) Share of each call spent building the neighbour list against system size. (b) Throughput against batch size at a fixed 108-atom cell. (c) Cost against the number of edges in the graph at the \qty{5.5}{\angstrom} cutoff for two geometries: bulk fcc Pt at 42 edges per atom (filled) and the Pt(111) slab at 33 (open), whose surface atoms are missing a quarter of their neighbours. (d) Ratio of CPU to GPU wall time, per architecture and per size.}
    \label{fig:si_cost_breakdown}
\end{figure}

\begin{figure}[H]
    \centering
    \includegraphics[width=\textwidth]{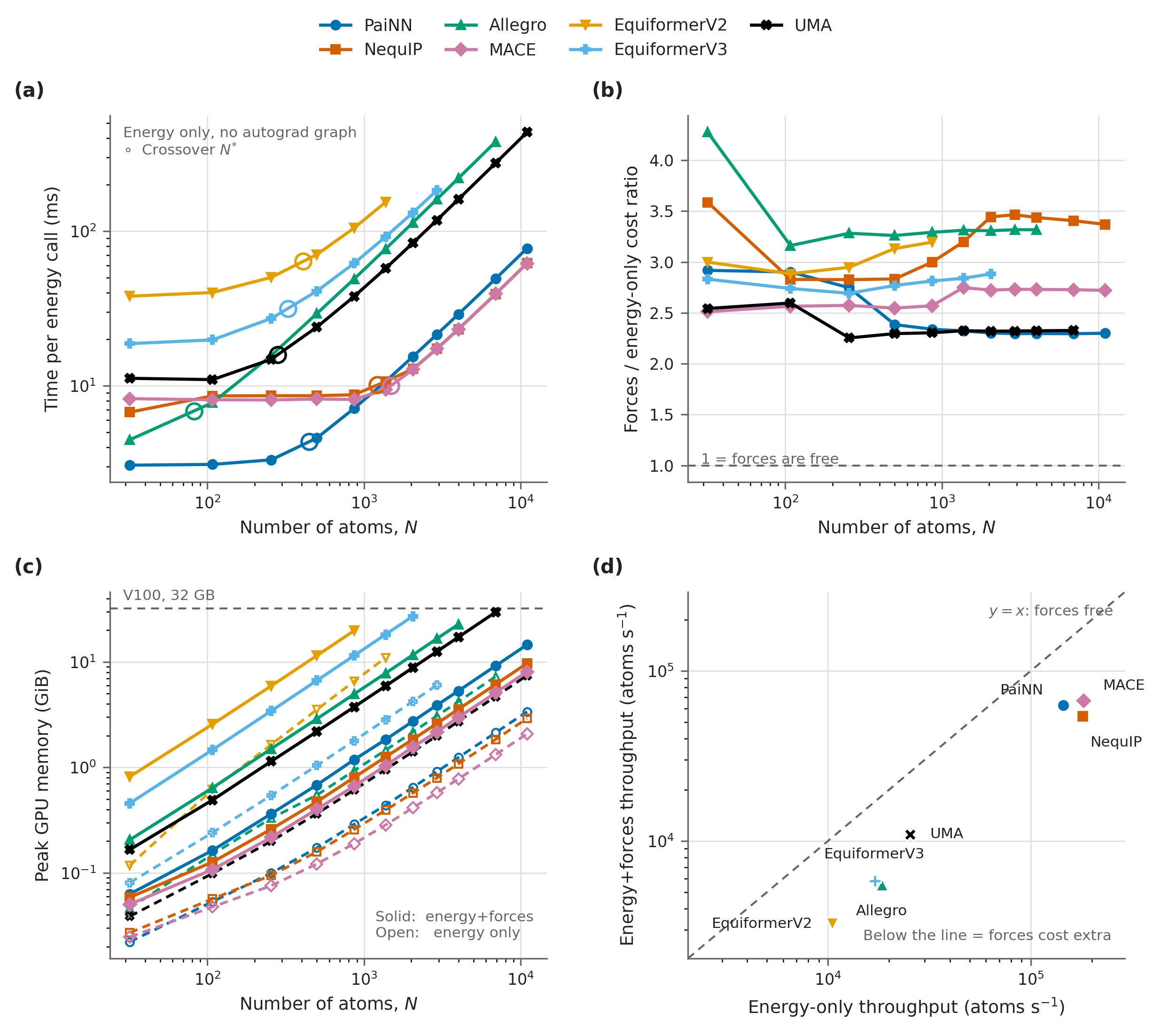}
    \caption{Energy-only inference versus energy+forces inference. (a) Time per energy-only call against system size, timed under no autograd graph is built. (b) Forces cost, as the ratio of the two timing modes. (c) Peak GPU memory in both modes. Solid is for the energy+forces mode and open is for the energy-only mode. (d) Throughput in the two modes against each other. The corresponding numbers are in Table~\ref{tab:si_energy}.}
    \label{fig:si_cost_energy_only}
\end{figure}

\section{Parallel training details}

\begin{table}[H]
\centering
\caption{Training efficiency of different architectures on MPtrj on 1 to 16 NVIDIA A100-SXM4-80GB GPUs. $B$ is each run's own per-rank batch size, so the global batch is $P \times B$ and the work per GPU is fixed; $X$ is the global throughput in structures per second and the efficiency is $X(P)/(P\,X(1))$. The all-reduce column is the part of the gradient reduction that is not hidden behind the backward pass.}
\label{tab:si_train_a100}
  \setlength{\tabcolsep}{2.5pt}
  \small
\begin{tabular}{lrrrrrrrr}
\toprule
Architecture & $B$ per & Global $B$ & $X(1)$ & $X(16)$ & Speedup & Efficiency & All-reduce & Mem/GPU \\
& rank & at $P$=16 & (struct./s) & (struct./s) & & (\%) & (\% step) & (GiB) \\
\midrule
PaiNN & 16 & 256 & 305.4 & 3892.5 & 12.75 & 79.7 $\pm$ 0.7 & 19.5 & 3.22 \\
NequIP & 16 & 256 & 218.6 & 2919.7 & 13.36 & 83.6 $\pm$ 0.2 & 8.1 & 1.62 \\
Allegro & 16 & 256 & 70.3 & 756.5 & 10.77 & 67.3 $\pm$ 1.0 & 41.1 & 9.24 \\
MACE & 16 & 256 & 208.8 & 2869.0 & 13.74 & 86.0 $\pm$ 0.5 & 7.5 & 2.17 \\
EquiformerV2 & 8 & 128 & 30.8 & 323.2 & 10.50 & 65.7 $\pm$ 1.6 & 43.2 & 24.27 \\
EquiformerV3 & 16 & 256 & 61.3 & 727.5 & 11.86 & 74.1 $\pm$ 0.3 & 29.1 & 19.99 \\
UMA & 16 & 256 & 95.4 & 1082.1 & 11.34 & 71.1 $\pm$ 0.4 & 37.2 & 12.75 \\
\bottomrule
\end{tabular}
\end{table}

\begin{table}[H]
\centering
\small
\caption{Training global throughput $X(P)$ in
structures per second, and parallel efficiency $X(P)/(P\,X(1))$, at every
allocation from 1 to 16~A100 GPUs. $B$ is the per-rank batch size, so the
global batch at each column is $P\times B$. Every world size times an identical
set of structures.}
\label{tab:si_train_sweep}
  \setlength{\tabcolsep}{4pt}
\begin{tabular}{l r rrrrr rrrr}
\toprule
 & & \multicolumn{5}{c}{$X(P)$ (structures\,s$^{-1}$)}
 & \multicolumn{4}{c}{Parallel efficiency (\%)} \\
\cmidrule(lr){3-7} \cmidrule(lr){8-11}
Architecture & $B$ & 1 & 2 & 4 & 8 & 16 & 2 & 4 & 8 & 16 \\
\midrule
PaiNN & 16 & 305 & 577 & 1106 & 2042 & 3892 & 94.3 & 90.5 & 83.5 & 79.7 \\
NequIP & 16 & 219 & 431 & 798 & 1509 & 2920 & 98.5 & 91.3 & 86.5 & 83.6 \\
Allegro & 16 & 70 & 118 & 223 & 410 & 756 & 84.4 & 79.7 & 73.0 & 67.3 \\
MACE & 16 & 209 & 409 & 775 & 1470 & 2869 & 98.0 & 92.8 & 88.1 & 86.0 \\
EquiformerV2 & 8 & 31 & 55 & 98 & 176 & 323 & 89.0 & 79.5 & 71.3 & 65.7 \\
EquiformerV3 & 16 & 61 & 113 & 209 & 385 & 727 & 91.9 & 85.0 & 78.5 & 74.1 \\
UMA & 16 & 95 & 174 & 316 & 575 & 1082 & 91.4 & 82.8 & 75.4 & 71.1 \\
\bottomrule
\end{tabular}
\end{table}

\begin{figure}[H]
    \centering
    \includegraphics[width=0.9\textwidth]{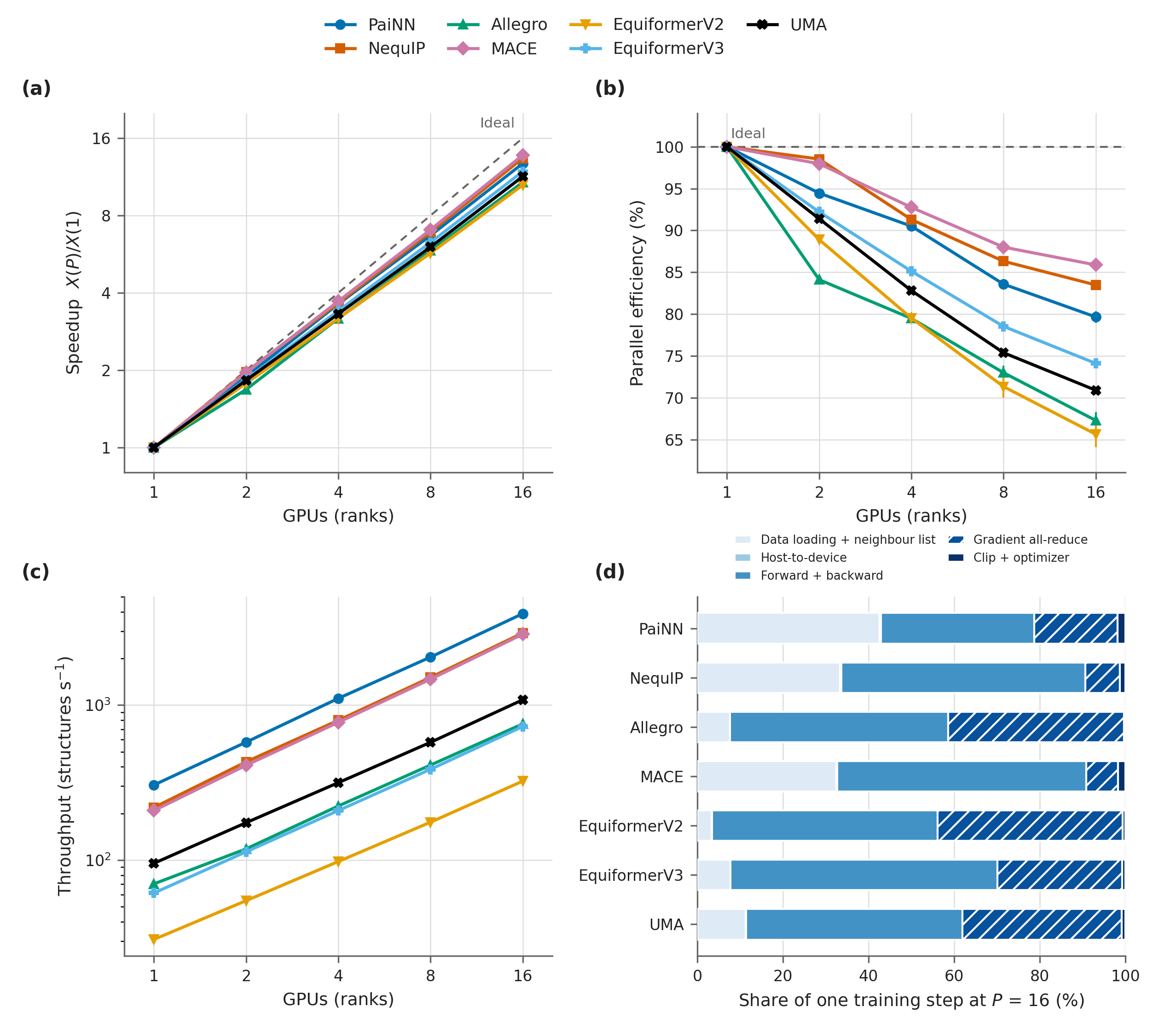}
    \caption{Parallel training on 1--16 A100 GPUs, batch 16 structures
    per rank, on the cleaned MPtrj database. (a) Speedup
    $X(P)/X(1)$ against rank count, with the ideal line. (b) Parallel
    efficiency. (c) Global throughput in structures per second.
    (d) Where one training step goes at $P=16$: data loading and
    neighbour-list construction, host-to-device transfer, forward and backward,
    the gradient all-reduce, and gradient clipping plus the optimizer update.
    The all-reduce share is what separates the curves in (b).}
    \label{fig:si_train_a100}
\end{figure}

\clearpage

\section{Parallel inference details}
\label{sec:lammps_inference}

\begin{table}[H]
\centering
\caption{The performance of the IANN LAMMPS interface of \texttt{pair\_style iann/multi\_gpu} on NVIDIA A100-80GB GPUs in Figure~7 in the manuscript is tabulated here. The step time per step on 1 to 8 GPUs, the corresponding efficiencies derived from them, and the $Y_8$ throughput at eight GPUs are listed at each system size. For 8 GPUs, four GPUs sit in one node (NVLink), and eight span two (Slingshot). The last column is the exact serial run of LAMMPS interface with \texttt{pair\_style iann} on a single GPU.}
\label{tab:si_lmp_scaling}
\small
\setlength{\tabcolsep}{4pt}
  \renewcommand{\arraystretch}{1.15}
\begin{tabular}{lrrrr>{\raggedleft\arraybackslash}p{2.35em}>{\raggedleft\arraybackslash}p{2.35em}>{\raggedleft\arraybackslash}p{2.35em}rr}
\toprule
 & \multicolumn{4}{c}{Time per step (s)} & \multicolumn{3}{c}{Parallel efficiency (\%)} &  &  \\
\cmidrule(lr){2-5} \cmidrule(lr){6-8}
$N$ & 1 GPU & 2 GPUs & 4 GPUs & 8 GPUs & $E_2$ & $E_4$ & $E_8$ & $Y_8$ (atom step s$^{-1}$) & serial (s) \\
\midrule
864 & 0.0896 & 0.0732 & 0.0629 & 0.0854 & 61 & 36 & 13 & 10,115 & 0.0151 \\
2,048 & 0.0990 & 0.0690 & 0.0471 & 0.1240 & 72 & 53 & 10 & 16,513 & 0.0330 \\
5,324 & 0.1905 & 0.1506 & 0.1085 & 0.1236 & 63 & 44 & 19 & 43,074 & 0.0789 \\
6,912 & 0.2286 & 0.1824 & 0.1480 & 0.1327 & 63 & 39 & 22 & 52,099 & 0.0994 \\
10,976 & 0.3307 & 0.2022 & 0.1633 & 0.1488 & 82 & 51 & 28 & 73,763 & 0.1587 \\
13,500 & 0.3876 & 0.2322 & 0.2073 & 0.1285 & 83 & 47 & 38 & 105,058 & 0.1930 \\
16,384 & 0.4501 & 0.2724 & 0.1982 & 0.1272 & 83 & 57 & 44 & 128,825 & 0.2300 \\
19,652 & 0.5186 & 0.3349 & 0.1866 & 0.1509 & 77 & 69 & 43 & 130,275 & 0.2849 \\
23,328 & 0.6017 & 0.3579 & 0.2099 & 0.1636 & 84 & 72 & 46 & 142,592 & 0.3337 \\
27,436$^{\ddagger2}$ & \textsc{oom} & 0.4049 & 0.2347 & 0.1447 & ref & 86 & 70 & 189,567 & 0.3904 \\
32,000$^{\ddagger2}$ & \textsc{oom} & 0.4563 & 0.2863 & 0.1585 & ref & 80 & 72 & 201,829 & 0.4523 \\
42,592$^{\ddagger2}$ & \textsc{oom} & 0.5788 & 0.3357 & 0.2185 & ref & 86 & 66 & 194,884 & 0.6036 \\
48,668$^{\ddagger2}$ & \textsc{oom} & 0.6408 & 0.3675 & 0.2343 & ref & 87 & 68 & 207,690 & \textsc{oom} \\
55,296$^{\ddagger4}$ & \textsc{oom} & \textsc{oom} & 0.4170 & 0.2775 & n/a & ref & 75 & 199,243 & \textsc{oom} \\
70,304$^{\ddagger4}$ & \textsc{oom} & \textsc{oom} & 0.5033 & 0.3118 & n/a & ref & 81 & 225,463 & \textsc{oom} \\
87,808$^{\ddagger4}$ & \textsc{oom} & \textsc{oom} & 0.5955 & 0.3549 & n/a & ref & 84 & 247,402 & \textsc{oom} \\
108,000$^{\ddagger8}$ & \textsc{oom} & \textsc{oom} & \textsc{oom} & 0.4060 & n/a & n/a & ref & 266,003 & \textsc{oom} \\
131,072$^{\ddagger8}$ & \textsc{oom} & \textsc{oom} & \textsc{oom} & 0.4550 & n/a & n/a & ref & 288,102 & \textsc{oom} \\
157,216$^{\ddagger8}$ & \textsc{oom} & \textsc{oom} & \textsc{oom} & 0.5152 & n/a & n/a & ref & 305,149 & \textsc{oom} \\
186,624$^{\ddagger8}$ & \textsc{oom} & \textsc{oom} & \textsc{oom} & 0.6181 & n/a & n/a & ref & 301,917 & \textsc{oom} \\
\bottomrule
\end{tabular}

\vspace{10pt}
\begin{minipage}{\linewidth}\footnotesize \textsc{oom}: CUDA out of memory on the 80\,GB device -- measured, not inferred, and reproduced in all three repeats. n/a: not defined, because the run it would be derived from \textsc{oom}ed. $^{\ddagger P}$: this size does not fit on one GPU, so its efficiencies are referenced to $P$ GPUs rather than one. Every entry is the median of three independent jobs. Runs use \texttt{PYTORCH\_CUDA\_ALLOC\_CONF=expandable\_segments:True} and a warm-up run before each timed run; without these the large cells are 16--29\,\% slower and their efficiencies correspondingly overestimated. $t_P$ is the wall-clock step time. The model-evaluation component alone scales smoothly at every size.\end{minipage}
\end{table}
\vspace{10pt}

\section{Distributed submission scripts}

The parallelization training submission script on a SLURM machine with four GPUs per node:

\begin{lstlisting}[language=bash]
#!/bin/bash
#SBATCH -N 2
#SBATCH -C gpu
#SBATCH -q regular
#SBATCH -t 04:00:00
#SBATCH --gpus-per-node=4
#SBATCH --ntasks-per-node=4
#SBATCH --cpus-per-task=32

export GPUS_PER_NODE=$SLURM_GPUS_ON_NODE
export NNODES=$SLURM_NNODES

srun -N $NNODES -n $((NNODES*GPUS_PER_NODE)) python train.py
\end{lstlisting}

\noindent Rank, world size, per-node local rank and the rendezvous address are read from
the scheduler environment by the trainer. Under PBS with \texttt{mpirun}, the
first host of \texttt{\$PBS\_NODEFILE} is used as the rendezvous address unless
\texttt{MASTER\_ADDR} is exported explicitly and forwarded to every rank.

\section{LAMMPS interface}
\label{sec:lammps}

A trained checkpoint is compiled to TorchScript with:

\begin{lstlisting}
from iann.plugins.converter import convert_model_for_lammps

convert_model_for_lammps(model_path="output/model.pt",
                         output_path="model_lmp.pt")
\end{lstlisting}

\noindent Several checkpoints are combined into one ensemble module with
\texttt{convert\_models\_for\_lammps}. The model type is read from the
checkpoint metadata and does not have to be supplied. The exported file is used
in a LAMMPS input script as:

\begin{lstlisting}[language=C]
units metal
atom_style atomic
boundary p p p
read_data initial.data

pair_style iann painn model_lmp.pt 5.5
pair_coeff * *

neighbor 0.5 bin
neigh_modify every 1 delay 0 check yes

compute variance all iann/variance
thermo_style custom step pe ke etotal temp press &
             c_variance[1] c_variance[2] c_variance[3] c_variance[4]
thermo 100

timestep 0.001
fix 1 all nvt temp 300.0 300.0 0.1
run 100000
\end{lstlisting}

\noindent where the four components of \texttt{c\_variance} are the mean energy variance,
the mean force variance, the maximum energy variance and the maximum force
variance over the ensemble. Replacing \texttt{pair\_style iann} by
\texttt{pair\_style iann/multi\_gpu} assigns one GPU to each MPI rank, so that
device placement follows the LAMMPS domain decomposition.

\section{Uncertainty quantification}

\begin{lstlisting}
from iann.calculators import AtomicEnsembleCalculator

calc = AtomicEnsembleCalculator(
    model_paths=["m1.pt", "m2.pt", "m3.pt"], device="cuda")
atoms.calc = calc
energy = atoms.get_potential_energy()
unc = calc.get_ensemble()
print(unc["energy_var"], unc["atomic_energy_var"])
\end{lstlisting}

\texttt{EnsembleCalculator} returns not only average energy and forces but also the variance of the total energy and of the
forces; \texttt{AtomicEnsembleCalculator} additionally returns the variance of the per-atom energy decomposition, which localises the uncertainty on individual atoms. 



\section{Appendix: Periodic-table statistics for each database}
\label{sec:si_ptables}

\begin{figure}[H]
  \centering
  \includegraphics[width=\textwidth]{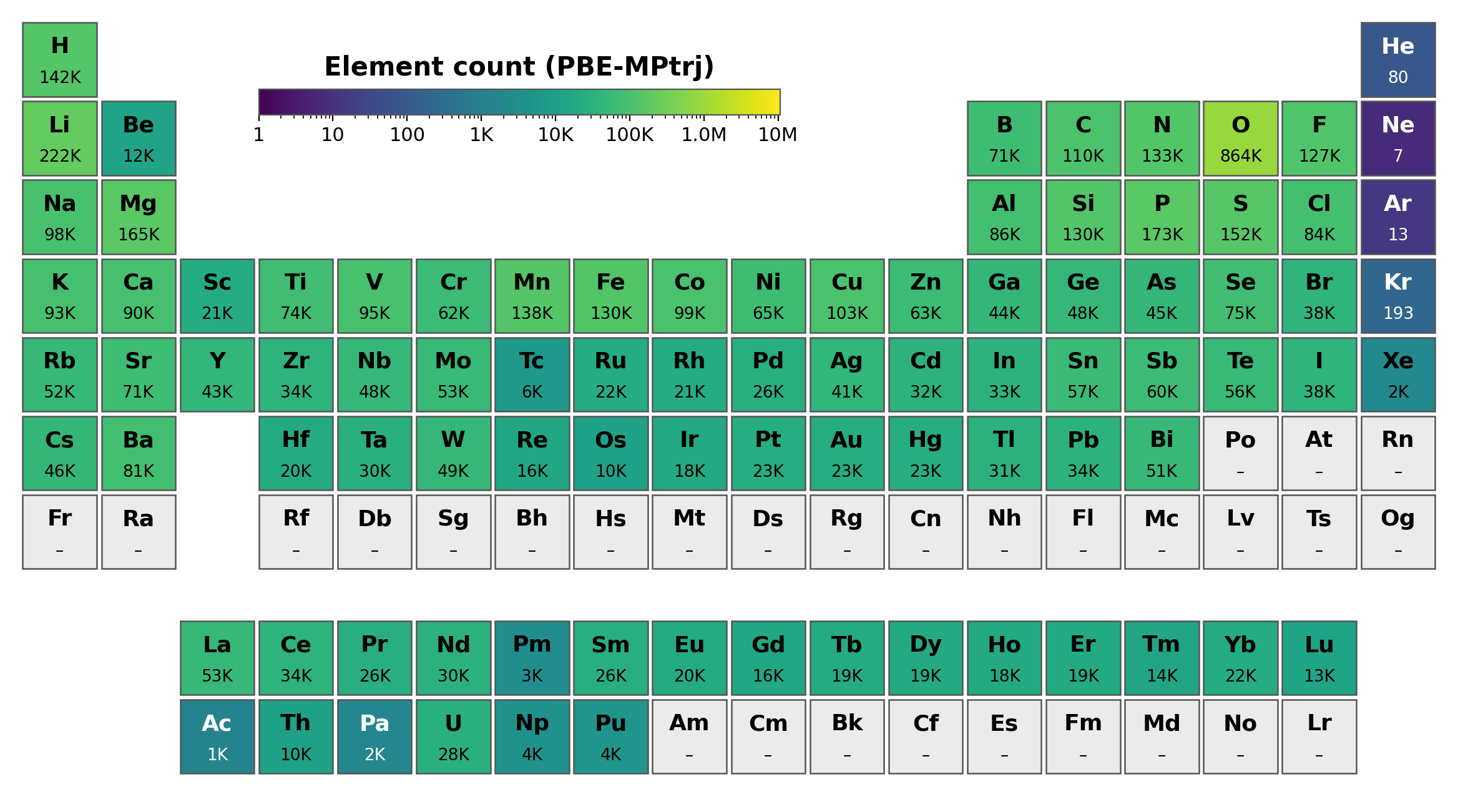}
  \caption{PBE-MPtrj: number of structures containing each element. 1,579,091 structures, 49,231,714 atoms, 89 elements. Color scales are shared across the databases of a functional, so tiles are directly comparable between the maps of this appendix.}
  \label{fig:app-pbe-mptrj-structures}
\end{figure}

\begin{figure}[H]
  \centering
  \includegraphics[width=\textwidth]{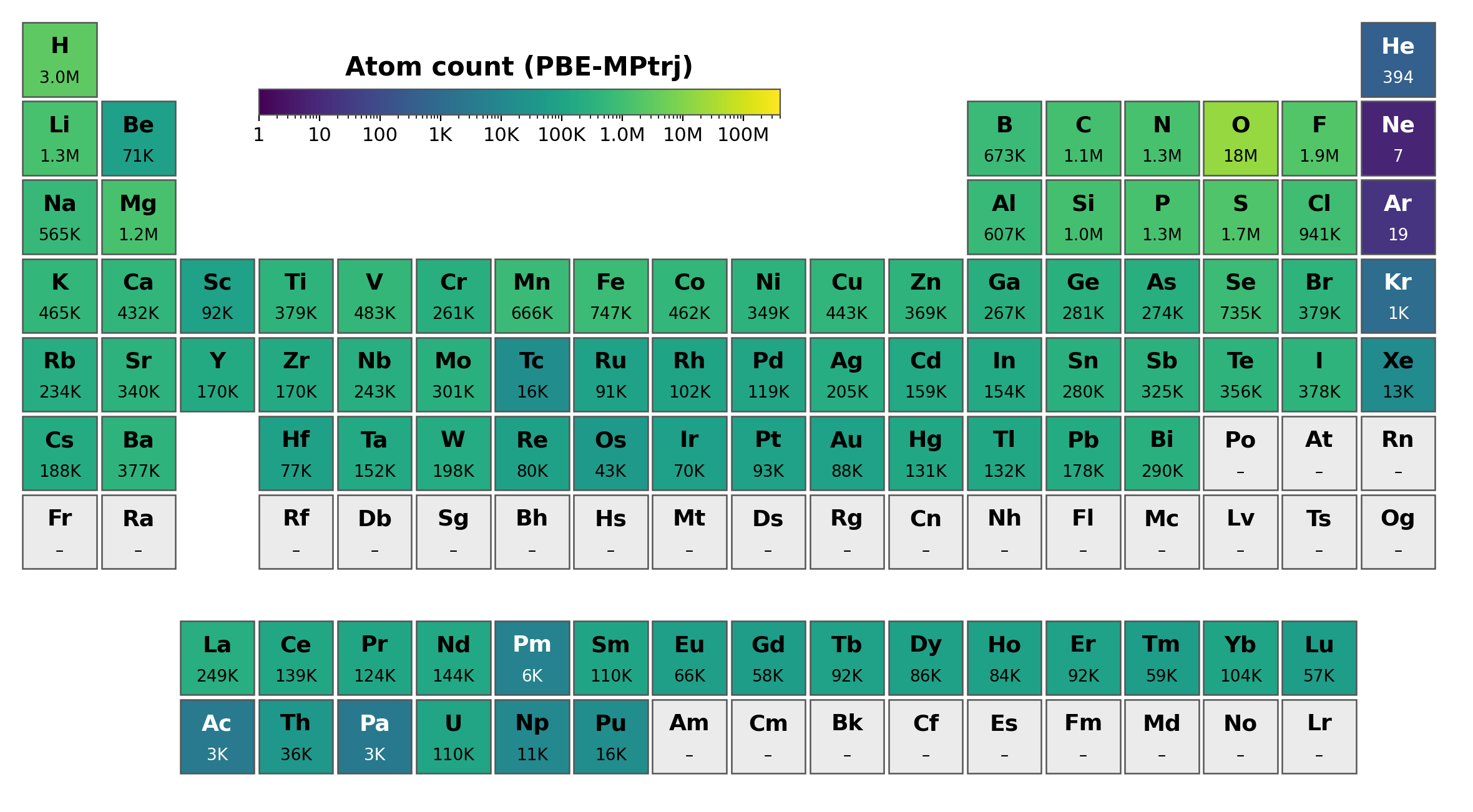}
  \caption{PBE-MPtrj: total number of atoms of each element. A structure with many atoms of one element counts once in the map above and once per atom here, so the two differ wherever the stoichiometry is skewed. Color scales are shared across the databases of a functional, so tiles are directly comparable between the maps of this appendix.}
  \label{fig:app-pbe-mptrj-atoms}
\end{figure}

\begin{figure}[H]
  \centering
  \includegraphics[width=\textwidth]{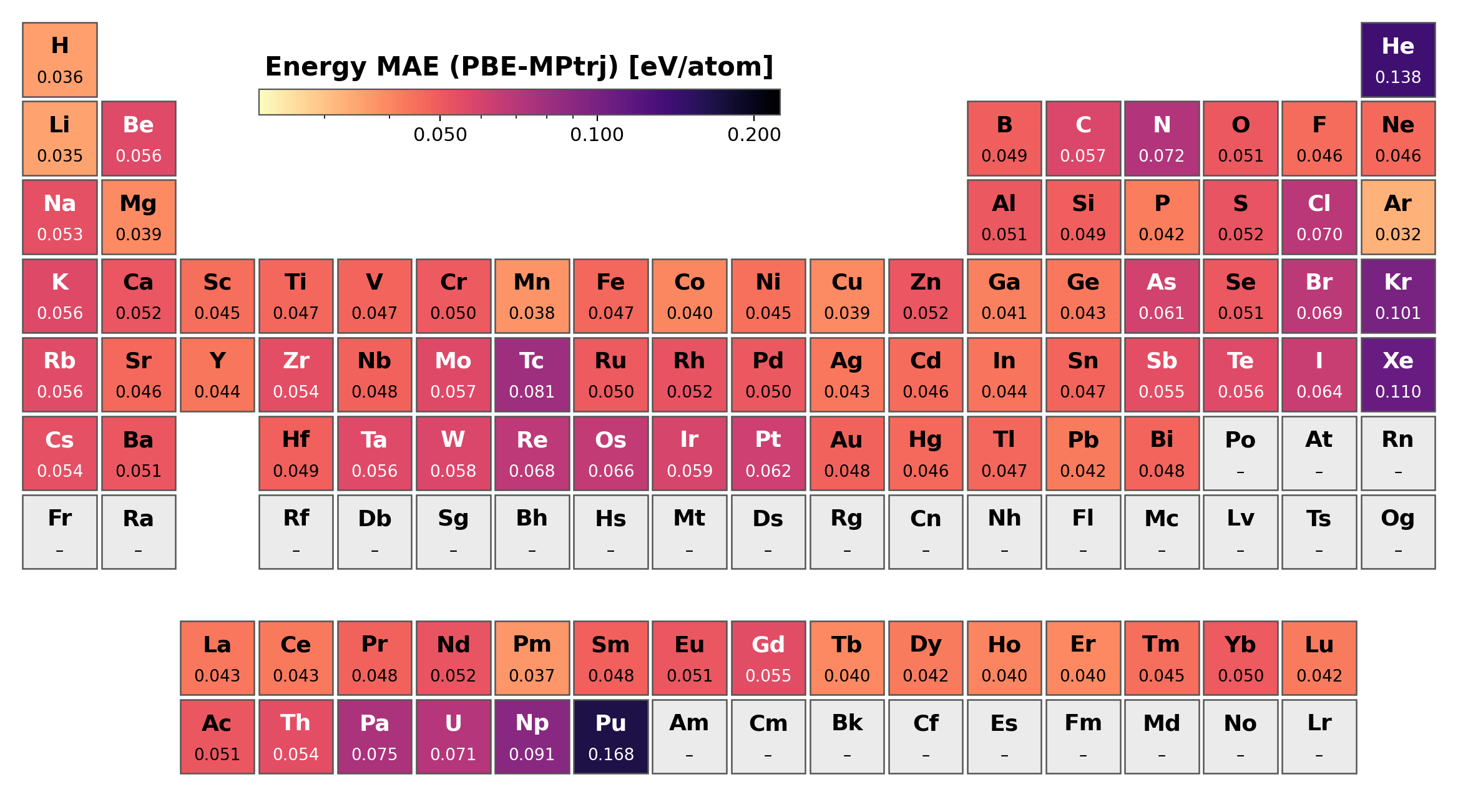}
  \caption{PBE-MPtrj: per-element energy MAE (eV/atom) of the model trained on this database alone, which reaches 0.0517~eV/atom overall. Color scales are shared across the databases of a functional, so tiles are directly comparable between the maps of this appendix.}
  \label{fig:app-pbe-mptrj-energy_mae}
\end{figure}

\begin{figure}[H]
  \centering
  \includegraphics[width=\textwidth]{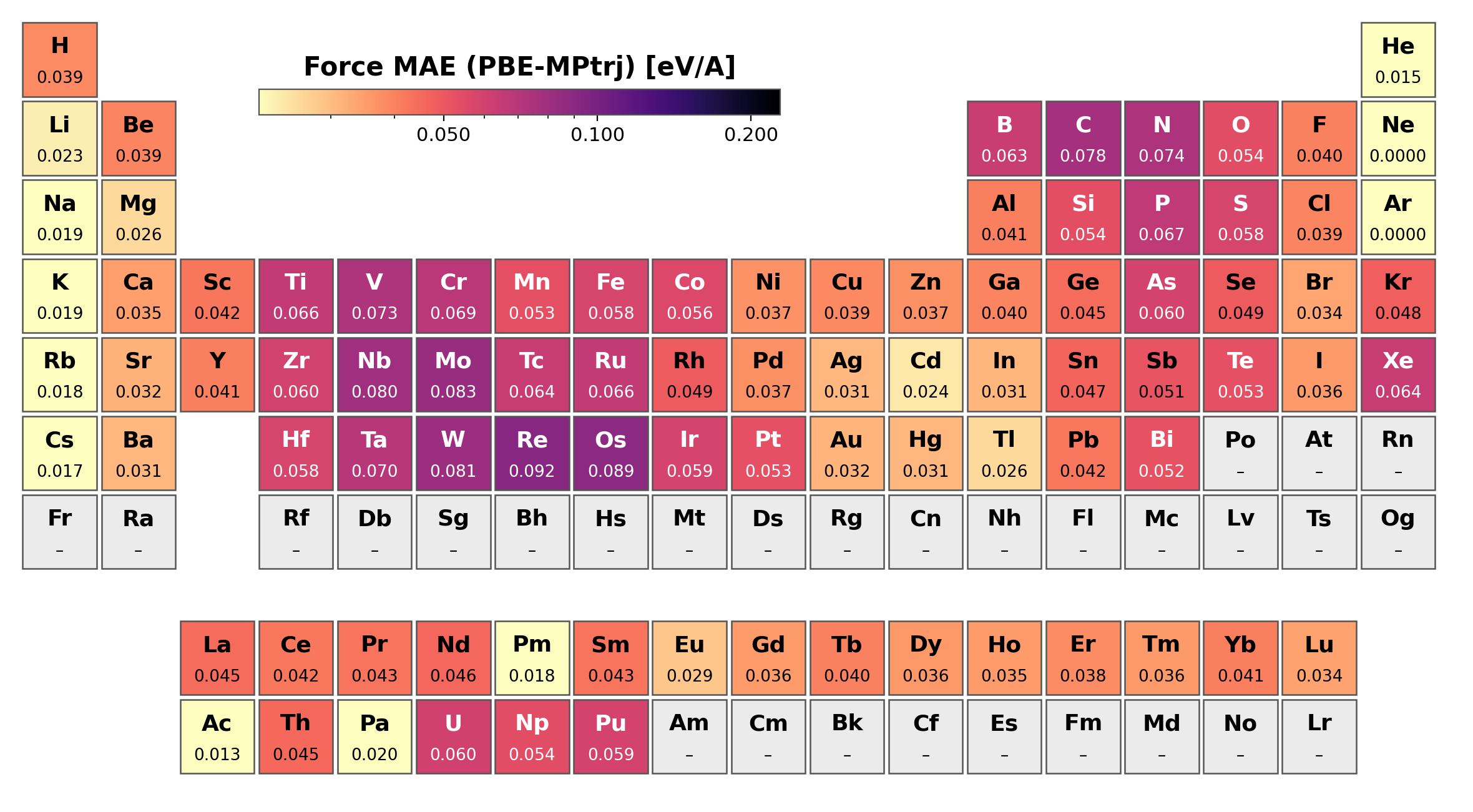}
  \caption{PBE-MPtrj: per-element force MAE (eV/\AA) of the model trained on this database alone, which reaches 0.0498~eV/\AA{} overall. Color scales are shared across the databases of a functional, so tiles are directly comparable between the maps of this appendix.}
  \label{fig:app-pbe-mptrj-force_mae}
\end{figure}

\begin{figure}[H]
  \centering
  \includegraphics[width=\textwidth]{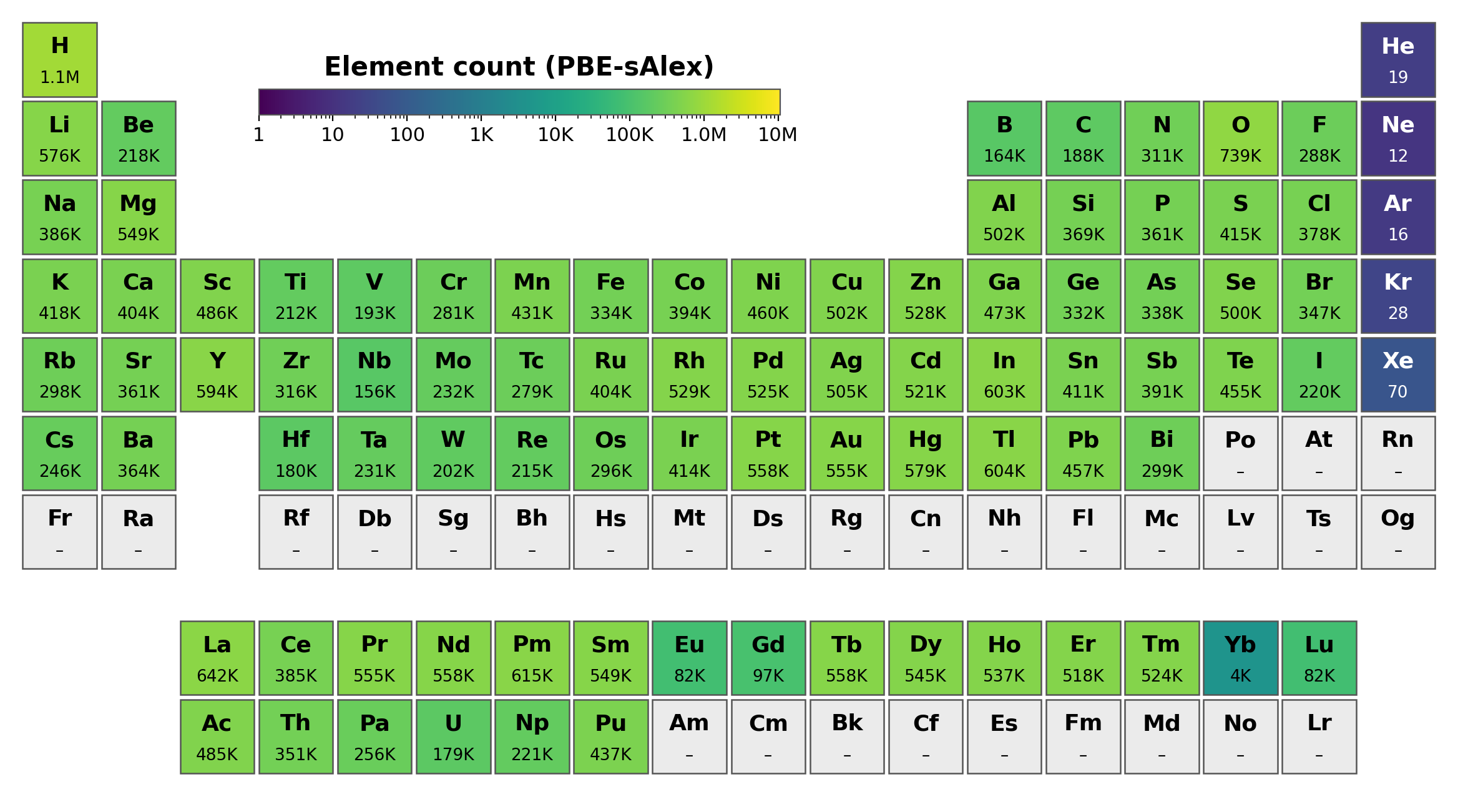}
  \caption{PBE-sAlex: number of structures containing each element. 10,423,284 structures, 107,455,806 atoms, 89 elements. Color scales are shared across the databases of a functional, so tiles are directly comparable between the maps of this appendix.}
  \label{fig:app-pbe-salex-structures}
\end{figure}

\begin{figure}[H]
  \centering
  \includegraphics[width=\textwidth]{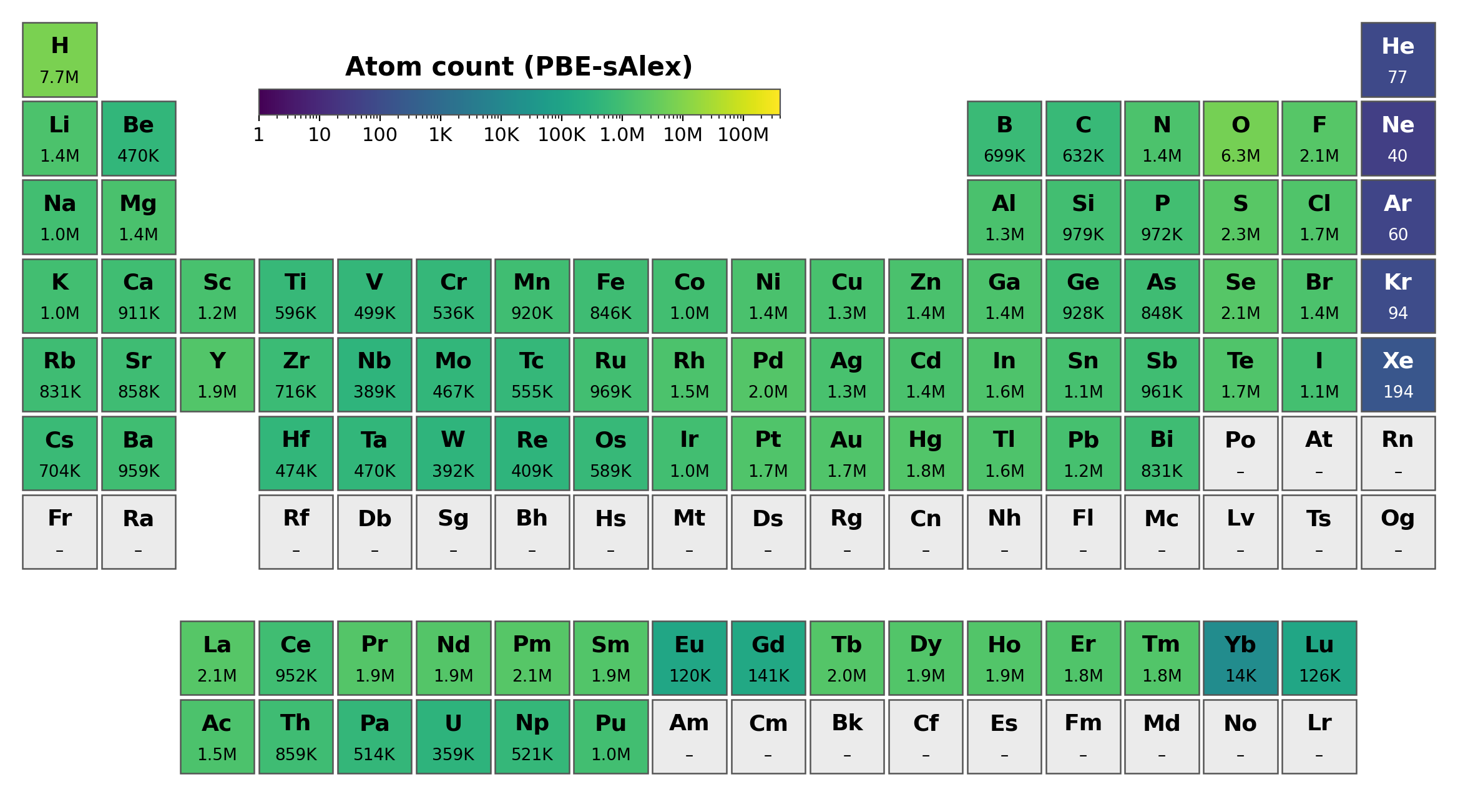}
  \caption{PBE-sAlex: total number of atoms of each element. A structure with many atoms of one element counts once in the map above and once per atom here, so the two differ wherever the stoichiometry is skewed. Color scales are shared across the databases of a functional, so tiles are directly comparable between the maps of this appendix.}
  \label{fig:app-pbe-salex-atoms}
\end{figure}

\begin{figure}[H]
  \centering
  \includegraphics[width=\textwidth]{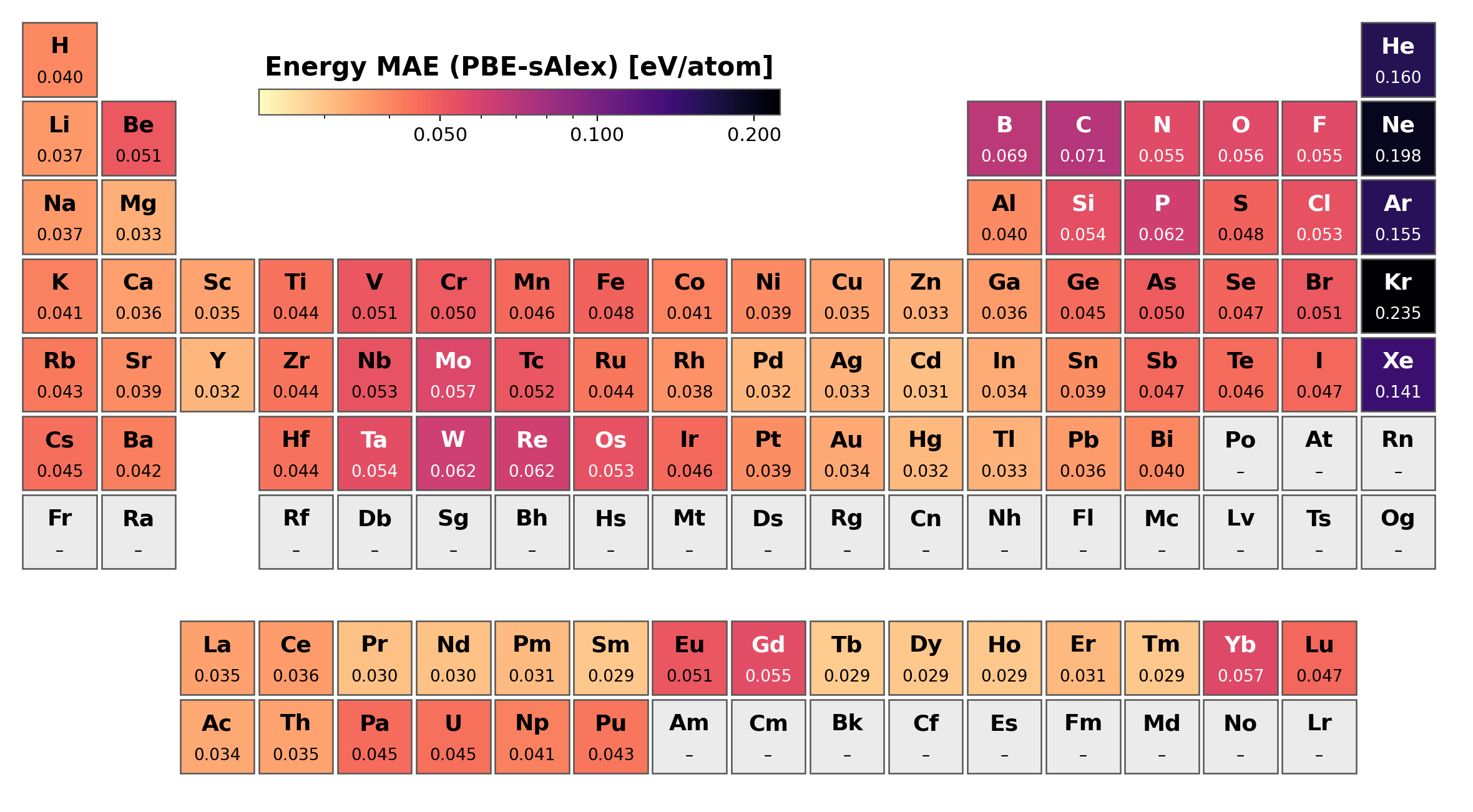}
  \caption{PBE-sAlex: per-element energy MAE (eV/atom) of the model trained on this database alone, which reaches 0.0408~eV/atom overall. Color scales are shared across the databases of a functional, so tiles are directly comparable between the maps of this appendix.}
  \label{fig:app-pbe-salex-energy_mae}
\end{figure}

\begin{figure}[H]
  \centering
  \includegraphics[width=\textwidth]{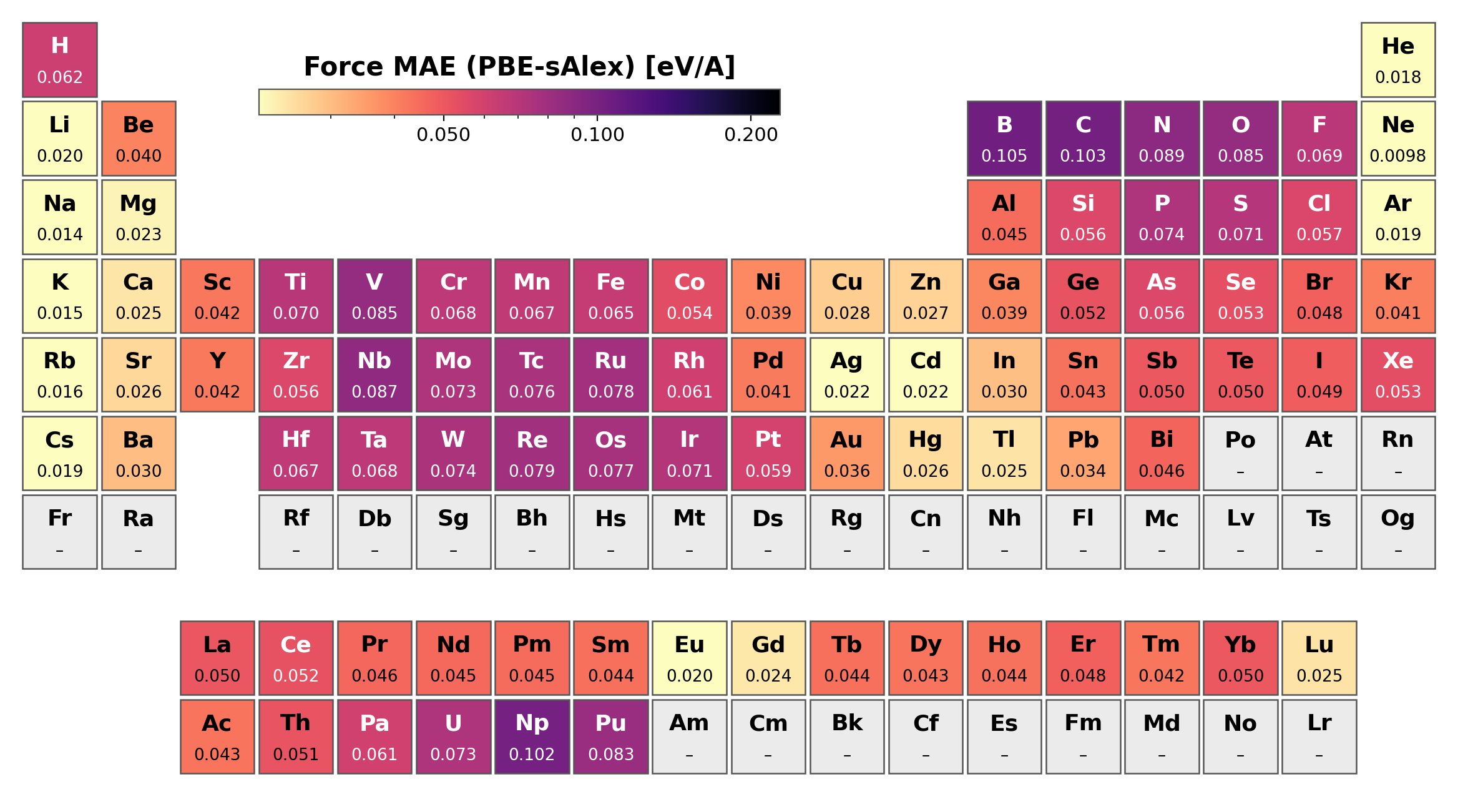}
  \caption{PBE-sAlex: per-element force MAE (eV/\AA) of the model trained on this database alone, which reaches 0.0513~eV/\AA{} overall. Color scales are shared across the databases of a functional, so tiles are directly comparable between the maps of this appendix.}
  \label{fig:app-pbe-salex-force_mae}
\end{figure}

\begin{figure}[H]
  \centering
  \includegraphics[width=\textwidth]{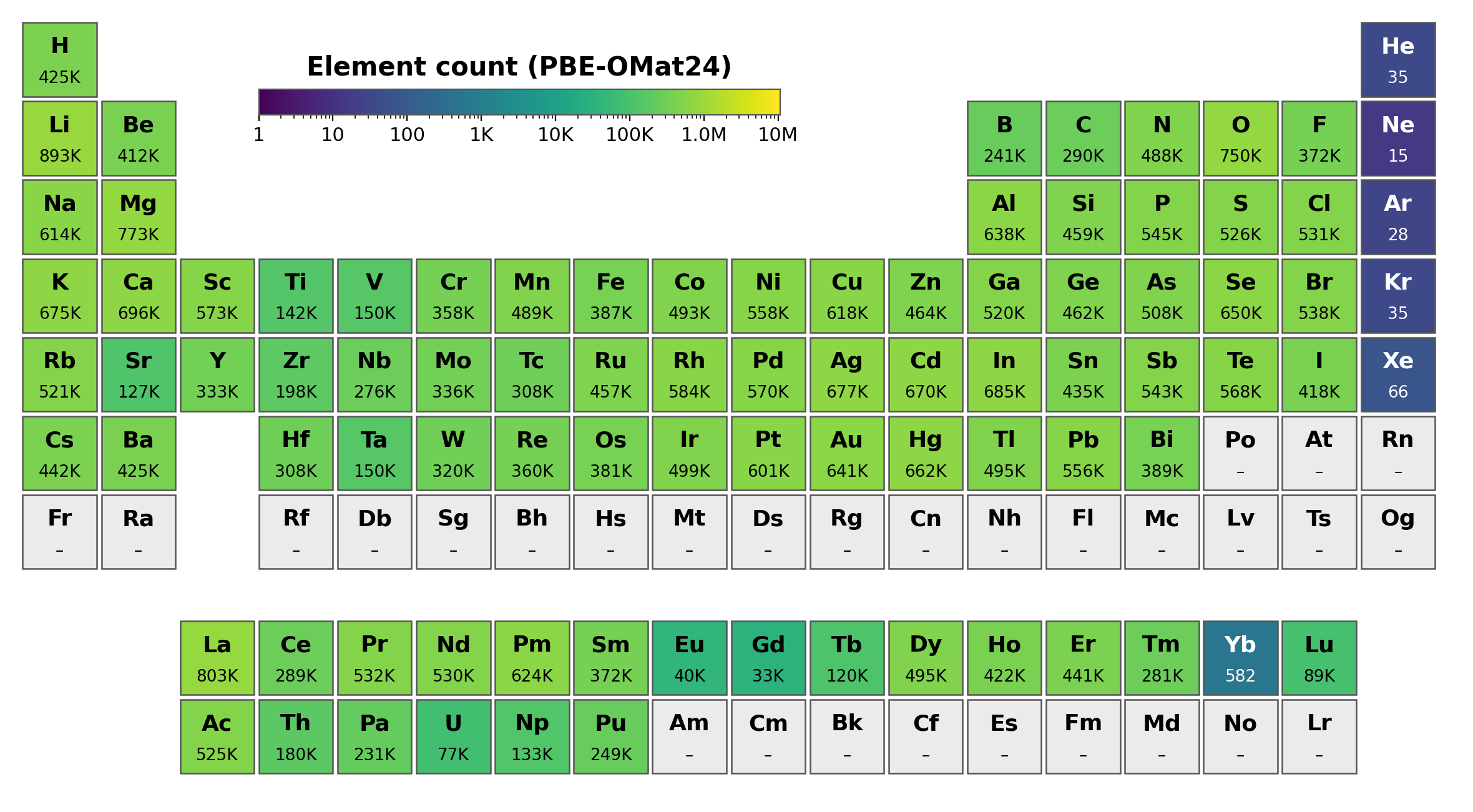}
  \caption{PBE-OMat24: number of structures containing each element. 11,274,650 structures, 157,069,839 atoms, 89 elements. Color scales are shared across the databases of a functional, so tiles are directly comparable between the maps of this appendix.}
  \label{fig:app-pbe-omat24-structures}
\end{figure}

\begin{figure}[H]
  \centering
  \includegraphics[width=\textwidth]{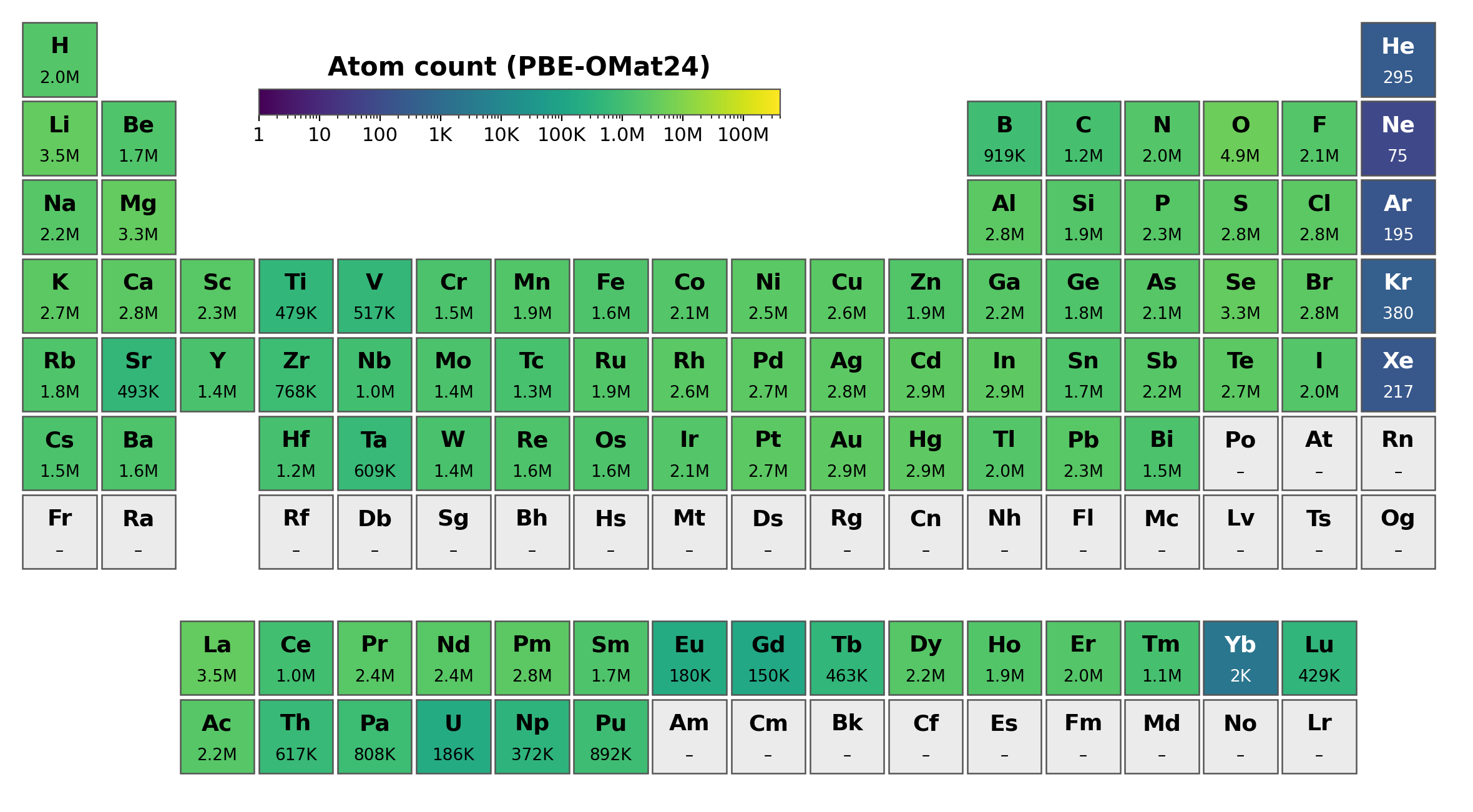}
  \caption{PBE-OMat24: total number of atoms of each element. A structure with many atoms of one element counts once in the map above and once per atom here, so the two differ wherever the stoichiometry is skewed. Color scales are shared across the databases of a functional, so tiles are directly comparable between the maps of this appendix.}
  \label{fig:app-pbe-omat24-atoms}
\end{figure}

\begin{figure}[H]
  \centering
  \includegraphics[width=\textwidth]{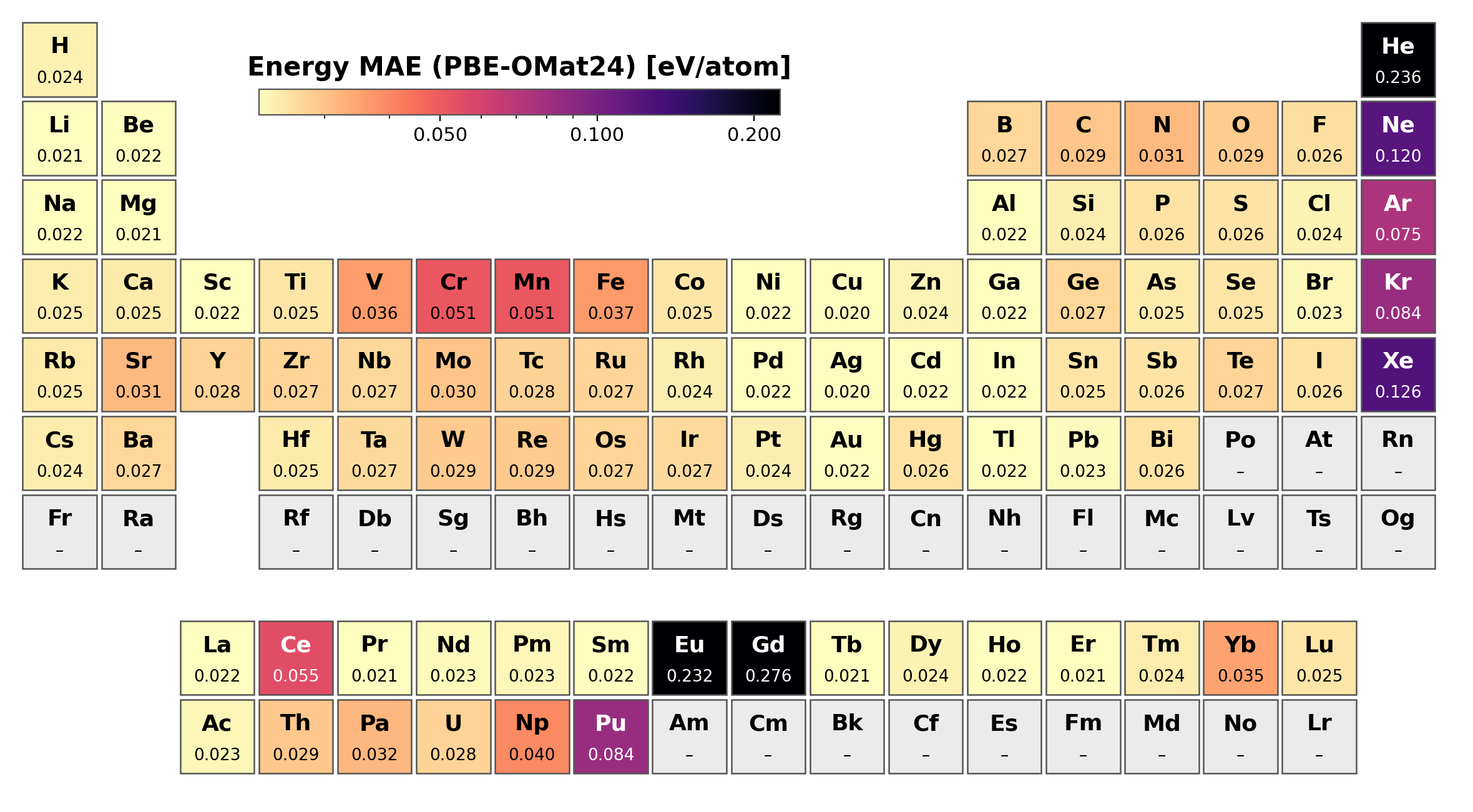}
  \caption{PBE-OMat24: per-element energy MAE (eV/atom) of the model trained on this database alone, which reaches 0.0264~eV/atom overall. Color scales are shared across the databases of a functional, so tiles are directly comparable between the maps of this appendix.}
  \label{fig:app-pbe-omat24-energy_mae}
\end{figure}

\begin{figure}[H]
  \centering
  \includegraphics[width=\textwidth]{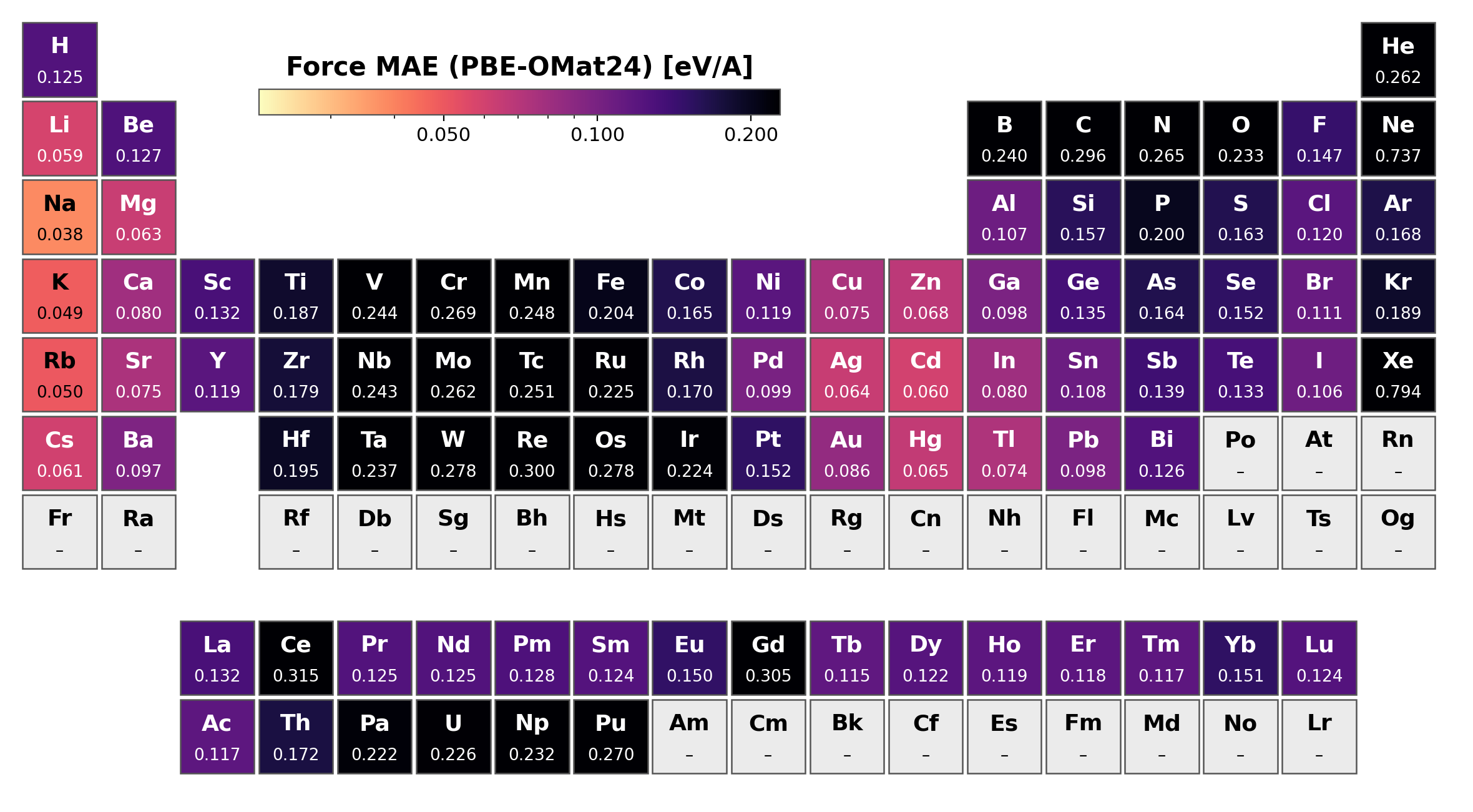}
  \caption{PBE-OMat24: per-element force MAE (eV/\AA) of the model trained on this database alone, which reaches 0.1391~eV/\AA{} overall. Color scales are shared across the databases of a functional, so tiles are directly comparable between the maps of this appendix.}
  \label{fig:app-pbe-omat24-force_mae}
\end{figure}

\begin{figure}[H]
  \centering
  \includegraphics[width=\textwidth]{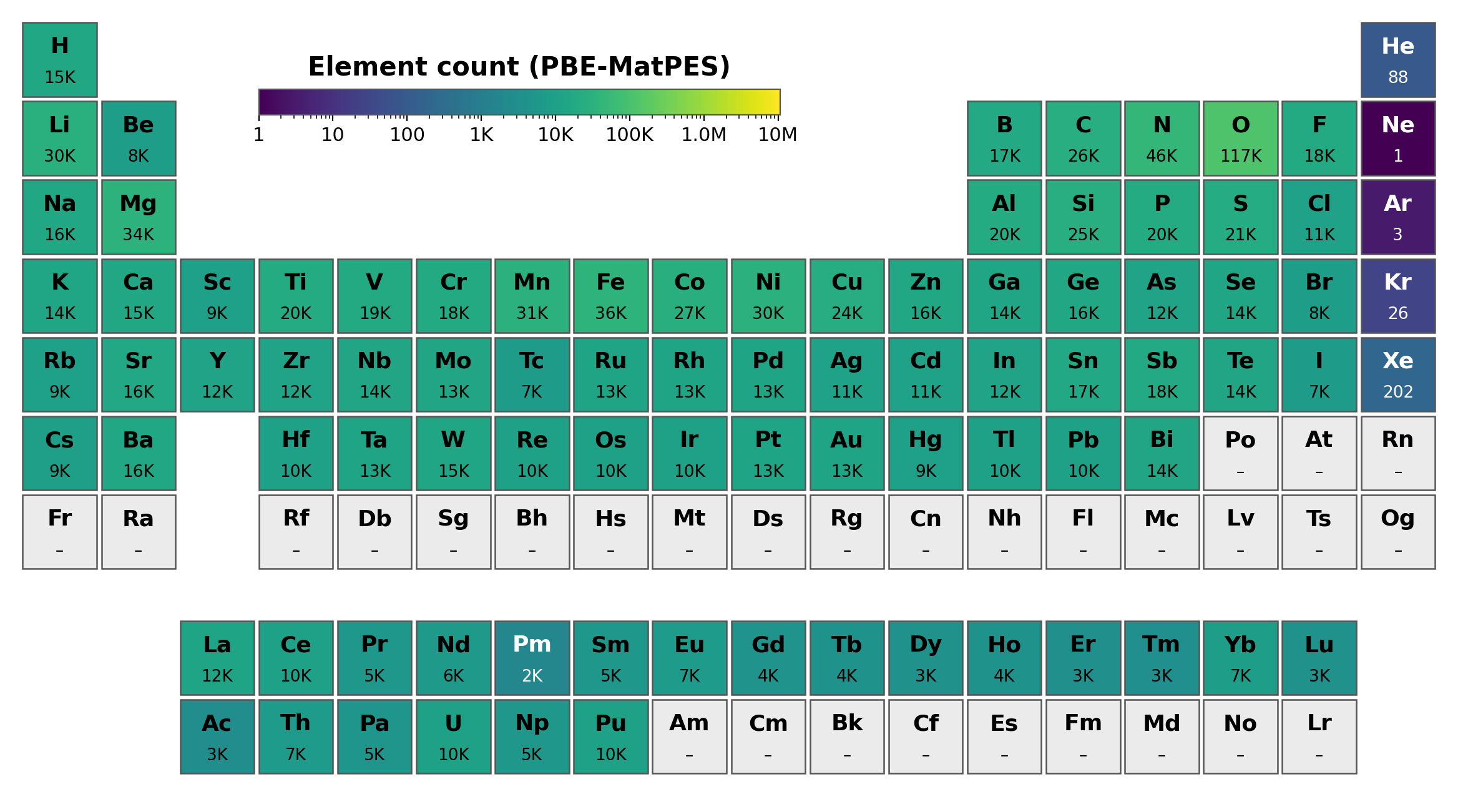}
  \caption{PBE-MatPES: number of structures containing each element. 434,092 structures, 3,865,819 atoms, 89 elements. Color scales are shared across the databases of a functional, so tiles are directly comparable between the maps of this appendix.}
  \label{fig:app-pbe-matpes-structures}
\end{figure}

\begin{figure}[H]
  \centering
  \includegraphics[width=\textwidth]{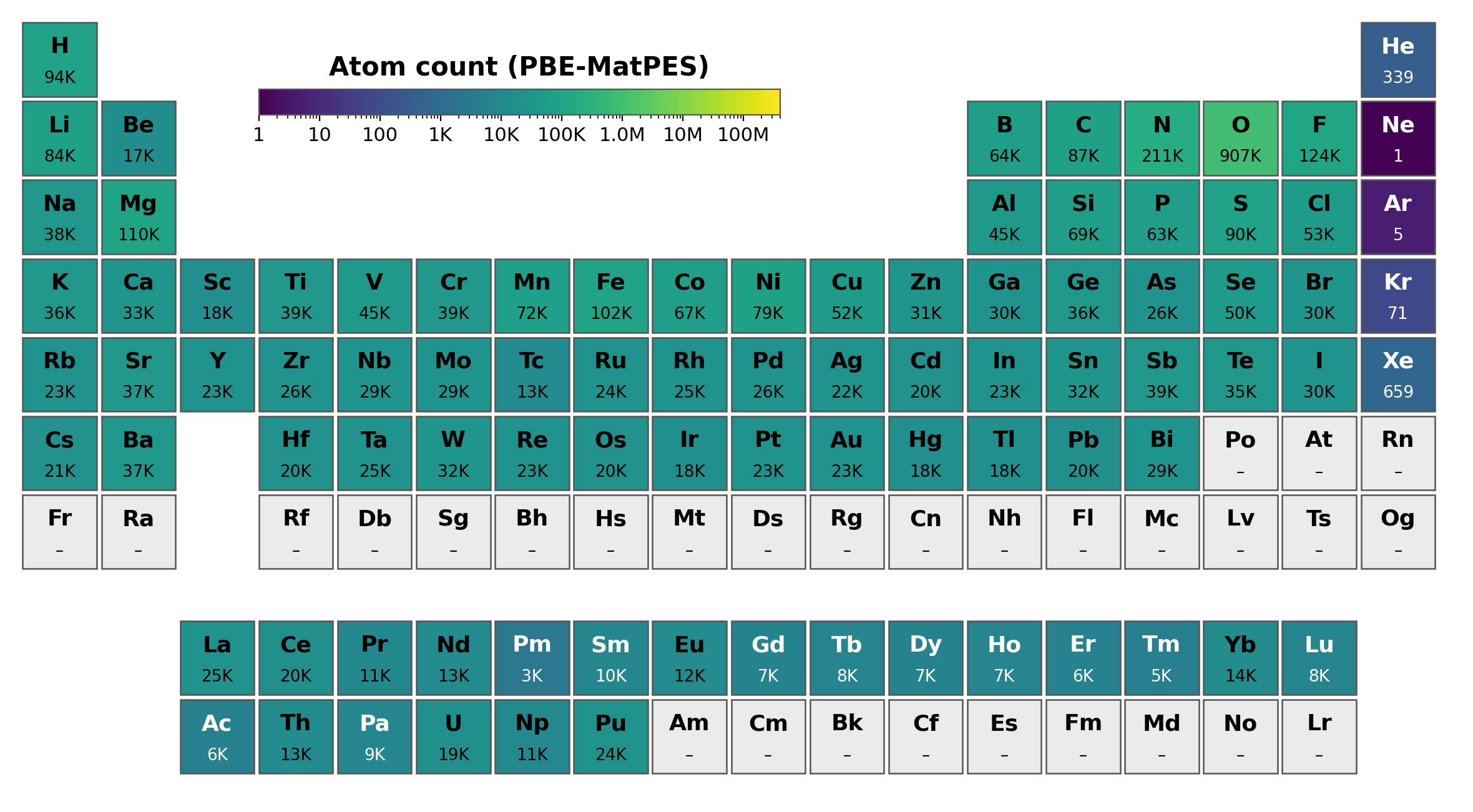}
  \caption{PBE-MatPES: total number of atoms of each element. A structure with many atoms of one element counts once in the map above and once per atom here, so the two differ wherever the stoichiometry is skewed. Color scales are shared across the databases of a functional, so tiles are directly comparable between the maps of this appendix.}
  \label{fig:app-pbe-matpes-atoms}
\end{figure}

\begin{figure}[H]
  \centering
  \includegraphics[width=\textwidth]{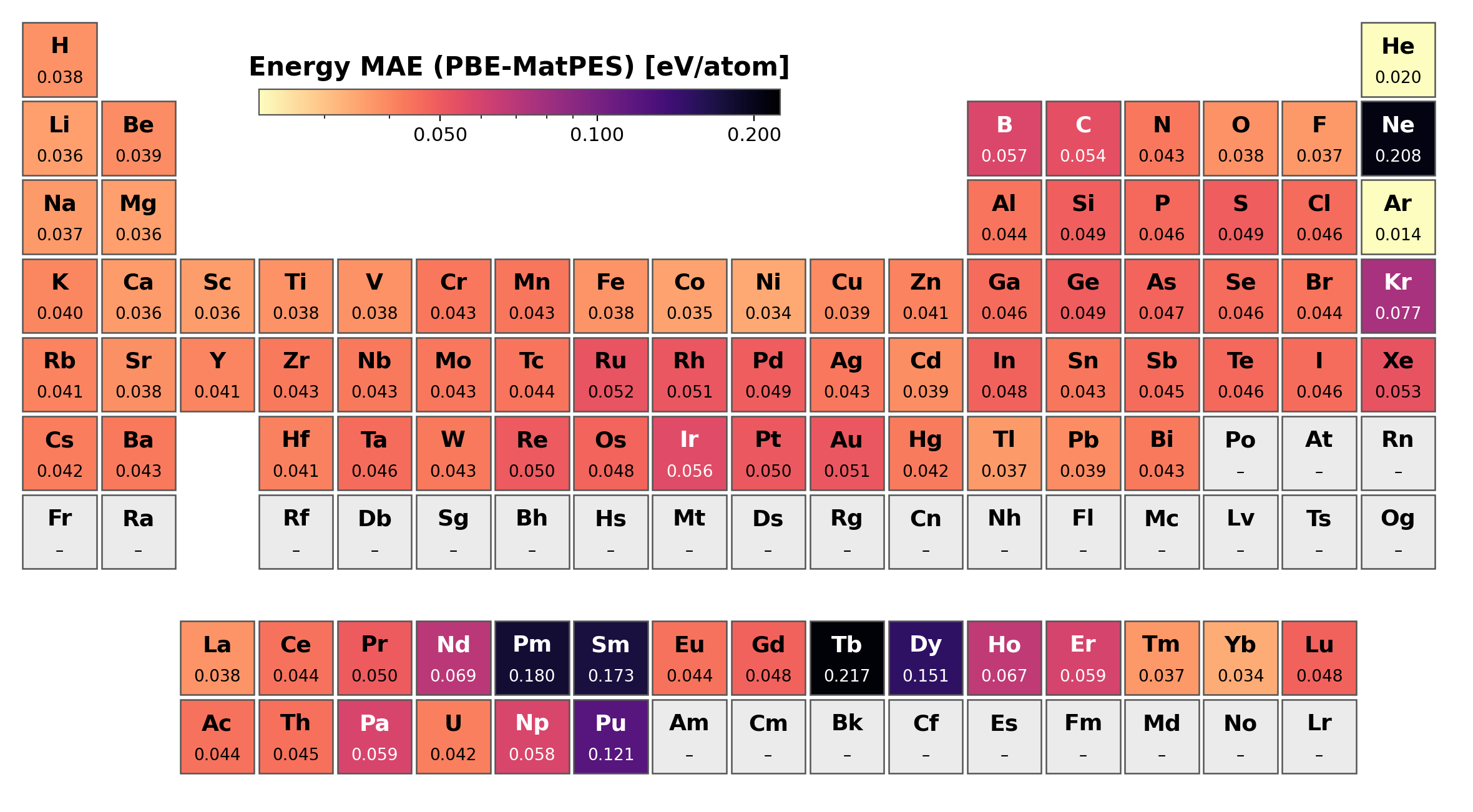}
  \caption{PBE-MatPES: per-element energy MAE (eV/atom) of the model trained on this database alone, which reaches 0.0463~eV/atom overall. Color scales are shared across the databases of a functional, so tiles are directly comparable between the maps of this appendix.}
  \label{fig:app-pbe-matpes-energy_mae}
\end{figure}

\begin{figure}[H]
  \centering
  \includegraphics[width=\textwidth]{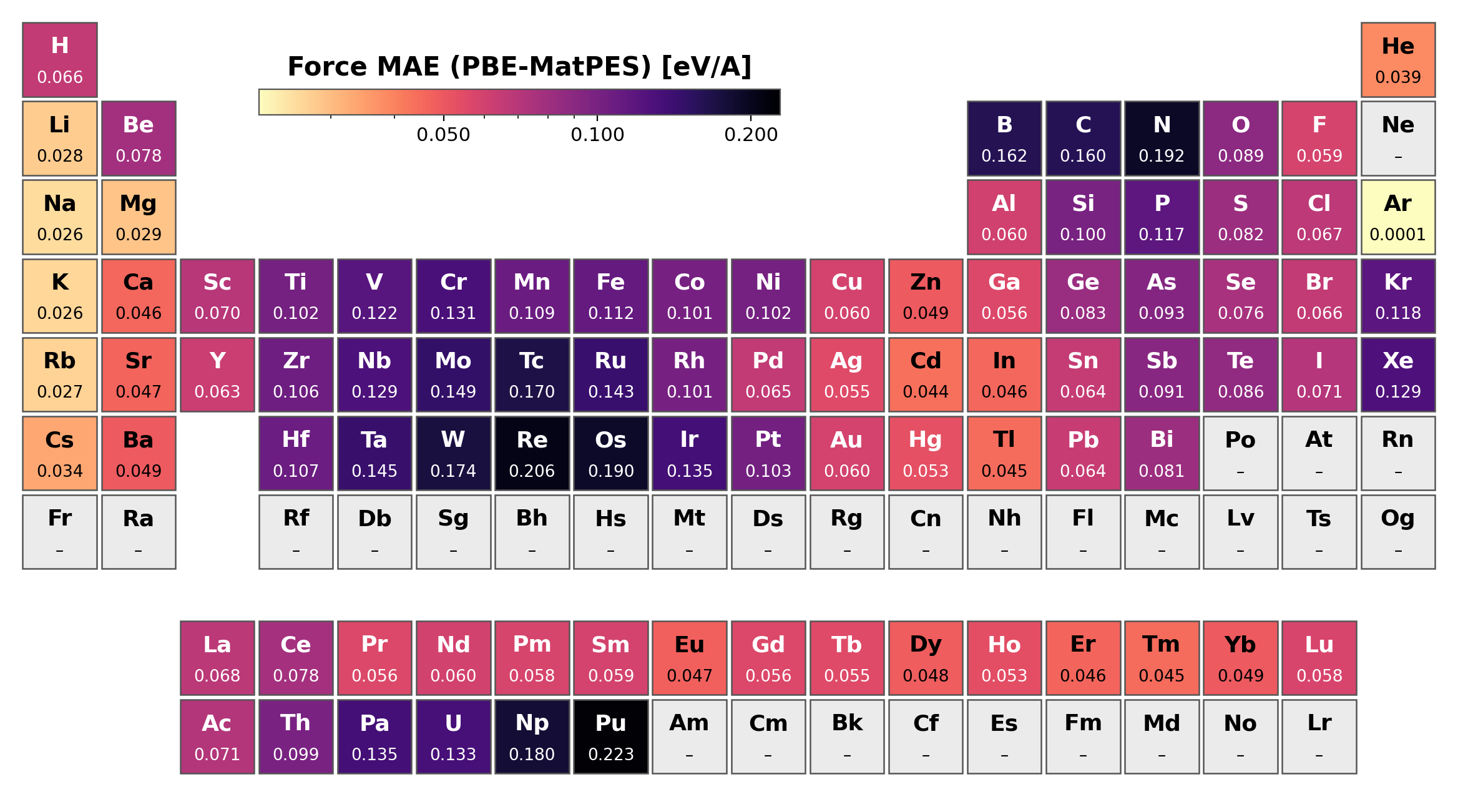}
  \caption{PBE-MatPES: per-element force MAE (eV/\AA) of the model trained on this database alone, which reaches 0.0930~eV/\AA{} overall. Color scales are shared across the databases of a functional, so tiles are directly comparable between the maps of this appendix.}
  \label{fig:app-pbe-matpes-force_mae}
\end{figure}

\begin{figure}[H]
  \centering
  \includegraphics[width=\textwidth]{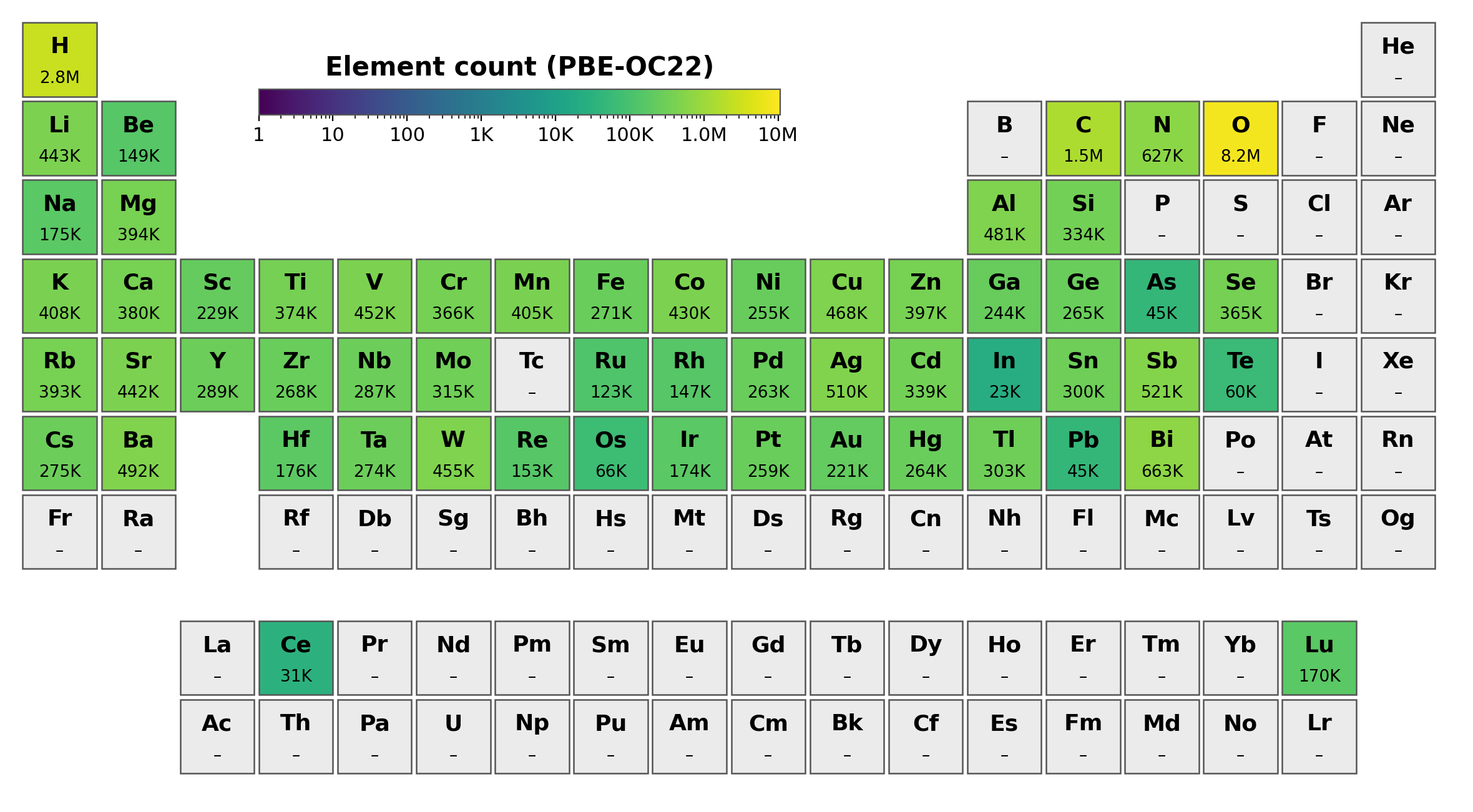}
  \caption{PBE-OC22: number of structures containing each element. 8,199,197 structures, 654,340,594 atoms, 57 elements. Color scales are shared across the databases of a functional, so tiles are directly comparable between the maps of this appendix.}
  \label{fig:app-pbe-oc22-structures}
\end{figure}

\begin{figure}[H]
  \centering
  \includegraphics[width=\textwidth]{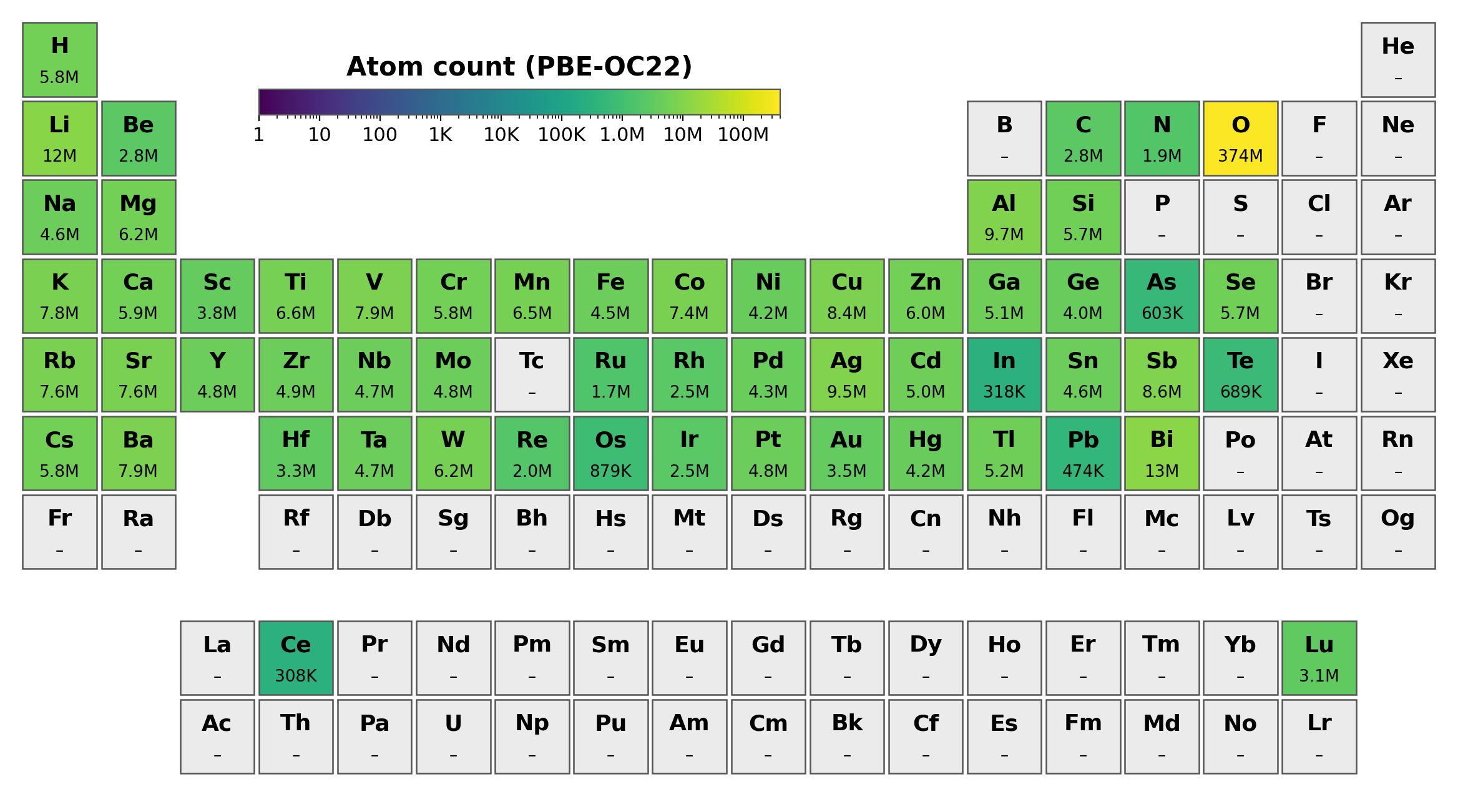}
  \caption{PBE-OC22: total number of atoms of each element. A structure with many atoms of one element counts once in the map above and once per atom here, so the two differ wherever the stoichiometry is skewed. Color scales are shared across the databases of a functional, so tiles are directly comparable between the maps of this appendix.}
  \label{fig:app-pbe-oc22-atoms}
\end{figure}

\begin{figure}[H]
  \centering
  \includegraphics[width=\textwidth]{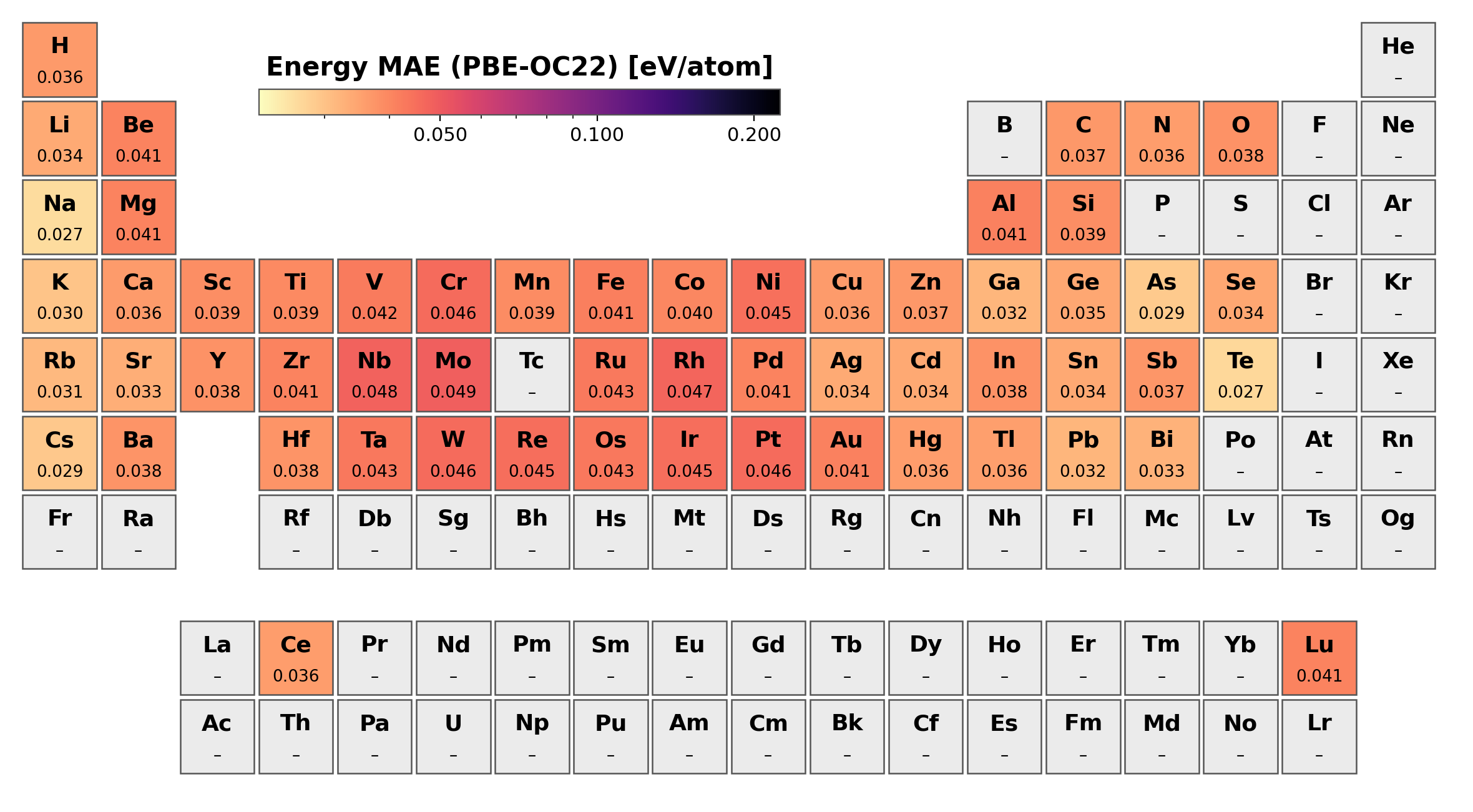}
  \caption{PBE-OC22: per-element energy MAE (eV/atom) of the model trained on this database alone, which reaches 0.0381~eV/atom overall. Color scales are shared across the databases of a functional, so tiles are directly comparable between the maps of this appendix.}
  \label{fig:app-pbe-oc22-energy_mae}
\end{figure}

\begin{figure}[H]
  \centering
  \includegraphics[width=\textwidth]{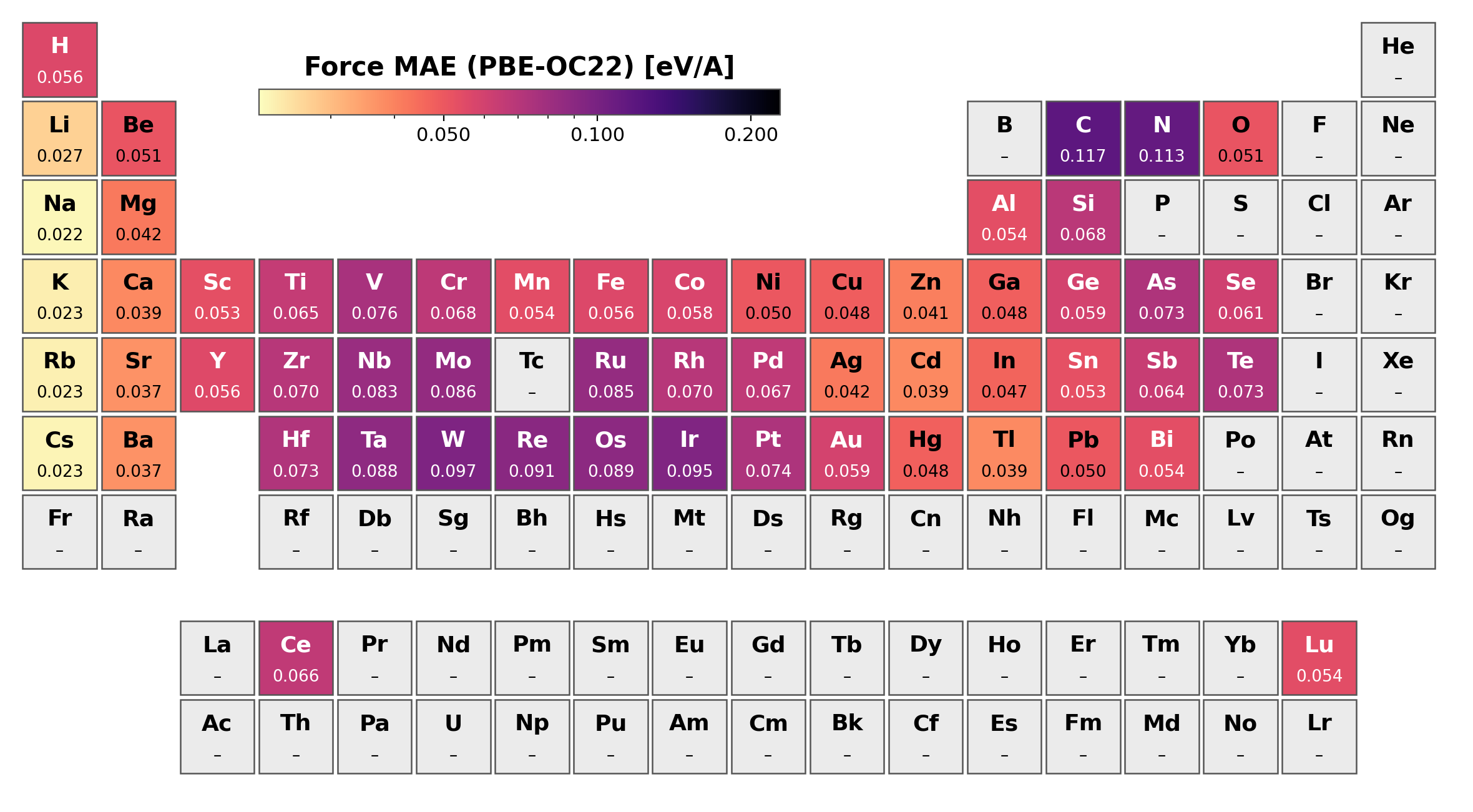}
  \caption{PBE-OC22: per-element force MAE (eV/\AA) of the model trained on this database alone, which reaches 0.0528~eV/\AA{} overall. Color scales are shared across the databases of a functional, so tiles are directly comparable between the maps of this appendix.}
  \label{fig:app-pbe-oc22-force_mae}
\end{figure}

\begin{figure}[H]
  \centering
  \includegraphics[width=\textwidth]{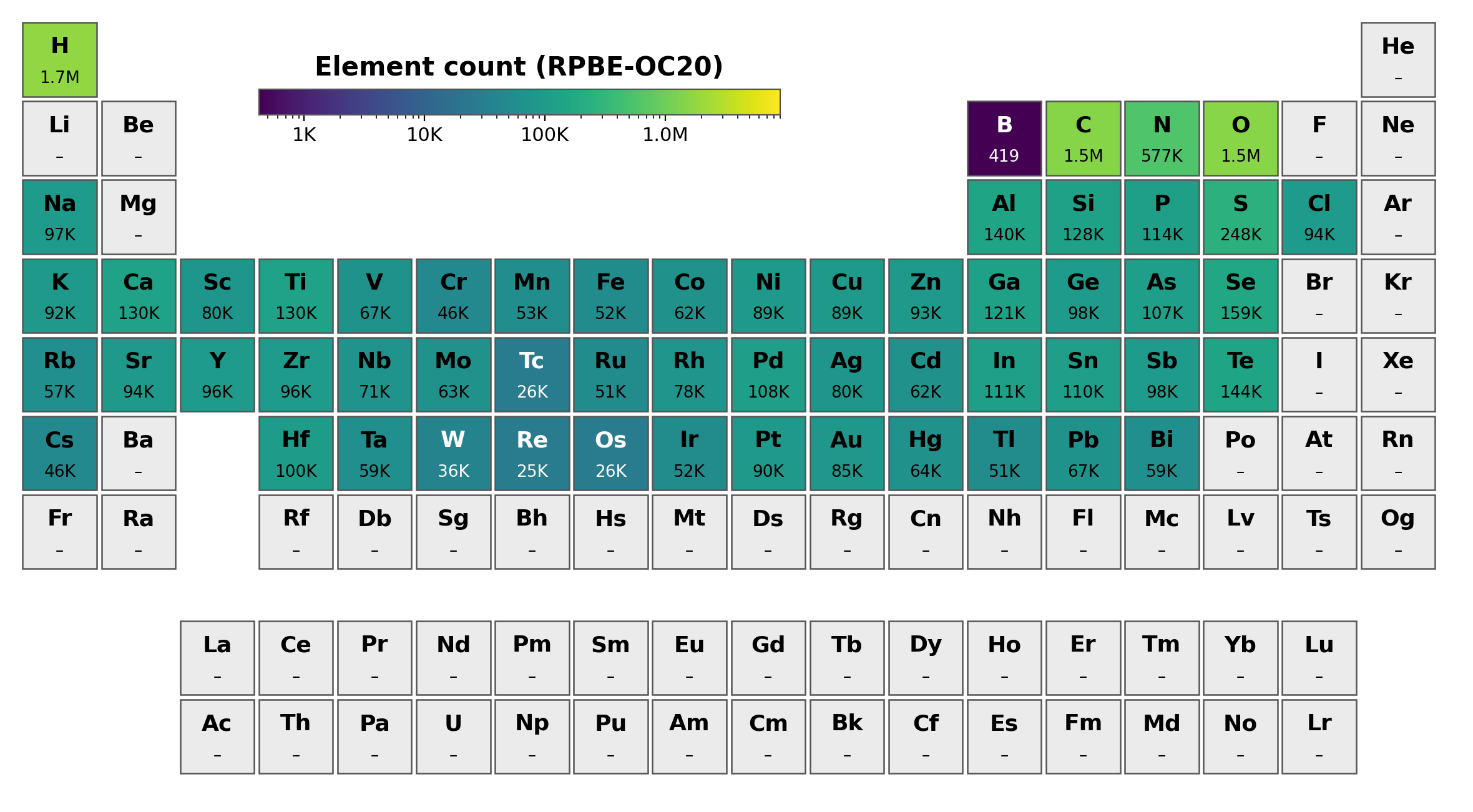}
  \caption{RPBE-OC20: number of structures containing each element. 1,999,572 structures, 146,467,030 atoms, 56 elements. Color scales are shared across the databases of a functional, so tiles are directly comparable between the maps of this appendix.}
  \label{fig:app-rpbe-oc20-structures}
\end{figure}

\begin{figure}[H]
  \centering
  \includegraphics[width=\textwidth]{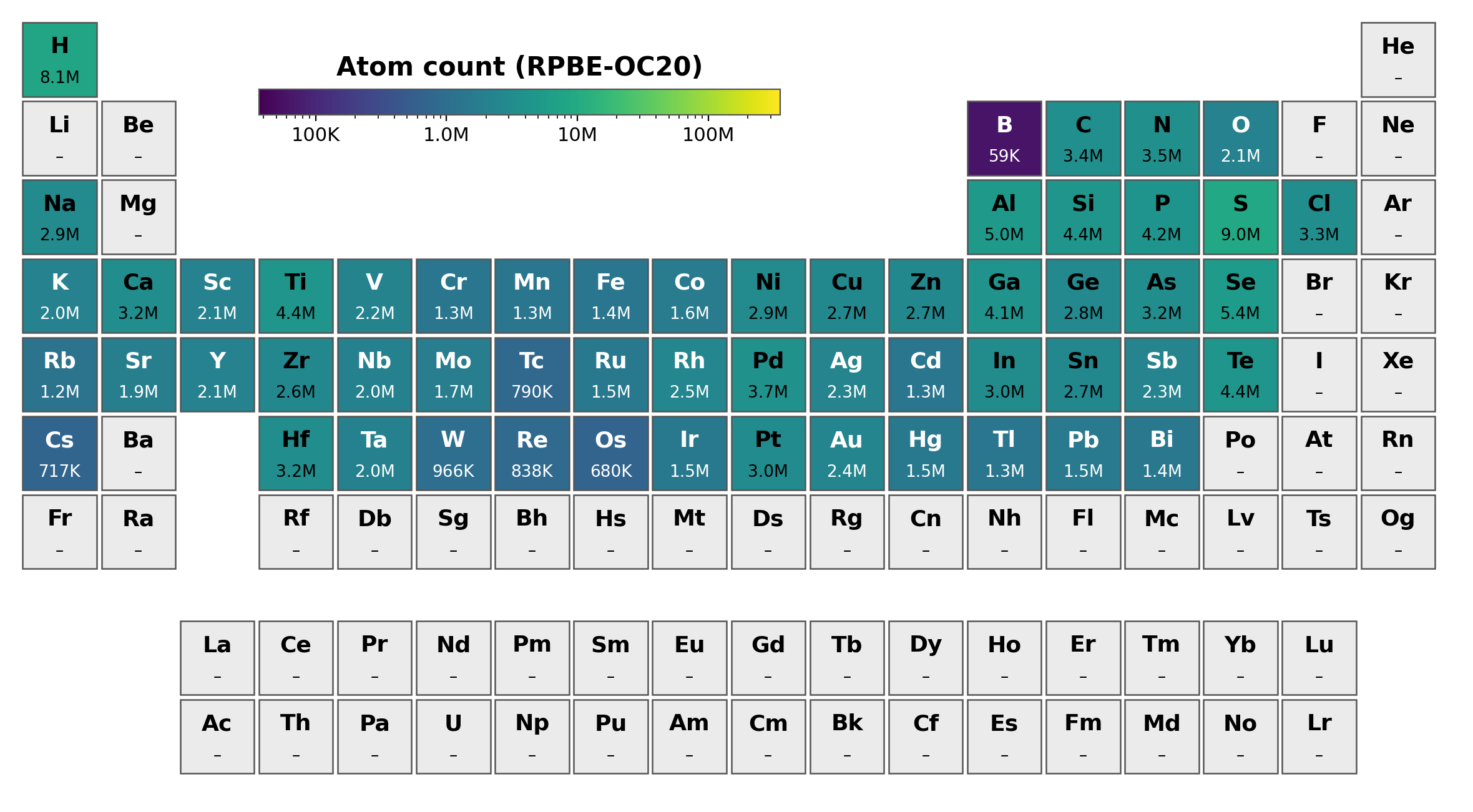}
  \caption{RPBE-OC20: total number of atoms of each element. A structure with many atoms of one element counts once in the map above and once per atom here, so the two differ wherever the stoichiometry is skewed. Color scales are shared across the databases of a functional, so tiles are directly comparable between the maps of this appendix.}
  \label{fig:app-rpbe-oc20-atoms}
\end{figure}

\begin{figure}[H]
  \centering
  \includegraphics[width=\textwidth]{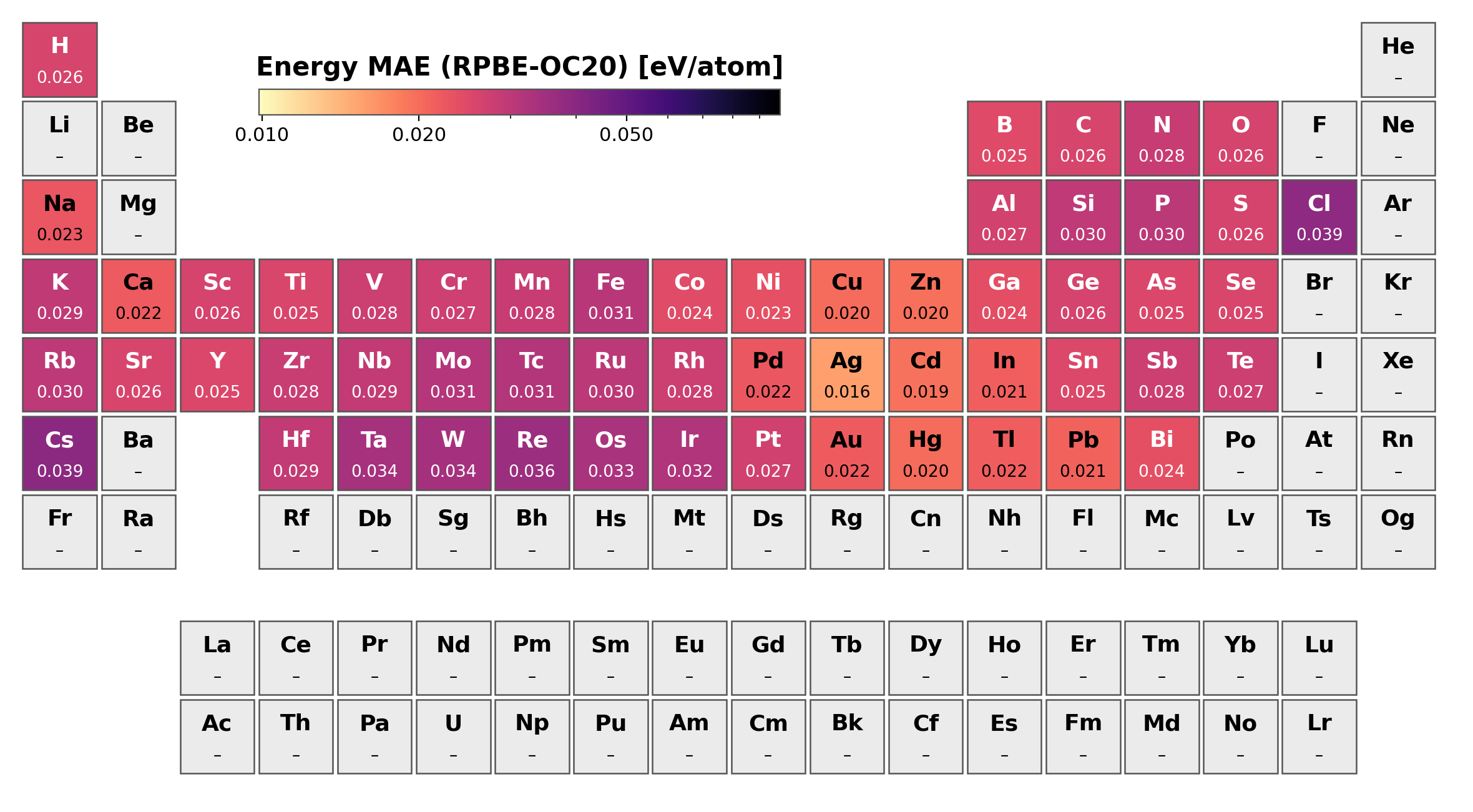}
  \caption{RPBE-OC20: per-element energy MAE (eV/atom) of the model trained on this database alone, which reaches 0.0263~eV/atom overall. Color scales are shared across the databases of a functional, so tiles are directly comparable between the maps of this appendix.}
  \label{fig:app-rpbe-oc20-energy_mae}
\end{figure}

\begin{figure}[H]
  \centering
  \includegraphics[width=\textwidth]{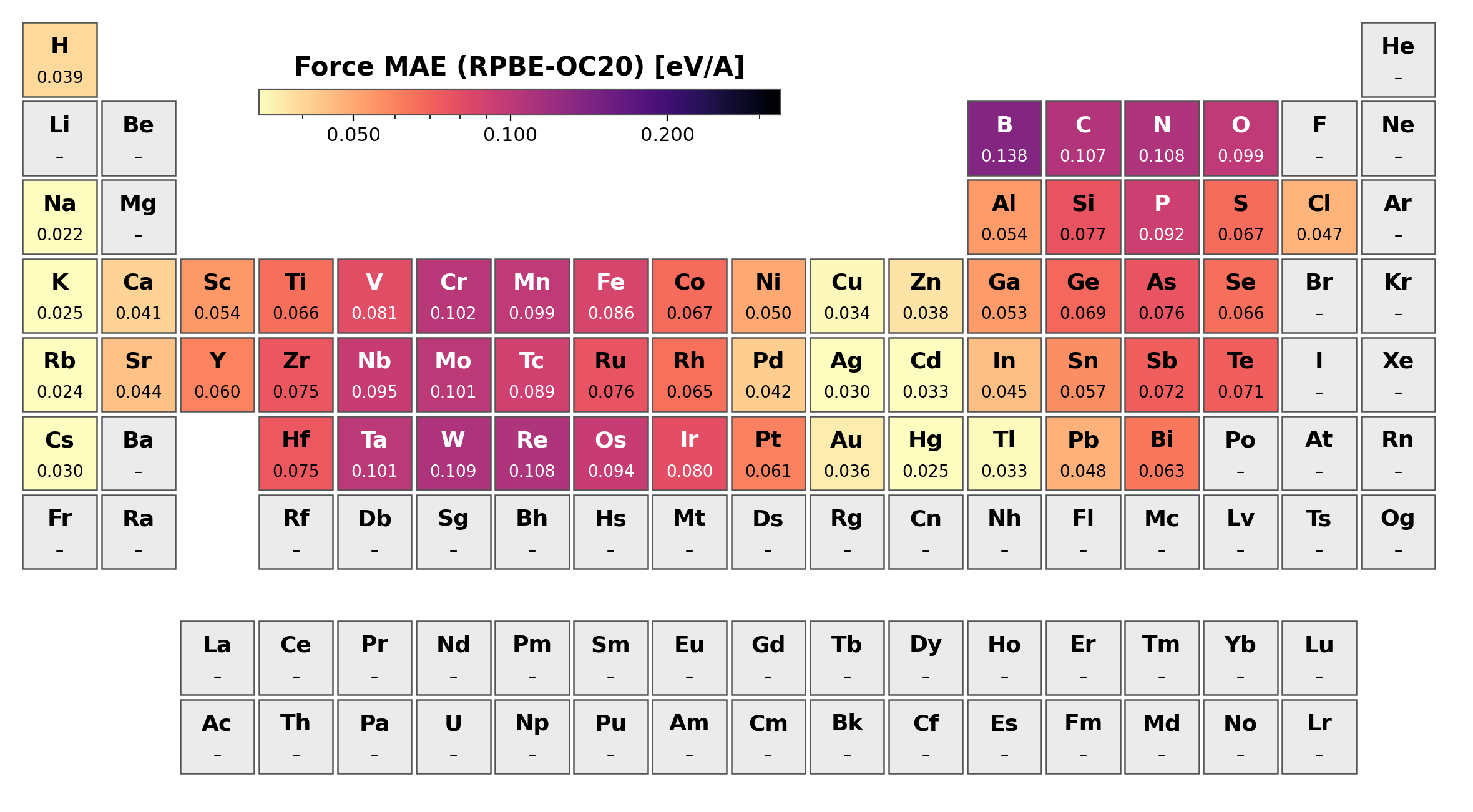}
  \caption{RPBE-OC20: per-element force MAE (eV/\AA) of the model trained on this database alone, which reaches 0.0630~eV/\AA{} overall. Color scales are shared across the databases of a functional, so tiles are directly comparable between the maps of this appendix.}
  \label{fig:app-rpbe-oc20-force_mae}
\end{figure}

\begin{figure}[H]
  \centering
  \includegraphics[width=\textwidth]{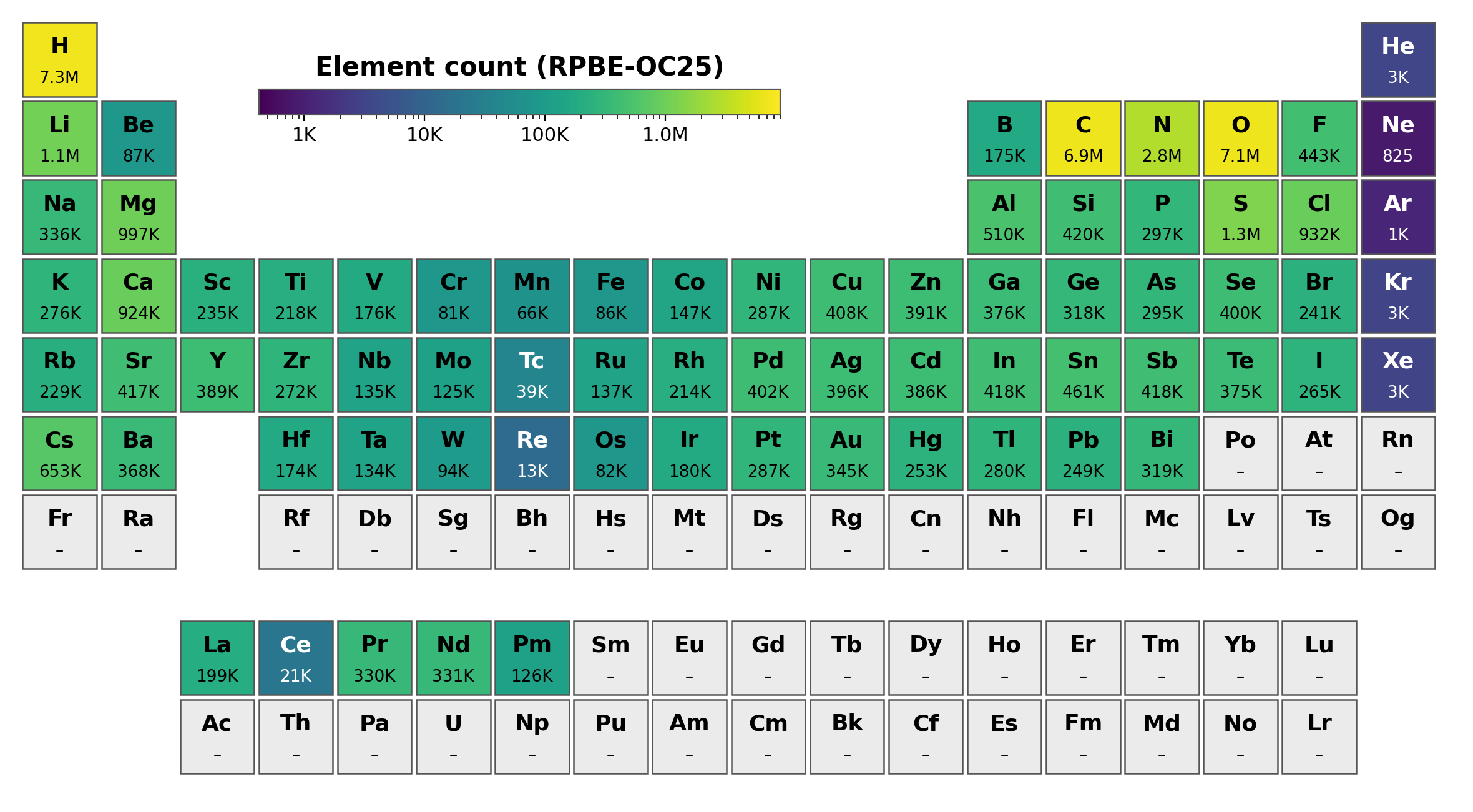}
  \caption{RPBE-OC25: number of structures containing each element. 7,370,461 structures, 1,064,330,294 atoms, 73 elements. Color scales are shared across the databases of a functional, so tiles are directly comparable between the maps of this appendix.}
  \label{fig:app-rpbe-oc25-structures}
\end{figure}

\begin{figure}[H]
  \centering
  \includegraphics[width=\textwidth]{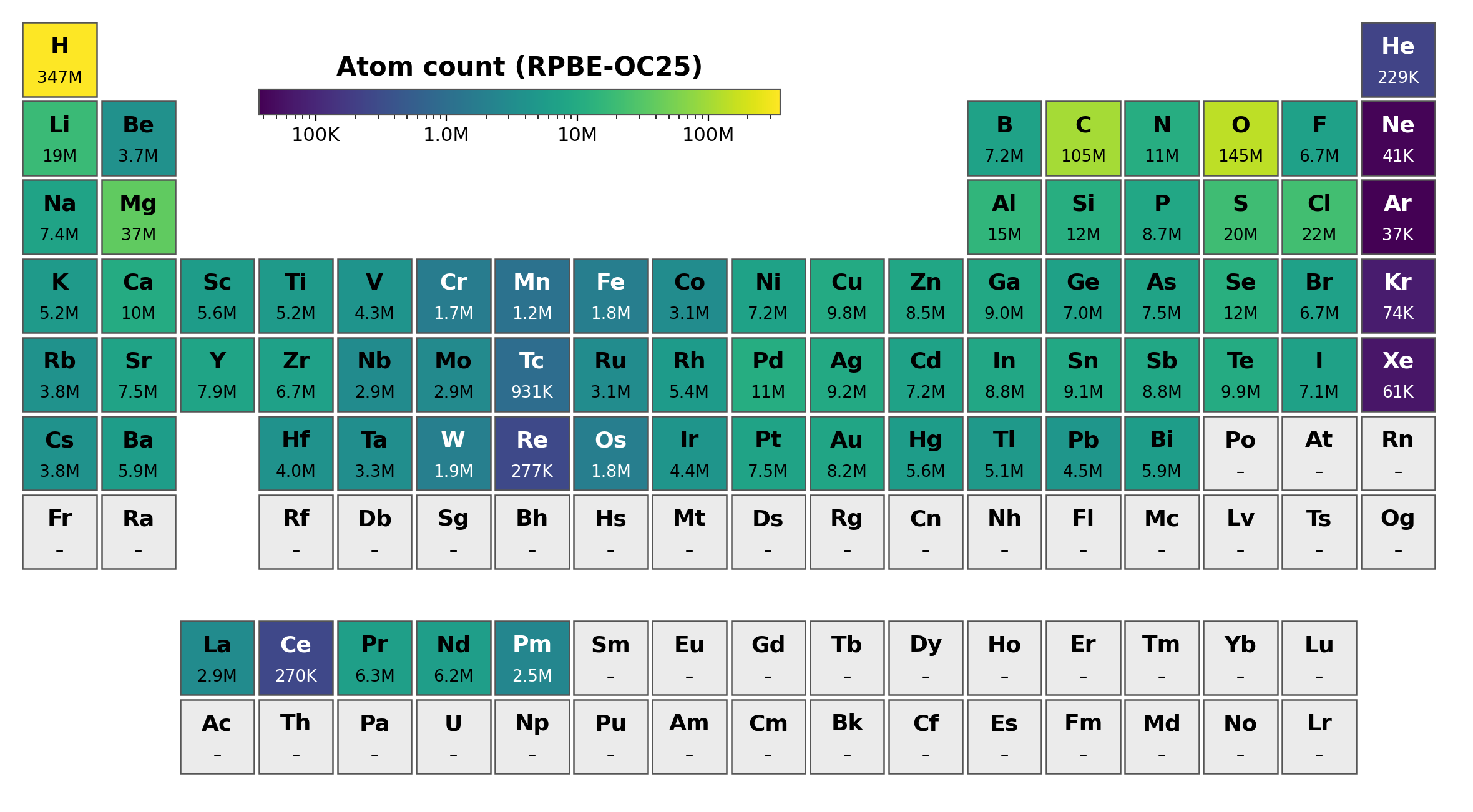}
  \caption{RPBE-OC25: total number of atoms of each element. A structure with many atoms of one element counts once in the map above and once per atom here, so the two differ wherever the stoichiometry is skewed. Color scales are shared across the databases of a functional, so tiles are directly comparable between the maps of this appendix.}
  \label{fig:app-rpbe-oc25-atoms}
\end{figure}

\begin{figure}[H]
  \centering
  \includegraphics[width=\textwidth]{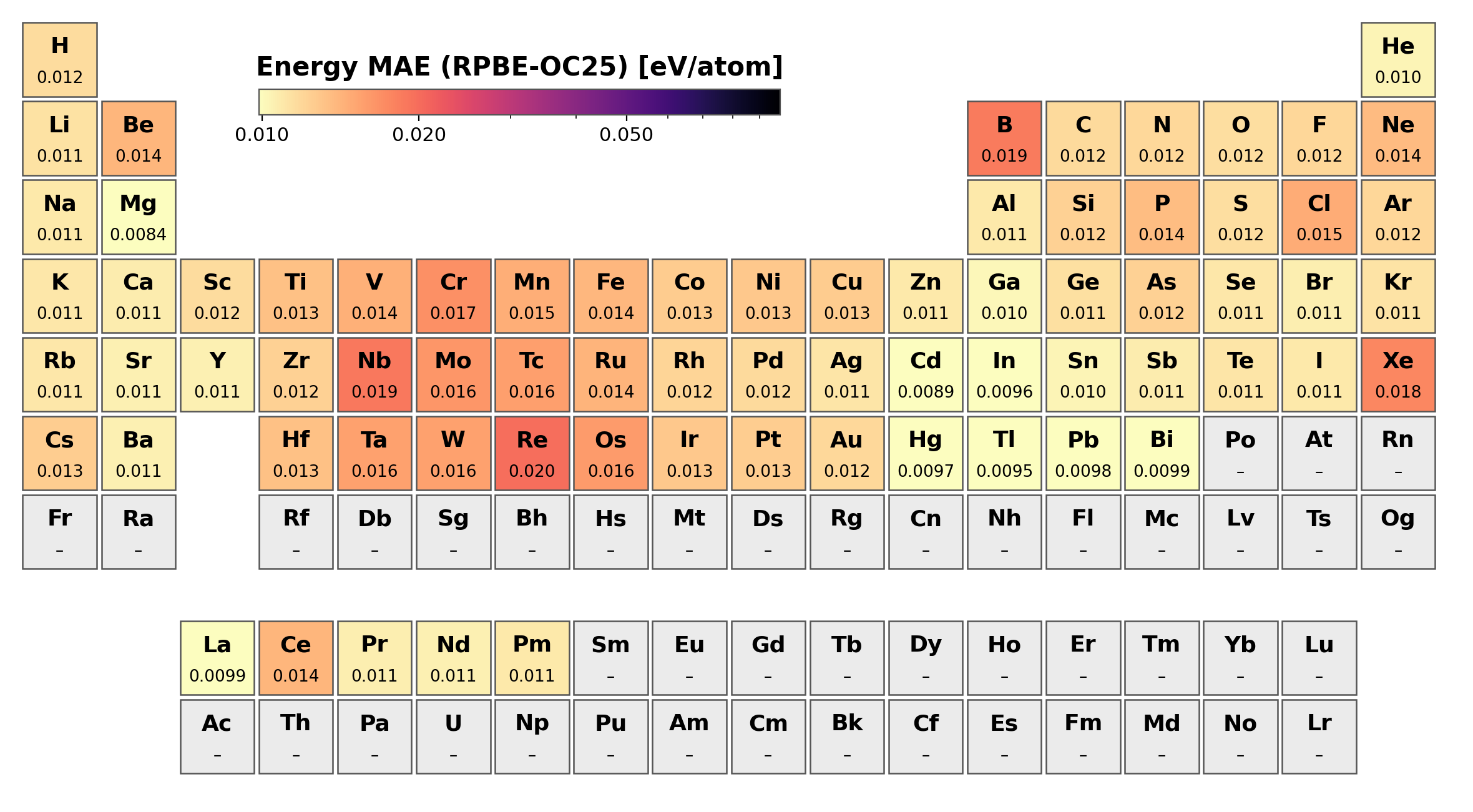}
  \caption{RPBE-OC25: per-element energy MAE (eV/atom) of the model trained on this database alone, which reaches 0.0116~eV/atom overall. Color scales are shared across the databases of a functional, so tiles are directly comparable between the maps of this appendix.}
  \label{fig:app-rpbe-oc25-energy_mae}
\end{figure}

\begin{figure}[H]
  \centering
  \includegraphics[width=\textwidth]{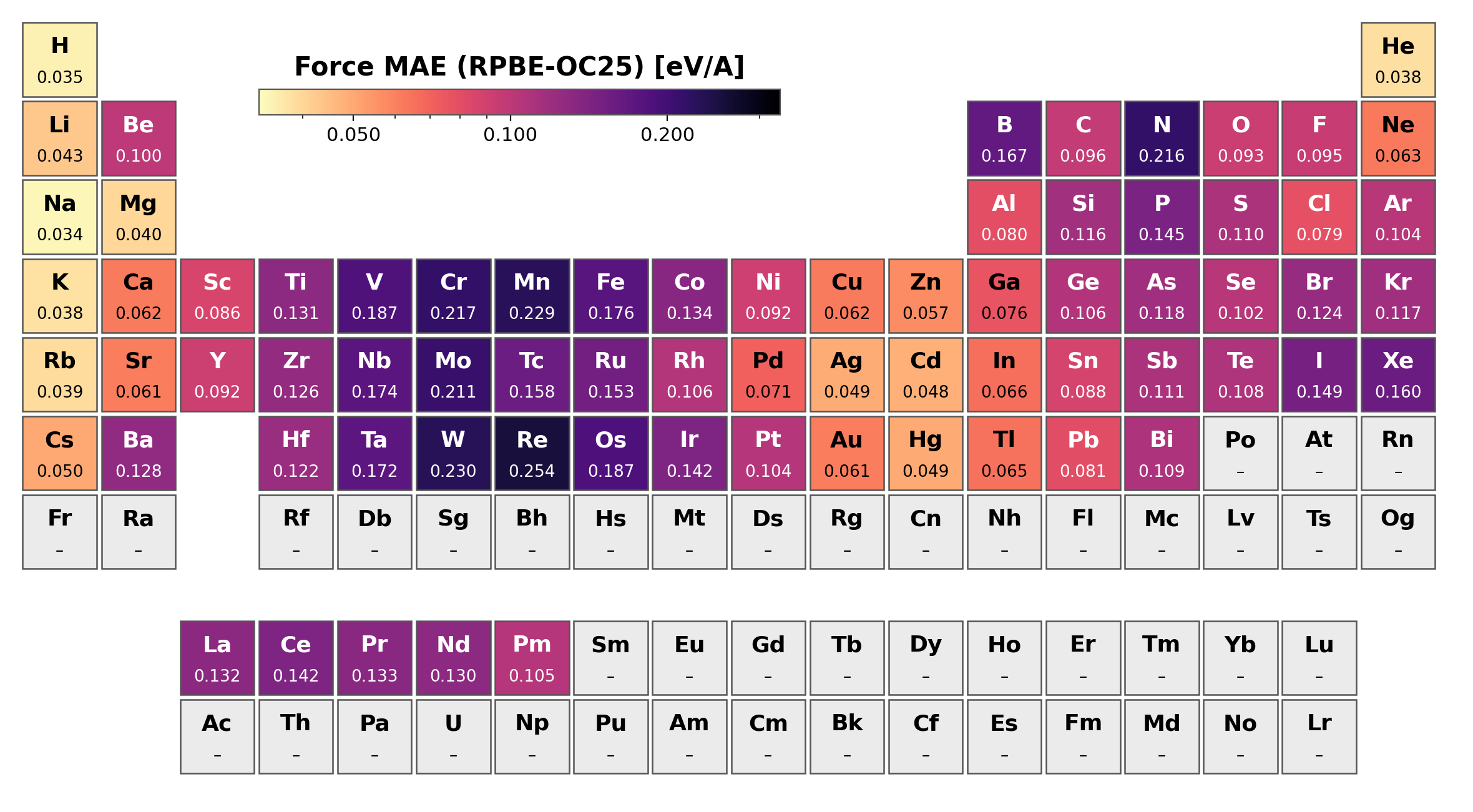}
  \caption{RPBE-OC25: per-element force MAE (eV/\AA) of the model trained on this database alone, which reaches 0.0753~eV/\AA{} overall. Color scales are shared across the databases of a functional, so tiles are directly comparable between the maps of this appendix.}
  \label{fig:app-rpbe-oc25-force_mae}
\end{figure}

\begin{figure}[H]
  \centering
  \includegraphics[width=\textwidth]{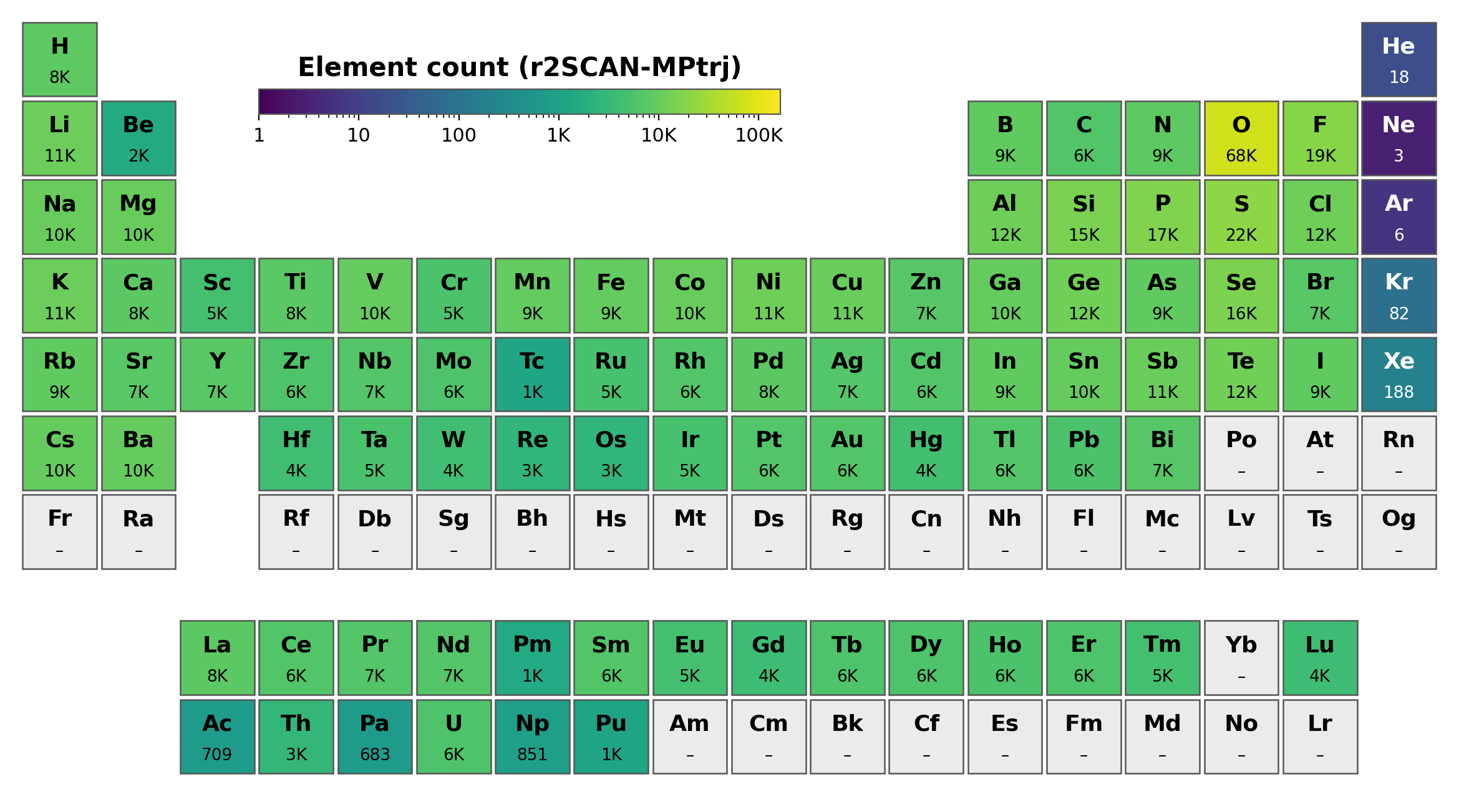}
  \caption{r2SCAN-MPtrj: number of structures containing each element. 238,241 structures, 5,022,951 atoms, 88 elements. Color scales are shared across the databases of a functional, so tiles are directly comparable between the maps of this appendix.}
  \label{fig:app-r2scan-mptrj-structures}
\end{figure}

\begin{figure}[H]
  \centering
  \includegraphics[width=\textwidth]{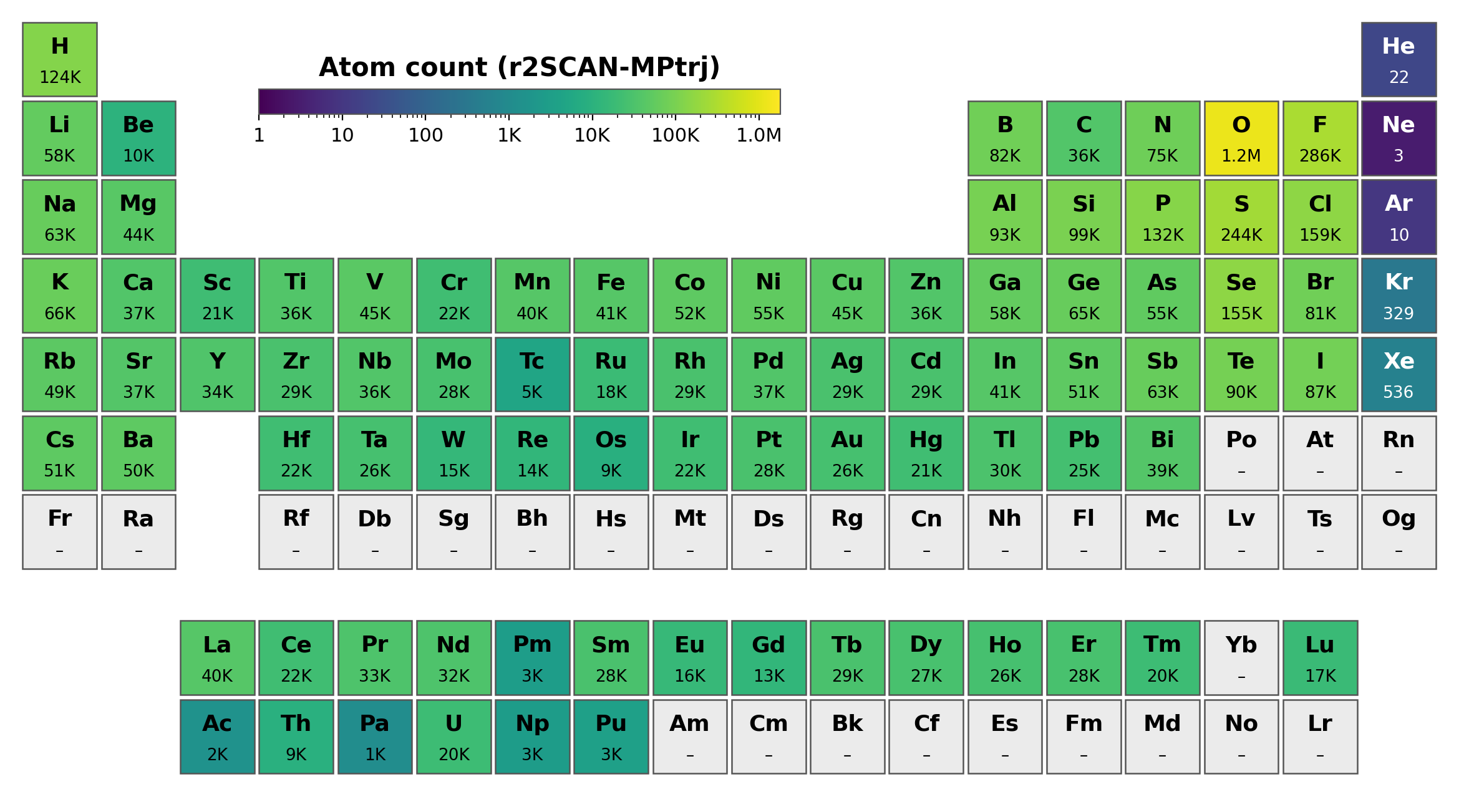}
  \caption{r2SCAN-MPtrj: total number of atoms of each element. A structure with many atoms of one element counts once in the map above and once per atom here, so the two differ wherever the stoichiometry is skewed. Color scales are shared across the databases of a functional, so tiles are directly comparable between the maps of this appendix.}
  \label{fig:app-r2scan-mptrj-atoms}
\end{figure}

\begin{figure}[H]
  \centering
  \includegraphics[width=\textwidth]{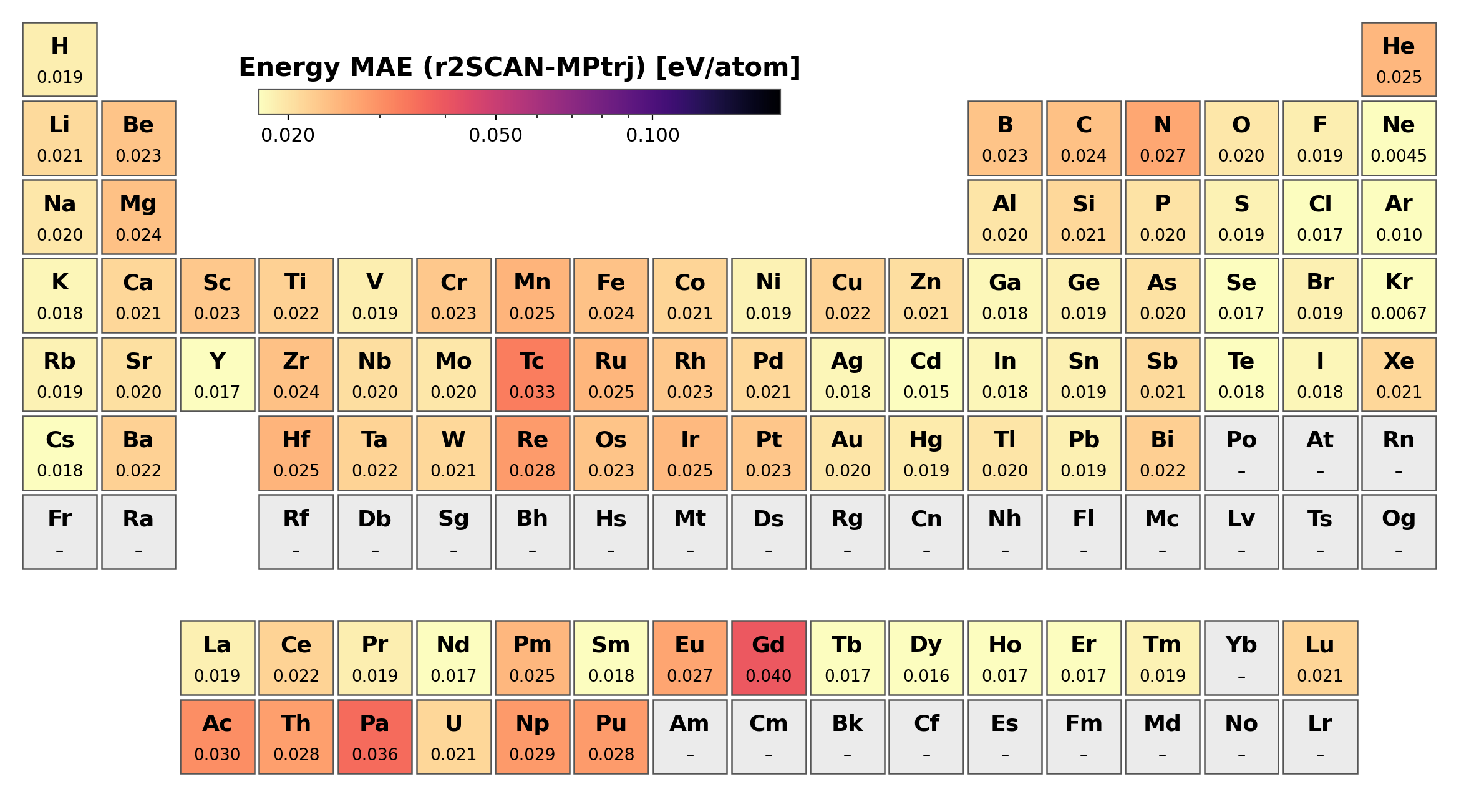}
  \caption{r2SCAN-MPtrj: per-element energy MAE (eV/atom) of the model trained on this database alone, which reaches 0.0206~eV/atom overall. Color scales are shared across the databases of a functional, so tiles are directly comparable between the maps of this appendix.}
  \label{fig:app-r2scan-mptrj-energy_mae}
\end{figure}

\begin{figure}[H]
  \centering
  \includegraphics[width=\textwidth]{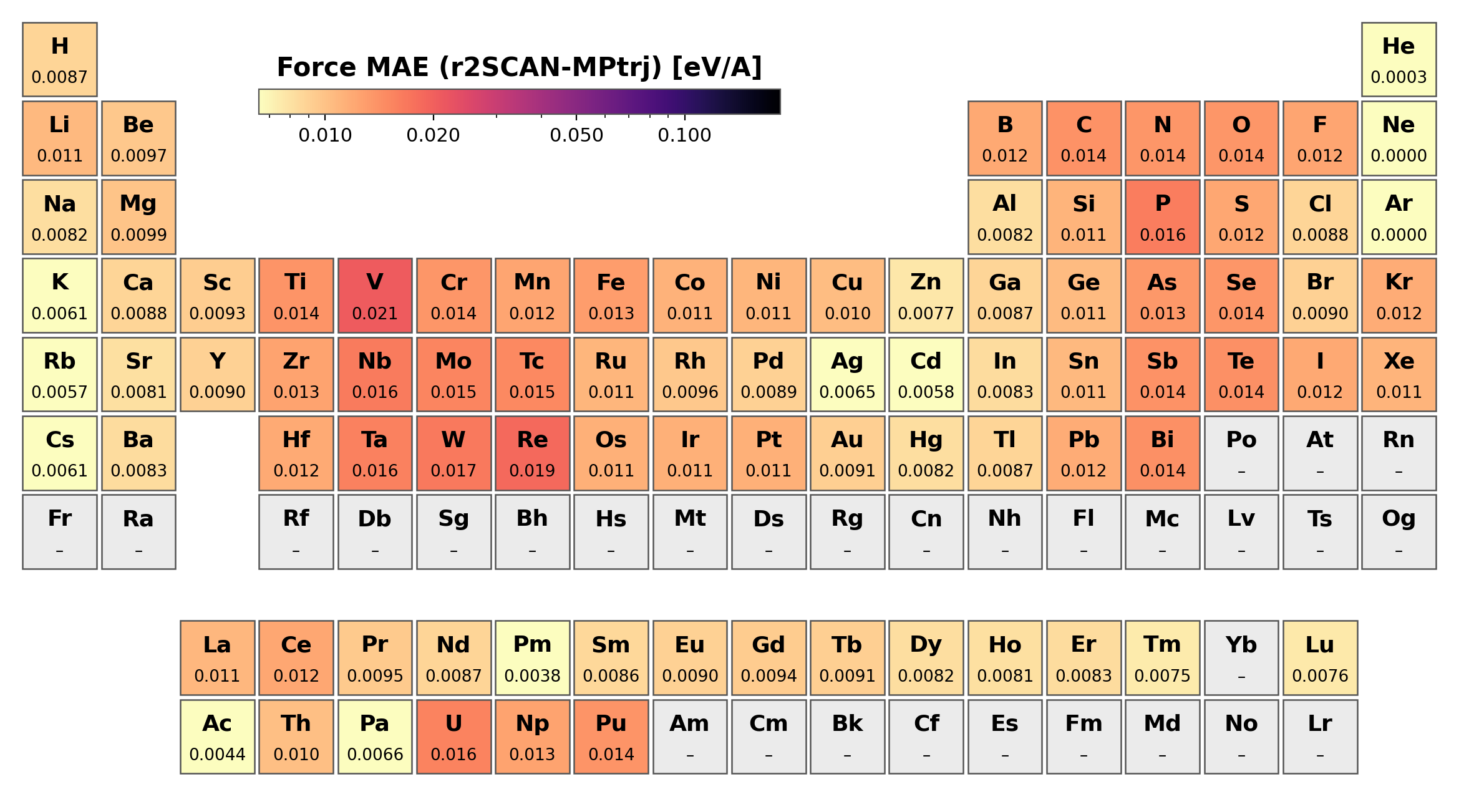}
  \caption{r2SCAN-MPtrj: per-element force MAE (eV/\AA) of the model trained on this database alone, which reaches 0.0117~eV/\AA{} overall. Color scales are shared across the databases of a functional, so tiles are directly comparable between the maps of this appendix.}
  \label{fig:app-r2scan-mptrj-force_mae}
\end{figure}

\begin{figure}[H]
  \centering
  \includegraphics[width=\textwidth]{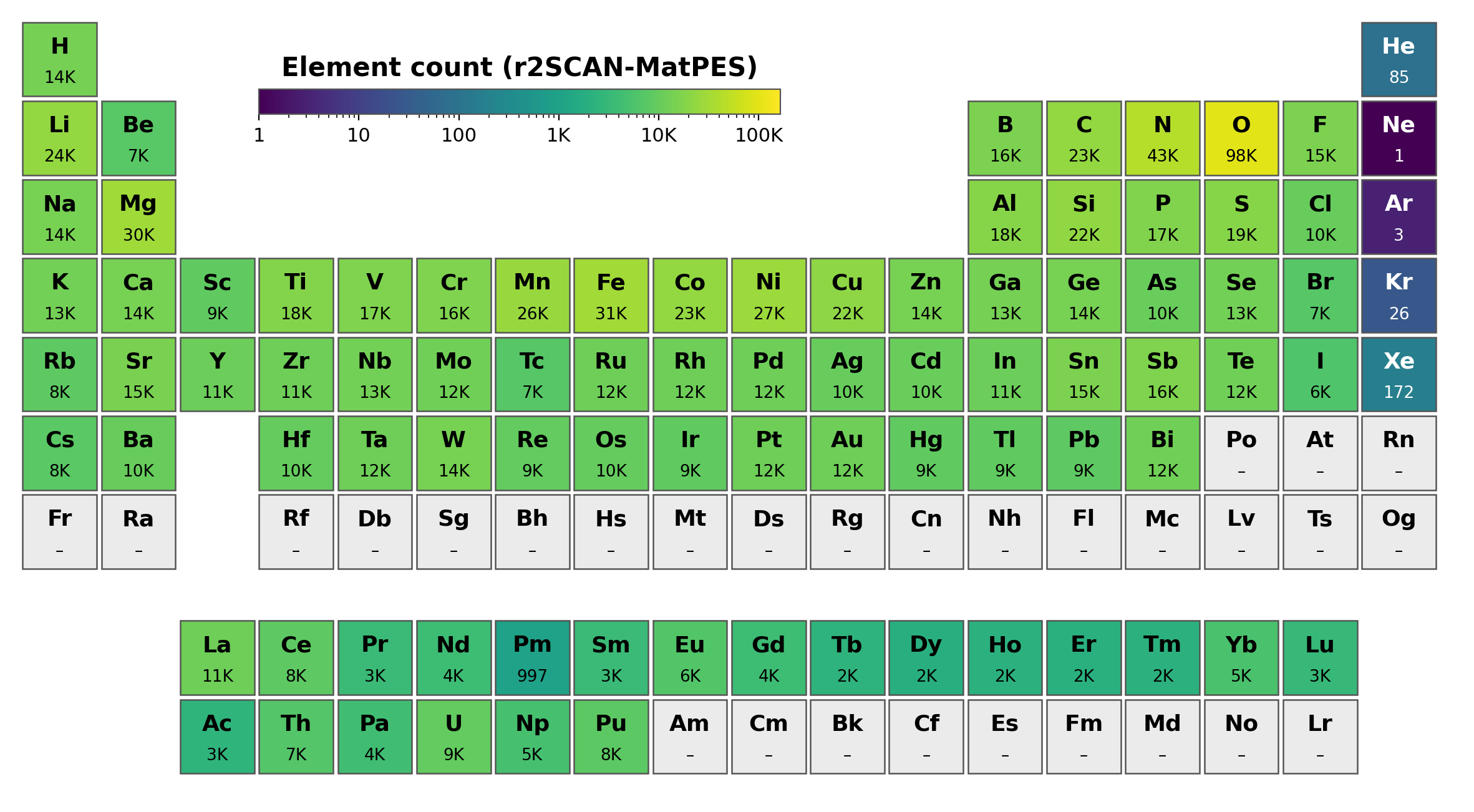}
  \caption{r2SCAN-MatPES: number of structures containing each element. 387,306 structures, 3,040,410 atoms, 89 elements. Color scales are shared across the databases of a functional, so tiles are directly comparable between the maps of this appendix.}
  \label{fig:app-r2scan-matpes-structures}
\end{figure}

\begin{figure}[H]
  \centering
  \includegraphics[width=\textwidth]{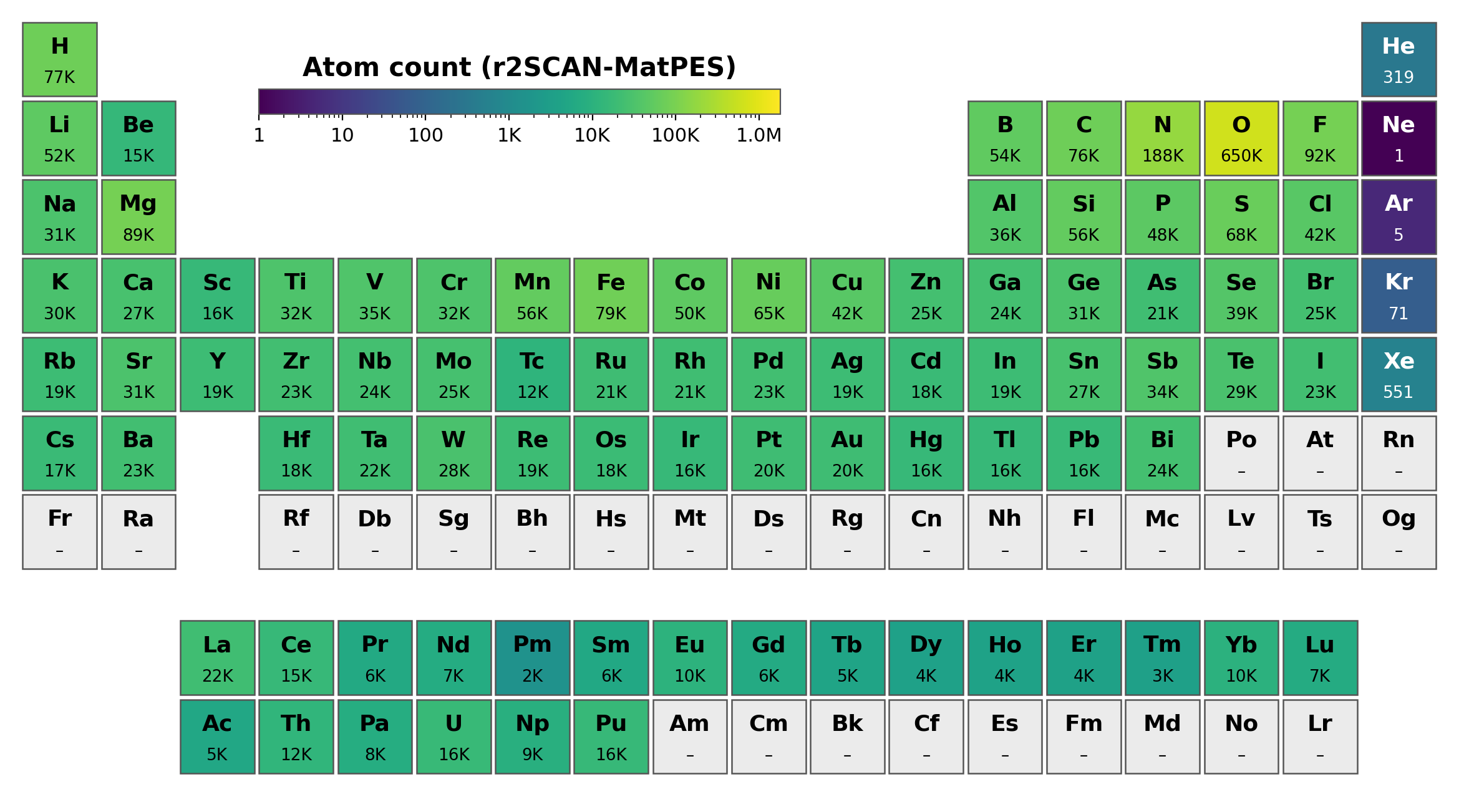}
  \caption{r2SCAN-MatPES: total number of atoms of each element. A structure with many atoms of one element counts once in the map above and once per atom here, so the two differ wherever the stoichiometry is skewed. Color scales are shared across the databases of a functional, so tiles are directly comparable between the maps of this appendix.}
  \label{fig:app-r2scan-matpes-atoms}
\end{figure}

\begin{figure}[H]
  \centering
  \includegraphics[width=\textwidth]{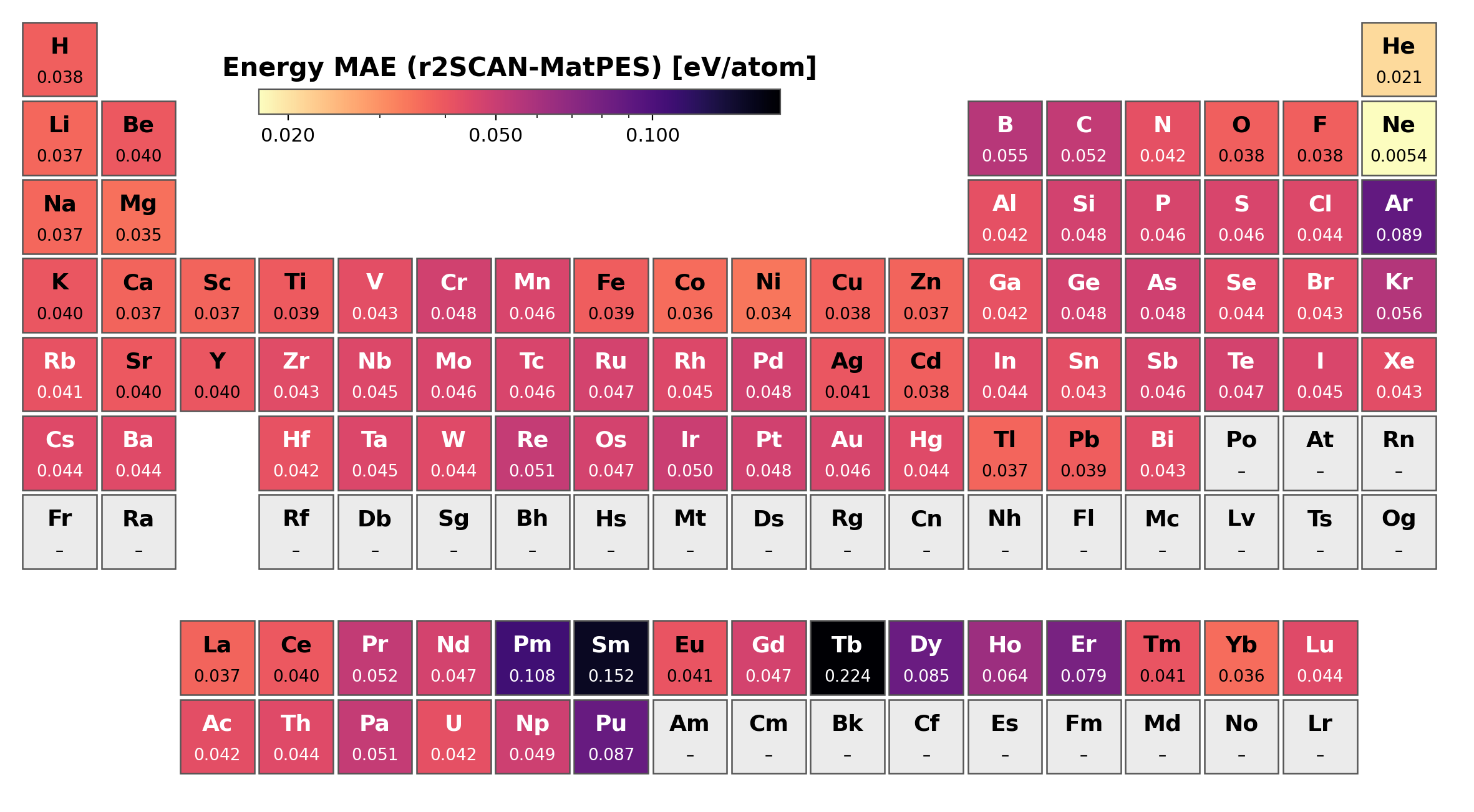}
  \caption{r2SCAN-MatPES: per-element energy MAE (eV/atom) of the model trained on this database alone, which reaches 0.0446~eV/atom overall. Color scales are shared across the databases of a functional, so tiles are directly comparable between the maps of this appendix.}
  \label{fig:app-r2scan-matpes-energy_mae}
\end{figure}

\begin{figure}[H]
  \centering
  \includegraphics[width=\textwidth]{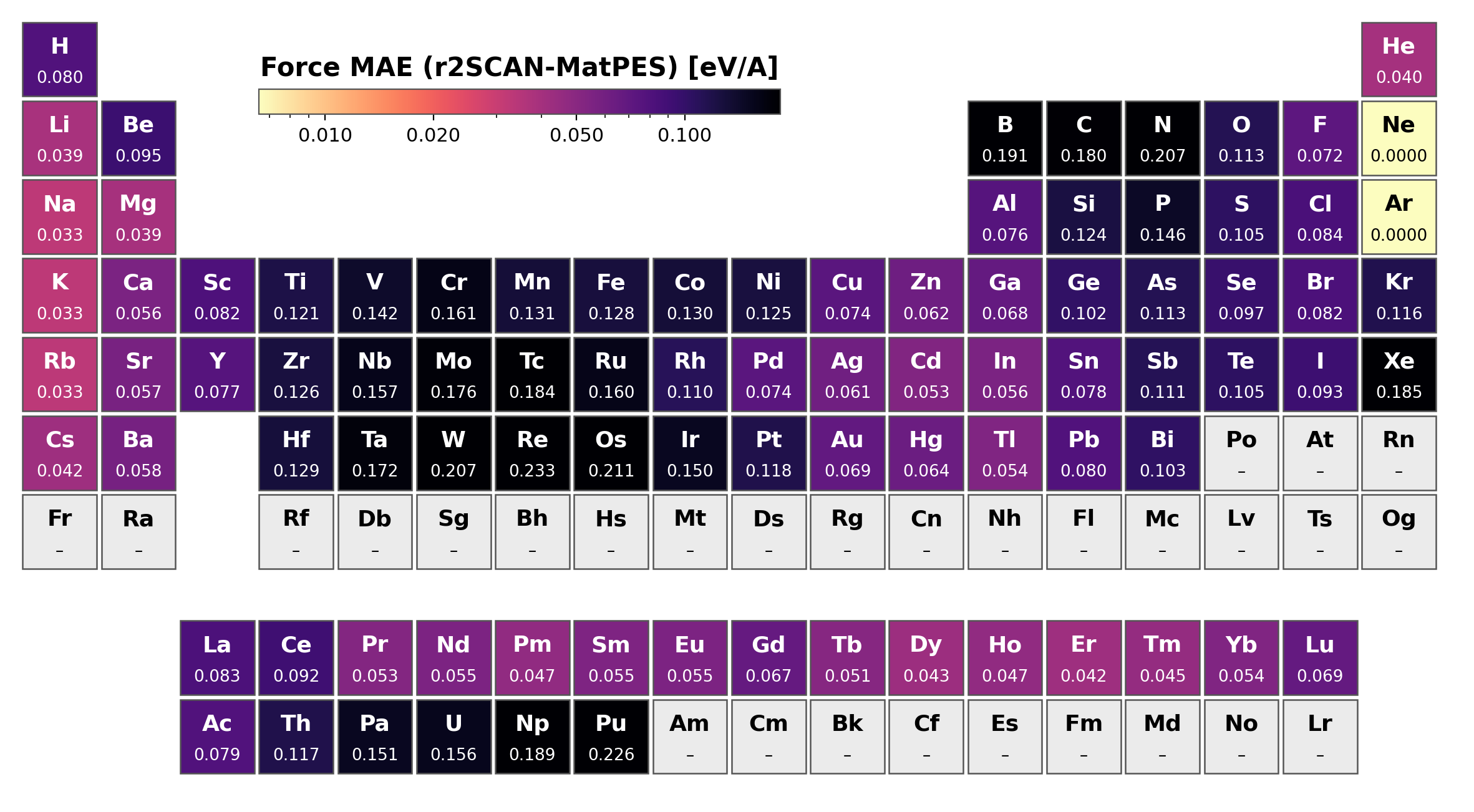}
  \caption{r2SCAN-MatPES: per-element force MAE (eV/\AA) of the model trained on this database alone, which reaches 0.1129~eV/\AA{} overall. Color scales are shared across the databases of a functional, so tiles are directly comparable between the maps of this appendix.}
  \label{fig:app-r2scan-matpes-force_mae}
\end{figure}

Figures~\ref{fig:app-pbe-mptrj-structures}--\ref{fig:app-r2scan-matpes-force_mae} repeat the element-resolved maps
for each source database on its own, using the model trained on that database
alone. 

\clearpage  

\clearpage  

\clearpage  

\clearpage  

\clearpage  

\clearpage  

\clearpage  

\clearpage  

\clearpage  

\clearpage
\printbibliography